\documentclass[a4paper,11pt]{article}

\usepackage{jheppub}
\usepackage[table]{xcolor}
\usepackage{slashbox}
\allowdisplaybreaks
\usepackage{graphicx}
\usepackage{epstopdf}

\newcommand{\be}{\begin{equation}}
	\newcommand{\ee}{\end{equation}}
\newcommand{\bea}{\begin{eqnarray}}
	\newcommand{\eea}{\end{eqnarray}}
\newcommand{\nn}{\nonumber}
\newcommand{\bdm}{\begin{displaymath}}
	\newcommand{\edm}{\end{displaymath}}

\title{From the EFT of Spinning Gravitating Objects 
	 \\to Poincar\'e and Gauge Invariance at the
	 \\4.5PN Precision Frontier}

\author[a]{Mich\`ele Levi,}
\author[b]{Roger Morales,}
\author[c]{Zhewei Yin}

\affiliation[a]{Mathematical Institute, University of Oxford, 
	\\Woodstock Road, Oxford OX2 6GG, United Kingdom}
\affiliation[b]{Niels Bohr Institute, University of Copenhagen,
	\\Blegdamsvej 17, 2100 Copenhagen, Denmark}
\affiliation[c]{Department of Physics and Astronomy, Uppsala University, 
	\\Box 516, 75120 Uppsala, Sweden}

\emailAdd{levi@maths.ox.ac.uk}
\emailAdd{roger.morales@nbi.ku.dk}
\emailAdd{zhewei.yin@physics.uu.se}

\abstract{We confirm the generalized actions of the complete NLO cubic-in-spin interactions for 
generic compact binaries which were first tackled via an extension of the EFT of spinning gravitating 
objects. 
We first reduce these generalized actions to standard actions with spins, where the interaction 
potentials are found to consist of $6$ independent sectors, including a new unique sector that is 
proportional to the square of the quadrupolar deformation parameter, $C_{\text{ES}^2}$. 
We derive the general Hamiltonians in an arbitrary reference frame, and for generic kinematic 
configurations. 
With these most general Hamiltonians we construct the full Poincar\'e algebra of all the sectors at 
the fourth and a half post-Newtonian (4.5PN) order, including the third subleading spin-orbit sector, 
recently accomplished uniquely via our framework, thus proving the Poincar\'e invariance of all 
relevant sectors. 
We then derive the binding energies with gauge-invariant relations useful for gravitational-wave 
applications. 
Finally, we also derive the extrapolated scattering angles in the aligned-spins configuration for the 
scattering problem. Yet, as made clear already as of quadratic-in-spin sectors, 
the aligned-spins simplification inherent to the scattering-angle observable, entails a great loss of 
physical information, that is only growing with higher-spin sectors.
Our completion of the full Poincar\'e algebra at the present $4.5$PN order provides strong confidence 
that this new precision frontier in PN theory has now been established. 
}

\begin{document}
	

\preprint{UUITP-48/22} 
	
\maketitle
	
\flushbottom

\section{Introduction}

Within few years of gravitational-wave (GW) astronomy, we already have an impressive catalogue of 
measurements of $90$ accumulated confirmed GW signals
\cite{LIGOScientific:2018mvr,LIGOScientific:2020ibl,LIGOScientific:2021djp}. This data collected 
by current second-generation GW detectors Advanced LIGO \cite{LIGOScientific:2014pky}, 
Advanced VIRGO \cite{VIRGO:2014yos}, and KAGRA \cite{KAGRA:2020tym}, includes as sources inspirals of 
compact binaries of black holes (BHs) \cite{Abbott:2016blz}, neutron stars (NSs)
\cite{TheLIGOScientific:2017qsa}, and even of mixed binaries of a BH and a NS. 
\cite{LIGOScientific:2021qlt} as sources. These binaries evolve virtually all of their lifetime 
through non-relativistic (NR) motion in weak gravity, and thus their evolution has been studied 
analytically using the post-Newtonian (PN) approximation of General Relativity (GR), see Blanchet's 
Living Review for the comprehensive progress and status of PN theory 
\cite{Blanchet:2013haa}. Building on PN theory, and interpolating over the swift phase of merger, 
in which the binary is subject to strong gravity fields, the effective-one-body (EOB) approach 
\cite{Buonanno:1998gg} has consistently enabled the generation of theoretical gravitational 
waveforms from such sources, against which measured signals are compared.  

To that end efforts have been increasing in recent years to push the state of the art of PN 
theory. For the conservative dynamics of generic compact binaries the high fifth PN ($5$PN) accuracy
in the two-body potentials has been tackled in both traditional GR 
\cite{Bini:2019nra,Bini:2020wpo,Bini:2020uiq}, and effective field theory (EFT) approaches 
\cite{Goldberger:2004jt,Blumlein:2020pyo,Goldberger:2022ebt}. These $5$PN precision 
results have been obtained in the point-mass sector, where finite-size effects, and thus the 
internal structure of the individual components of the binary first kicks in at this high order. 
The third subleading spin-orbit sector at the $4.5$PN order (the PN counting of sectors with spin is 
always evaluated for maximally-rotating objects) was also 
approached following the approach in \cite{Bini:2019nra,Bini:2020wpo} via traditional GR 
\cite{Antonelli:2020aeb,Antonelli:2020ybz}, but was fully accomplished including general 
Hamiltonians via the EFT of spinning gravitating objects in 
\cite{Levi:2015msa,Levi:2017kzq,Levi:2020kvb,Kim:2022pou}. 
Similar to the point-mass sector the spin-orbit sector, which is linear in the spins, is also uniquely 
simple in that finite-size effects first enter at an even higher PN order than the point-mass sector. 

Yet, such finite-size effects hold valuable 
information to better our understanding of strong gravity and QCD theories.
For higher-spin sectors, namely as of the quadratic order in the spins of rotating compact 
components of the binary, such finite-size effects enter already at the $2$PN order 
\cite{Barker:1975ae}, and thus have to be consistently tackled in order to push the PN  
precision frontier. This has required to formulate a theory for higher-spin orders in gravity, which 
was introduced in \cite{Levi:2014gsa,Levi:2015msa}, 
and has fascinating links to higher-spin field theories, see e.g.~\cite{Bekaert:2022poo}. 
In particular for the $4.5$PN accuracy, the next-to-leading (NLO) cubic-in spin sectors for generic 
compact binaries also need to be completed. 

This paper is aimed at the completion of such results in the NLO cubic-in spin sectors at the $4.5$PN 
order, following on our EFT computation of the generalized action of these sectors 
\cite{Levi:2019kgk}. 
The latter work has built on the EFT of spinning gravitating objects 
\cite{Levi:2015msa,Levi:2018nxp}, the \texttt{EFTofPNG} public code \cite{Levi:2017kzq}, and a 
series of works in this approach that accomplished the present state of the art in sectors with spins 
at the $4$PN order \cite{Levi:2011eq,Levi:2014gsa,Levi:2015uxa,Levi:2015ixa,Levi:2016ofk} including
their general Hamiltonians. 
Recently, this EFT approach has also enabled the completion of the third subleading (N$^3$LO) 
quadratic-in-spin sectors at the $5$PN order in 
\cite{Levi:2020uwu,Kim:2021rfj,Kim:2022bwv,Levi:2022rrq}, including the general Hamiltonians in 
\cite{Kim:2022bwv} (and then also in \cite{Mandal:2022ufb}), and their full Poincar\'e in 
\cite{Levi:2022rrq}.
In this paper we reduce the generalized actions of the NLO cubic-in spin sectors from 
\cite{Levi:2019kgk} to standard actions with spins, and derive the general Hamiltonians for an 
arbitrary reference frame, and for generic kinematic configurations. From them we derive the 
binding energies with gauge-invariant relations useful for GW applications.

This paper is also aimed at validating the results of all conservative sectors at the $4.5$PN 
accuracy in order to establish this as the new precision frontier. We accomplish this objective 
through the construction of the full Poincar\'e algebra at this PN order, which includes: 
1.~The NLO cubic-in-spin sectors from \cite{Levi:2019kgk}, with the general Hamiltonians obtained in 
the present paper. 
2.~The N$^3$LO spin-orbit sector with the general Hamiltonians obtained in our \cite{Kim:2022pou}. 
The Poincar\'e algebra in phase space provides the most stringent consistency check for the full 
general PN Hamiltonians by way of proving their Poincar\'e invariance, 
and it is sensitive to the smallest deviations from proper canonical Hamiltonians. 
Thus in addition to providing the global Poincar\'e invariants of the system, the completion of the 
Poincar\'e algebra at the $4.5$PN order provides a powerful validation of this new PN precision 
frontier. 

In the scattering problem the NLO cubic-in-spin sectors in the weak-field or so-called 
``post-Minkowskian'' (PM) approximation have also been approached. 
In \cite{Guevara:2018wpp} scattering angles for BHs in the aligned-spins case were first approached. 
In \cite{Chen:2021qkk} the NLO PM Hamiltonians in the center-of-mass (COM) frame for BHs were 
presented, and in \cite{Bern:2022kto} similar NLO COM Hamiltonians for generic compact binaries were 
approached.
All of these scattering studies built on the higher-spin theory introduced and formulated in our EFT 
of spinning objects \cite{Levi:2014gsa,Levi:2015msa} as their basis, and thus they all put 
forward derivations that are inherently dependent on our framework. 
Moreover, scattering angles simply make for poor input at higher-spin sectors, since they are always 
inherently restricted to the aligned-spins simplification, where there is a growing loss of physical 
information that is only increasing with spin orders, as of the quadratic order in spins. 
Furthermore, when Hamiltonians are provided in these scattering studies, which thus far seems to be 
feasible only at low loop orders and spin orders, which are already known in PN theory, they are 
always restricted to the COM frame. 
This fact also does not allow to study the Poincar\'e algebra of the system, which could in turn also 
provide a critical check for the validity of such results. 

In these scattering-amplitudes derivations, quantum degrees of freedom (DOFs) are unnecessarily 
invoked, that then need to be laboriously removed from the meaningful classical results. 
Moreover the scattering results should be linked to the bound inspiral setup, which requires more 
work, and becomes an obstacle as of third subleading loop or spin orders. In our EFT approach, there 
are only classical DOFs, directly set up in the bound problem, and thus our approach readily gets at 
the necessary results for GW measurements. Moreover, our approach provides the general arbitrary 
reference-frame Hamiltonians, which form part of the full Poincar\'e algebra. Our EFT approach is thus 
instrumental to high-precision GW measurements, as well as critical to guide such efforts to attempt 
at diverse derivations in the related scattering problem. To that end, we also derive from our generic 
PN Hamiltonians the extrapolated scattering angles in the aligned-spins configuration for the 
scattering problem. 

This paper is organized as follows. 
In section \ref{higherspin} we review our EFT of higher-spin 
in gravity, which contains two main formal ingredients in the theory, that contribute to all 
orders in spin: spin-gauge invariance and spin-induced couplings 
\cite{Levi:2014gsa,Levi:2015msa,Levi:2018nxp}. 
In section \ref{actions} we confirm the generalized actions that were evaluated via an EFT computation 
in \cite{Levi:2019kgk}, and reduce them to the final actions of $6$ independent subsectors from which 
the equations of motion (EOMs) for both the position and spin can be obtained directly and simply. 
In section \ref{allhams} we derive the full general Hamiltonians in an arbitrary reference frame, and 
then gradually specialize to the COM frame, and to the aligned-spins configuration, where it is shown 
how significant is the loss of physical information in these simplifications, notably growing with  
higher orders in spin. 
In section \ref{Poincare} we construct the full Poincar\'e algebra with the general Hamiltonians of
the NLO cubic-in-spin sectors, and N$^3$LO spin-orbit sector, that make up the $4.5$PN precision, thus 
proving their Poincar\'e invariance and establishing this PN order as the new precision frontier. 
In section \ref{obs} we derive useful observables and gauge-invariant relations for GW applications. 
We also derive the extrapolated scattering angles in the aligned-spins configuration of the scattering 
problem for guidance of scattering-amplitudes derivations.
Finally in our appendices we include: A brief note on typos in \cite{Levi:2019kgk} in Appendix 
\ref{wherewerewe}, explicit results for the new redefinitions, final actions, and general 
Hamiltonians of the NLO cubic-in-spin sectors, in Appendices \ref{newredefs}, \ref{distilledaction}, 
and \ref{sixgenhams}, respectively, and the Poincar\'e COM generator of the N$^3$LO spin-orbit sector 
in Appendix \ref{thirdsubleadinggenwspin}. All the corresponding results are also provided in our 
ancillary files to this publication.

\section{EFT of Higher Spin in Gravity}
\label{higherspin}

To complete the precision frontier at the $4.5$PN order, we need to consider carefully the EFT of 
spinning gravitating objects \cite{Levi:2015msa}, which was originally formulated to establish the 
$3$PN order as the state of the art, and to then obtain the present state of the art at the $4$PN order, 
via an EFT approach for spins in gravity. In particular the $4.5$PN order requires to tackle the NLO 
sectors that are cubic in spin, and thus to extend the EFT at higher orders in spin with greater 
attention and rigour \cite{Levi:2019kgk,Levi:2020lfn}. Let us review our theory that was presented in 
\cite{Levi:2015msa,Levi:2018nxp}, which we then built on. 

For the conservative interactions of the compact binary inspiral we consider an effective 
action that captures a two-particle system in a weak gravity field at the orbital scale of 
binary separation \cite{Goldberger:2004jt}:    
\be 
\label{twopart}
S_{\text{eff}}=S_{\text{gr}}[g_{\mu\nu}]+\sum_{a=1}^{2}S_{\text{pp}}(\lambda_a).
\ee
$S_{\text{gr}}$ is the purely gravitational action in some classical theory of gravity, e.g.~GR, 
and it is supplemented by an infinite tower of interactions between the gravitational field and the 
worldline degrees of freedom (DOFs), representing each $a$-th compact object. These 
interactions make up the point-particle action, $S_{\text{pp}}$, localized on the worldlines, 
parametrized by $\lambda_a$. The challenge for rotating objects is then to bootstrap the effective 
action of a spinning particle.

First, for a spinning object the action of a point-particle can be written as 
\cite{Hanson:1974qy,Bailey:1975fe,Porto:2005ac,Levi:2015msa}:
\begin{align}
\label{sppspinold}
S_{\text{pp}}\left[g_{\mu\nu},y^\mu,e^{\mu}_{A}\right]
= \int d \lambda \left[ -m \sqrt{u^2}
	- \frac{1}{2} S_{\mu\nu}\Omega^{\mu\nu}
	+ L_{\text{SI}}\left[u^{\mu}, S_{\mu\nu}, g_{\mu\nu}
	\left(y^\mu\right)\right]\right],
\end{align} 
where $u^\mu\equiv dy^\mu/d\lambda$, and $y^\mu$, $e^{\mu}_{A}$ are the particle worldline 
coordinate and tetrad DOFs, respectively. From the worldline tetrad,  
$\eta^{AB}e_A{}^\mu(\lambda)e_B{}^\nu(\lambda)=g^{\mu\nu}$, the angular velocity is defined as 
$\Omega^{\mu\nu}(\lambda)\equiv e^\mu_A\frac{De^{A\nu}}{D\lambda}$, and then its conjugate, the 
worldline spin, $S_{\mu\nu}(\lambda)\equiv-2\frac{\partial L}{\partial\Omega^{\mu\nu}}$, is added 
as another explicit DOF to the action. 
$L_{\text{SI}}$ denotes the non-minimal coupling part of the Lagrangian that is induced due to 
the presence of spin.

While the minimal coupling in the form of eq.~\eqref{sppspinold} is fixed only from the symmetries 
of general covariance and reparametrization invariance \cite{Hanson:1974qy,Bailey:1975fe}, it 
turns out that it conceals additional symmetries related with the rotational DOFs, the worldline 
tetrad and the spin: As we have shown in \cite{Levi:2015msa} this form of the action in fact 
already assumes the Tulczyjew gauge for the spin \cite{Tulczyjew:1959b}. Related with that hidden 
symmetry is the fact that the particle worldline coordinate can in general be shifted from the position 
that represents the rotating object's ``center''. As to the non-minimal coupling part of the action, 
the symmetries of parity and $SO(3)$ invariance play a major role in constraining it. 

Indeed, these were the $2$ fundamental challenges tackled successfully in bootstrapping the effective 
action of a spinning particle in our EFT formulated in \cite{Levi:2015msa}: 1.~Making the spin gauge 
invariance manifest in the action as of minimal coupling. Notice that this contributes to all 
orders in spin. 2.~Fixing the leading non-minimal couplings to all orders in spin. 
In sections \ref{sgi} and \ref{highernmc} below we go over these $2$ major formal developments 
accomplished in \cite{Levi:2015msa}.

\subsection{Spin Gauge Invariance}
\label{sgi}

The key observation here is the symmetries related with the worldline tetrad. There is an $SO(3)$ 
invariance of the worldline spatial triad, and then what we refer to as ``spin gauge invariance'', 
which is some freedom to complete the timelike component of the worldline spatial triad to a tetrad
\cite{Levi:2015msa}. This gauge choice will fix both tetrad and spin variables. To make the gauge 
freedom of the rotational variables manifest in the effective action, we applied a 4-dimensional 
covariant boost-like transformation on the worldline tetrad, introducing new gauge DOFs, 
$\hat{e}_{[0]\mu} = w_{\mu}$, for the timelike vector of the tetrad. This leads to a generic gauge 
condition for the spin (traditionally called ``SSC'') \cite{Levi:2015msa}:
\begin{equation}
\label{ssccurved}
\hat{S}^{\mu\nu} \left( p_{\nu} + \sqrt{p^2} \hat{e}_{[0]\nu} \right) = 0,
\end{equation}
which removes the redundant DOFs from both the angular velocity and the spin.
From the minimal coupling term in eq.~\eqref{sppspinold} we then obtain \cite{Levi:2015msa}:
\begin{align}
\label{thomasexposed}
\frac{1}{2} S_{\mu\nu} \Omega^{\mu\nu} &= 
\frac{1}{2} \hat{S}_{\mu\nu} \hat{\Omega}^{\mu\nu}
+ \frac{\hat{S}^{\mu\nu} p_{\nu}}{p^2} \frac{D p_{\mu}}{D \sigma},
\end{align}
where a new general term emerges in the action, which was not accounted for in past formulations 
of spin in gravity, including in Yee and Bander \cite{Yee:1993ya}, which was later adopted in 
\cite{Porto:2008jj}. This kinematic term, essentially Thomas precession as elaborated in 
\cite{Levi:2015msa}, originates from minimal coupling, and thus is clearly not preceded by any 
Wilson coefficient, though it contributes from leading order in spin -- to finite-size effects at all 
orders in spin. 

Using the worldline Lorentz matrices, 
$\eta^{AB}\Lambda_A{}^a(\lambda)\Lambda_B{}^b(\lambda)=\eta^{ab}$, we can write the locally-flat 
angular velocity, $\hat{\Omega}^{ab}_{\text{flat}} 
= \hat{\Lambda}^{Aa} \frac{d \hat{\Lambda}_A{}^b}{d \lambda}$, and the conjugate local 
spin, $\hat{S}_{ab}=\tilde{e}^{\mu}_{a}\tilde{e}^{\nu}_{b}\hat{S}_{\mu\nu}$, with the local tetrad 
field, $\eta^{ab}\tilde{e}_a{}^\mu(x)\tilde{e}_b{}^\nu(x)=g^{\mu\nu}(x)$. Then, using the Ricci 
rotation coefficients, $\omega_{\mu}{}^{ab} \equiv \tilde{e}^b{}_{\nu} D_{\mu}\tilde{e}^{a\nu}$, 
the first term on the RHS in eq.~\eqref{thomasexposed} can be rewritten as 
\cite{Levi:2010zu,Levi:2015msa}:
\begin{align} 
\label{ssplitfield}
\frac{1}{2} \hat{S}_{\mu\nu} \hat{\Omega}^{\mu\nu}&= 
\frac{1}{2} \hat{S}_{ab} \hat{\Omega}^{ab}_{\text{flat}}
+ \frac{1}{2} \hat{S}_{ab} \omega_{\mu}{}^{ab} u^{\mu}.      
\end{align}
At this point we note that we fix the gauge of the rotational variables so as to fully disentangle 
the field from the worldline DOFs, as was first put forward in \cite{Levi:2008nh}. We fix the gauge of  
rotational variables to the canonical gauge for curved spacetime that we generalized from flat 
spacetime Pryce-Newton-Wigner SSC \cite{Pryce:1948pf,Newton:1949cq}.

\subsection{Higher-Spin Coupling}
\label{highernmc}

Based on the full set of symmetries that we noted, the key element in bootstrapping the 
non-minimal coupling of spin to gravity was to invoke the classical analogue of the Pauli-Lubanski 
vector, $S^{\mu}$, as 
the building block for the action \cite{Levi:2014gsa,Levi:2015msa}. Focusing on parity and $SO(3)$ invariance 
and through a full rigorous analysis, the leading non-minimal couplings to all orders in spin were 
presented in \cite{Levi:2015msa}: 
\begin{align} 
\label{toinfinity}
L_{\text{SI}}=&\sum_{n=1}^{\infty} \frac{\left(-1\right)^n}{\left(2n\right)!}
\frac{C_{\text{ES}^{2n}}}{m^{2n-1}} D_{\mu_{2n}}\cdots D_{\mu_3}
\frac{E_{\mu_1\mu_2}}{\sqrt{u^2}} S^{\mu_1}S^{\mu_2}\cdots 
S^{\mu_{2n-1}}S^{\mu_{2n}}\nn\\
&+\sum_{n=1}^{\infty} \frac{\left(-1\right)^n}{\left(2n+1\right)!}
\frac{C_{\text{BS}^{2n+1}}}{m^{2n}} 
D_{\mu_{2n+1}}\cdots D_{\mu_3}\frac{B_{\mu_1\mu_2}}{\sqrt{u^2}} 
S^{\mu_1}S^{\mu_2}\cdots 
S^{\mu_{2n-1}}S^{\mu_{2n}}S^{\mu_{2n+1}},
\end{align}
with a new infinite set of Wilson coefficients that correspond to ``multipolar deformation parameters'' 
in traditional GR. 
This infinite tower of operators contains definite-parity curvature components, either the electric or 
magnetic, $E_{\mu\nu}$ or $B_{\mu\nu}$, respectively. For the present sectors, we only need to 
pull out the first two terms of this infinite series \cite{Levi:2014gsa,Levi:2015msa}:
\begin{align} 
\label{nmcs2s3}
L_{\text{ES}^2} \equiv& -\frac{C_{\text{ES}^2}}{2m} \frac{E_{\mu\nu}}{\sqrt{u^2}} 
S^{\mu} S^{\nu},\\
L_{\text{BS}^3}\equiv&-\frac{C_{\text{BS}^3}}{6m^2}\frac{D_\lambda B_{\mu\nu}}
{\sqrt{u^2}}S^{\mu} S^{\nu} S^{\lambda}, 
\end{align}
which correspond to the quadrupolar \cite{Barker:1975ae,Porto:2005ac} and octupolar 
\cite{Levi:2014gsa} deformations.

Following the EFT of spinning gravitating objects introduced in \cite{Levi:2014gsa,Levi:2015msa}, 
and as reviewed above, various scattering-amplitudes approaches tackled the 
gravitational scattering problem with massive higher-spin particles, including 
\cite{Guevara:2018wpp,Chen:2021qkk,Bern:2022kto}. In particular, the infinite tower of $S^l$ couplings 
in eq.~\eqref{toinfinity} has been used for the corresponding $3$-point amplitudes with massive 
particles of spin $s=l/2$, that make up the building blocks to derive any scattering amplitude. 
Furthermore, all these approaches used input from implementing the EFT of spinning gravitating objects 
\cite{Levi:2014gsa,Levi:2015msa} for the specific case of BHs within traditional 
GR \cite{Guevara:2018wpp} -- as a critical guide to their derivations. 
In particular, the dependence of \cite{Guevara:2018wpp} in our worldline theory for higher-spin 
\cite{Levi:2014gsa,Levi:2015msa} should be noted here, as it was omitted in \cite{Guevara:2018wpp}.

As to non-minimal couplings that are quadratic in the curvature, an extension of the action that 
covers the cubic order in spin was introduced in \cite{Levi:2020lfn}. Similar to the spin-orbit 
sector, it is found that such cubic-in-spin operators enter only at the $6.5$PN order, and thus are 
not relevant to the present sectors. 
At this point it should be noted that the spin which is used in the construction of non-minimal 
couplings is in the Tulczyjew gauge to begin with, and thus in order to switch to a generic spin 
variable as in section \ref{sgi} above, the following relation should be used:
\begin{equation}
\label{canbegen}
S_{\mu\nu} = \hat{S}_{\mu\nu} - \frac{\hat{S}_{\mu\rho} p^{\rho} p_{\nu}}{p^2}
+ \frac{\hat{S}_{\nu\rho} p^{\rho} p_{\mu}}{p^2}.
\end{equation}

\section{Effective Actions}
\label{actions}

Using our effective theory for higher spin in gravity reviewed in the previous section, we carried 
out an EFT evaluation of the NLO cubic-in-spin sectors in \cite{Levi:2019kgk}.
The evaluation of the relevant interactions involved $53$ unique Feynman graphs 
\cite{Levi:2019kgk}. 
The printed values for $5$ of these graphs contained typos, which we note in 
appendix \ref{wherewerewe} below. 
These copying errors in the individual values of graphs in the printed 
manuscript are arbitrary typos, and did not affect the total sum of the graphs, that was 
provided in \cite{Levi:2019kgk}. 
The generalized actions of the NLO cubic-in-spin interactions are then written as 
\cite{Levi:2019kgk}: 
\begin{equation}
L^{\text{NLO}}_{\text{S}^3}=
L^{\text{NLO}}_{\text{S}_1^2\text{S}_2}
+L^{\text{NLO}}_{\text{S}_1^3}+(1\leftrightarrow 2),
\end{equation}
where we identify the following distinct pieces:
\begin{align}
L^{\text{NLO}}_{\text{S}_1^2\text{S}_2}=&\
+\frac{G^2}{r^5}L_{(1)}
+C_{1(\text{ES}^2)}\frac{G}{r^4}\frac{1}{m_1}L_{(2)}
+C_{1(\text{ES}^2)}\frac{G^2}{r^5}L_{(3)}
+ C_{1(\text{ES}^2)}\frac{G^2m_2}{r^5m_1}L_{(4)}
\nn&\\
&
+\frac{G^2}{r^4}L_{(5)}
+C_{1(\text{ES}^2)}\frac{G}{r^3}\frac{1}{m_1}L_{(6)}
+C_{1(\text{ES}^2)}\frac{G^2}{r^4}L_{(7)}
+C_{1(\text{ES}^2)}\frac{G^2m_2}{r^4m_1}L_{(8)}
\nn&\\
&
+C_{1(\text{ES}^2)}\frac{G}{r^2}\frac{1}{m_1}L_{(9)}
+C_{1(\text{ES}^2)}\frac{G}{r}\frac{1}{m_1}L_{(10)},
\end{align}
as well as the following ones:
\begin{align}
\label{genacts13}	
L^{\text{NLO}}_{\text{S}_1^3}=&\
C_{1(\text{ES}^2)}\frac{G^2m_2}{r^5m_1}L_{[1]}
+C_{1(\text{ES}^2)}\frac{G^2m_2^2}{r^5m_1^2}L_{[2]}
+C_{1(\text{BS}^3)}\frac{Gm_2}{r^4m_1^2}L_{[3]}
\nn&\\
&
+C_{1(\text{BS}^3)}\frac{G^2m_2}{r^5m_1}L_{[4]}
+C_{1(\text{BS}^3)}\frac{G^2m_2^2}{r^5m_1^2}L_{[5]}
+C_{1(\text{ES}^2)}\frac{Gm_2}{r^3m_1^2}L_{[6]}
\nn&\\
&
+C_{1(\text{ES}^2)}\frac{G^2m_2}{r^4m_1}L_{[7]}
+C_{1(\text{BS}^3)}\frac{Gm_2}{r^3m_1^2}L_{[8]}
+C_{1(\text{BS}^3)}\frac{Gm_2}{r^2m_1^2}L_{[9]},
\end{align}
where we also provide these generalized actions in machine-readable format in the ancillary files of 
this publication. Let us stress that the computer files of \cite{Levi:2019kgk}, also included in the 
ancillary files to the present publication, contain the correct results.
Note that in eq.~\eqref{genacts13} there can also exist in principle a piece of the form
$C_{1(\text{ES}^2)}m_2/(m_1^2 r^2)$, which is absorbed into $L_{[6]}$, eq.~(5.19) in 
\cite{Levi:2019kgk}, by a total time derivative.

\subsection{Redefinition of Actions}
\label{redeftostd}

As noted in \cite{Levi:2019kgk} the generalized actions that are obtained from the EFT computation 
need to be reduced to ``standard'' actions with spin variables, namely which do not contain 
higher-order time derivatives beyond velocity and spin. The reduction procedure via formal 
redefinitions that we show here was introduced in \cite{Levi:2014sba} to include rotational 
variables, and we build here on the derivations shown in \cite{Kim:2022pou,Kim:2022bwv}. 
Table \ref{tableredefsects} summarizes the redefinitions that need to be applied gradually to the 
relevant sectors that contribute to the present sectors, in increasing PN order, even those that do 
not require any reduction in themselves. 

\begin{table}[t]
\begin{center}
\begin{tabular}{|l|c|c|}
\hline
\backslashbox{\quad\boldmath{$l$}}{\boldmath{$n$}} & (N\boldmath{$^{0}$})LO
& N\boldmath{$^{(1)}$}LO 
\\
\hline
\boldmath{S$^0$} &  & 
\\
\hline
\boldmath{S$^1$} & + \qquad  & ++ 
\\
\hline
\boldmath{S$^2$} &  & ++ 
\\
\hline
\boldmath{S$^3$} 
& 
\quad+ 
& 
++ 
\\
\hline
\end{tabular}
\caption{The shorthand notation of sectors, $(n,l)$, and the general formula $n+l+\text{Parity}(l)/2$ 
for their PN counting was introduced in \cite{Levi:2019kgk}, where $n$ is the subleading order (or 
highest $n$-loop order), and $l$ is the highest order in spins of each of the sectors, and the parity
is $0$ or $1$ for even or odd $l$, respectively.
The $8$ sectors that contribute to the present NLO cubic-in-spin sectors, $(1,3)$, 
through redefinition of variables. 
"+" marks sectors that require only position shifts or redefinitions of rotational variables to be 
fixed, 
and "++" marks sectors that require redefinition of both position and rotational 
variables.}
\label{tableredefsects}
\end{center}
\end{table}

Based on the spin and PN power-counting of the various redefinitions \cite{Levi:2014sba}, we first 
note that similar to \cite{Levi:2015msa}, for higher-spin sectors as of the NLO, we need to apply 
position shifts beyond linear order. On 
the other hand, also according to the extension of the procedure beyond linear order in the 
rotational variables which we carried out in \cite{Kim:2022pou}, here we only need to apply 
redefinitions of the rotational variables to linear order. For the present sectors we need to take into 
account redefinitions that are fixed in $5$ sectors, as shown in table \ref{tableredefsects}, and we 
follow the detailed presentation in \cite{Kim:2022pou,Kim:2022bwv}. 

The redefinitions in $3$ of these sectors, below cubic in spin, are detailed in 
\cite{Kim:2022bwv,Kim:2022pou}, and thus here we need to further consider the $2$ sectors that are 
cubic in the spin, shown in tables \ref{los3redef}--\ref{nlos3redef}, whose structure was explained 
in \cite{Kim:2022pou}. 
The algorithm used for the reduction is similar to that we used in \cite{Kim:2022pou,Kim:2022bwv}, 
only that it implements higher-order position shifts as seen in table \ref{nlos3redef}. Thus, we 
now go through the relevant sectors according to their PN order, with the unreduced actions always 
computed with the \texttt{EFTofPNG} code \cite{Levi:2017kzq}. For the LO and NLO spin-orbit, and 
NLO quadratic-in-spin sectors, our unreduced actions and redefinitions can be found in 
\cite{Kim:2022pou}, and \cite{Kim:2022bwv}, respectively.

Now we can approach the LO cubic-in-spin sectors as shown in table \ref{los3redef}, where we will 
not conform to the choices of unreduced actions and redefinitions of our original derivation of 
these sectors in \cite{Levi:2014gsa}. 
Thus the unreduced potential is:
\bea
V^{\text{LO}}_{\text{S}^3} &=&	\frac{3 G C_{1\text{ES}^2}}{m_{1} r{}^4} \Big[ S_{1}^2 \vec{S}_{2}\times\vec{n}\cdot\vec{v}_{1} - 2 \vec{S}_{1}\cdot\vec{n} \vec{S}_{1}\times\vec{S}_{2}\cdot\vec{v}_{1} - S_{1}^2 \vec{S}_{2}\times\vec{n}\cdot\vec{v}_{2} \nn\\ 
&& + 2 \vec{S}_{1}\cdot\vec{n} \vec{S}_{1}\times\vec{S}_{2}\cdot\vec{v}_{2} - 5 \vec{S}_{2}\times\vec{n}\cdot\vec{v}_{1} ( \vec{S}_{1}\cdot\vec{n})^{2} + 5 \vec{S}_{2}\times\vec{n}\cdot\vec{v}_{2} ( \vec{S}_{1}\cdot\vec{n})^{2} \Big] \nn\\ 
&& - 	\frac{3 G C_{1\text{ES}^2}}{m_{1} r{}^3} \Big( \dot{\vec{S}}_{1}\cdot\vec{n} \vec{S}_{1}\times\vec{n}\cdot\vec{S}_{2} + \vec{S}_{1}\cdot\vec{n} \dot{\vec{S}}_{1}\times\vec{n}\cdot\vec{S}_{2} \Big) \nn\\
&&+ 	\frac{G C_{1\text{BS}^3} m_{2}}{m_{1}{}^2 r{}^4} \Big[ S_{1}^2 \vec{S}_{1}\times\vec{n}\cdot\vec{v}_{1} - S_{1}^2 \vec{S}_{1}\times\vec{n}\cdot\vec{v}_{2} - 5 \vec{S}_{1}\times\vec{n}\cdot\vec{v}_{1} ( \vec{S}_{1}\cdot\vec{n})^{2} \nn\\ 
&& + 5 \vec{S}_{1}\times\vec{n}\cdot\vec{v}_{2} ( \vec{S}_{1}\cdot\vec{n})^{2} \Big].
\eea
There is no new position shift in this sector, but a new redefinition for the rotational variables 
that is fixed as:
\bea
\left( \omega^{ij} \right)^{\text{LO}}_{\text{S}^3} =	
\frac{3 G C_{1\text{ES}^2}}{m_{1} r{}^3} \Big[ \vec{S}_{2}\cdot\vec{n} S_{1}^i n^j 
- 2 \vec{S}_{1}\cdot\vec{n} S_{2}^i n^j + S_{2}^i S_{1}^j \Big] - (i \leftrightarrow j).
\eea

\begin{table}[t]
\begin{center}
\begin{tabular}{|l|c|c|}
\hline
\backslashbox{from}{\boldmath{to}} 
& (0P)N & LO S$^2$
\\
\hline
LO S$^1$ &   & $\Delta \vec{x}$
\\
\hline
LO S$^3$ & $\Delta \vec{S}$  &
\\
\hline
\end{tabular}
\caption{Contributions to the LO S$^3$ sectors from position shifts and spin 
	redefinitions in lower-order sectors.}
\label{los3redef}
\end{center}
\end{table}

We can now consider the redefinitions at the present NLO cubic-in-spin sectors, as detailed in 
table \ref{nlos3redef}. 

\begin{table}[t]
\begin{center}
\begin{tabular}{|l|c|c|c|c|c|}
\hline
\backslashbox{from}{\boldmath{to}} 
& (0P)N & LO S$^1$ & LO S$^2$ & NLO S$^2$ & LO S$^3$
\\
\hline
LO S$^1$ & $(\Delta \vec{x})^3$ & $(\Delta \vec{x})^2$ &  & $\Delta \vec{x}$ &
\\
\hline
NLO S$^1$ & & & $\Delta \vec{x}$ & & $\Delta \vec{S}$
\\
\hline
NLO S$^2$ &  & $\Delta \vec{x}$ & $\Delta \vec{S}$ &  &
\\
\hline
LO S$^3$ &  & $\Delta \vec{S}$ & & &
\\
\hline
NLO S$^3$ & $\Delta \vec{x}$, $\Delta \vec{S}$ & & & &
\\
\hline
\end{tabular}
\caption{Contributions to the NLO S$^3$ sectors from position shifts and spin 
redefinitions in lower-order sectors.}
\label{nlos3redef}
\end{center}
\end{table}

\noindent The new position shifts and redefinitions of rotational variables fixed in the present sectors can be written as:
\bea
\left(\Delta \vec{x}_1\right)^{\text{NLO}}_{\text{S}^3} &=&\left(\Delta \vec{x}_1\right)^{\text{NLO}}_{\text{S}_1^3}+\left(\Delta \vec{x}_1\right)^{\text{NLO}}_{\text{S}_1^2 \text{S}_2} +\left(\Delta \vec{x}_1\right)^{\text{NLO}}_{\text{S}_1 \text{S}_2^2} +\left(\Delta \vec{x}_1\right)^{\text{NLO}}_{\text{S}_2^3},\\
\left(\omega^{ij}_1\right)^{\text{NLO}}_{\text{S}^3} &=&\left(\omega^{ij}_1\right)^{\text{NLO}}_{\text{S}_1^3}+\left(\omega^{ij}_1\right)^{\text{NLO}}_{\text{S}_1^2 \text{S}_2} +\left(\omega^{ij}_1\right)^{\text{NLO}}_{\text{S}_1 \text{S}_2^2}  - (i \leftrightarrow j),
\eea
where the explicit redefinitions are presented in appendix \ref{newredefs}, and we also provide them in machine-readable format in the ancillary files of this publication.

\subsection{Final Actions}
\label{finalactions}

As explained in \cite{Levi:2015msa} we can already obtain the EOMs for the position and spin from 
the generalized actions before reduction due to our use of the generalized canonical gauge 
formulated in \cite{Levi:2015msa}. However, it is much easier to derive the EOMs with the more 
compact actions obtained after the reduction that we have shown in the previous section. 
The final potentials that we obtain for the NLO cubic-in-sectors, comprise the following $6$ 
distinct sectors:
\begin{align}
V^{\text{NLO}}_{\text{S}^3} = &
V^{\text{NLO}}_{\text{S}_1^3 } 
+ C_{1\text{ES}^2} V^{\text{NLO}}_{(\text{ES}_1^2 ) \text{S}_1} 
+ C_{1\text{ES}^2}^2 V^{\text{NLO}}_{C_{\text{ES}_1^2}^2 \text{S}_1^3 }
+ C_{1\text{BS}^3} V^{\text{NLO}}_{\text{BS}_1^3 } 
+ V^{\text{NLO}}_{\text{S}_1^2 \text{S}_2}  
+ C_{1\text{ES}^2} V^{\text{NLO}}_{(\text{ES}_1^2 ) \text{S}_2} \nn\\
& + (1 \leftrightarrow 2),
\end{align}
where the explicit actions are presented in appendix \ref{distilledaction}, and we also provide them 
in machine-readable format in the ancillary files of this publication.

Notice that a new ``self-induced'' cubic-in-spin potential arises in eq.~\eqref{sqCES2}, that is 
proportional to the square of the quadrupolar-deformation parameter for generic compact binaries.
As we shall see in our construction of the Poincar\'e algebra of the sectors at this PN order in 
section \ref{Poincare} below, and more specifically from eq.~\eqref{nontrivialpoincare} there, this 
new sector is actually imposed by Poincar\'e invariance. 
It can be seen as arising from the precession of spin due to the leading quadrupolar deformation, as 
it enters and affects in turn on higher orders of the spin-induced quadrupolar potential. 
The interference of misaligned quadrupole effects effectively gives rise to this new self-induced 
octupole potential. 
Of course, when the simplification, that all spins are aligned, is assumed, then this new effect drops 
out. 
 
From these final potentials for the NLO cubic-in-sectors the consequent EOMs for the position and 
spin were derived in \cite{Morales:2021}. 


\section{Hamiltonians}
\label{allhams}

From the generalized canonical gauge for the rotational variables that is included in our 
formulation \cite{Levi:2015msa}, the derivation of full general Hamiltonians is straightforward, via a 
Legendre transform of the final actions only with respect to the position variables. This Legendre 
transform involves all the sectors that lead up to the present ones as noted in table 
\ref{tableredefsects}. It should also be highlighted that the Hamiltonians which we derive in our 
approach hold for an arbitrary reference frame, and thus are the most general ones, and in particular 
more general than various specialized Hamiltonians provided in all other methods, such as the COM, EOB, 
or aligned-spins Hamiltonians, which are all -- to begin with -- already restricted to the COM frame.

The general Hamiltonians already have important applications both 
formally and phenomenologically. 
With the general Hamiltonians the Poincar\'e algebra of the 
conserved integrals of motion can be uniquely uncovered, which also provides a stringent kinematic 
consistency-check for the validity of Hamiltonians obtained, as will be discussed in detail in 
section \ref{Poincare} below. Phenomenologically from the Hamiltonians one can also construct 
various possible EOB models for the present sectors, and study how they perform. Finally, the general 
Hamiltonians can be specialized to certain simplified kinematic configurations, see section 
\ref{shams} below, in which gauge-invariant observables, notably the binding energies as function of 
the GW frequencies, can be extracted, as will be detailed in section \ref{obs} below.

\subsection{Full Hamiltonians}
\label{fullhams}

Similar to the final potentials presented in section \ref{finalactions}, 
our full general Hamiltonian for the present NLO cubic-in-spin sectors is 
comprised of $6$ distinct sectors:
\begin{align}
H^{\text{NLO}}_{\text{S}^3} = & 
H^{\text{NLO}}_{\text{S}_1^3 } 
+ C_{1\text{ES}^2} H^{\text{NLO}}_{(\text{ES}_1^2 ) \text{S}_1} 
+ C_{1\text{ES}^2}^2 H^{\text{NLO}}_{C_{\text{ES}_1^2}^2 \text{S}_1^3 }
+ C_{1\text{BS}^3} H^{\text{NLO}}_{\text{BS}_1^3 } 
+ H^{\text{NLO}}_{\text{S}_1^2 \text{S}_2}  
+ C_{1\text{ES}^2} H^{\text{NLO}}_{(\text{ES}_1^2 ) \text{S}_2} \nn \\
& +  (1 \leftrightarrow 2),
\label{eq:hamcns3}
\end{align}
where the explicit Hamiltonians are presented in appendix \ref{sixgenhams}, and we also provide 
them in machine-readable format in the ancillary files of this publication.

\subsection{Specialized Hamiltonians}
\label{shams}

In order to express specialized Hamiltonians, we use various binary-mass conventions:
\begin{align}
m & \equiv m_1+m_2, 
\quad q \equiv m_1/m_2,  
\quad \mu \equiv m_1 m_2 / m, \\
\nu & \equiv m_1 m_2/m^2 = q/(1+q)^2 = \mu/m ,
\end{align}
where the latter is the dimensionless symmetric mass-ratio. We further transform all variables to 
be dimensionless using $Gm$ and $\mu$ to rescale length and mass, respectively, and denote all 
dimensionless variables with a tilde. 

First, the Hamiltonians are specified in the center-of-mass (COM) frame, with $\vec{p} 
\equiv \vec{p}_1 = - \vec{p}_2$. Using $\vec{p}$, the orbital angular momentum is defined as
$\vec{L} \equiv r \vec{n} \times \vec{p}$. Then the COM Hamiltonians of the NLO cubic-in-spin 
sectors are written in the form:
\begin{align}
\tilde{H}^{\text{NLO}}_{\text{S}^3} = &
\tilde{H}^{\text{NLO}}_{\text{S}_1^3 } 
+ C_{1\text{ES}^2} \tilde{H}^{\text{NLO}}_{(\text{ES}_1^2 ) \text{S}_1} 
+ C_{1\text{ES}^2}^2 \tilde{H}^{\text{NLO}}_{C_{\text{ES}_1^2}^2 \text{S}_1^3 }
+ C_{1\text{BS}^3} \tilde{H}^{\text{NLO}}_{\text{BS}_1^3 } 
+ \tilde{H}^{\text{NLO}}_{\text{S}_1^2 \text{S}_2} 
+ C_{1\text{ES}^2} \tilde{H}^{\text{NLO}}_{(\text{ES}_1^2 ) \text{S}_2} \nn \\
& +  (1 \leftrightarrow 2),
\end{align}
where
\bea
\tilde{H}^{\text{NLO}}_{\text{S}_1^3 } &=& \frac{\nu^2 \tilde{\vec{L}} \cdot \tilde{\vec{S}}_1 \tilde{S}_1^2}{\tilde{r}^6} \left[
-3 \nu -\frac{9  }{4}
-\frac{9  }{4}
\frac{\tilde{L}^2}{\tilde{r}}
+\tilde{p}_r^2 \tilde{r} \left(
\frac{33  }{16}-\frac{39 \nu }{8}
\right) \right.\nn\\
&&\left. +\frac{1}{\nu q} \left(
-3 \nu ^2-\frac{3 \nu }{2}+\frac{9}{4}
+ \frac{\tilde{L}^2}{\tilde{r}} \left(
\frac{9}{4}-\frac{9 \nu }{2}
\right)
+\tilde{p}_r^2 \tilde{r} \left(
-\frac{75 \nu ^2}{16}+9 \nu -\frac{33}{16}
\right) \right) \right]\nn\\
&&+  \frac{\nu^2 (\tilde{\vec{L}} \cdot \tilde{\vec{S}}_1)^3}{\tilde{r}^7} \left[ 
2 \nu +\frac{7}{8}
+\frac{1}{\nu q} \left(
\frac{15 \nu ^2}{8}-\frac{\nu }{4}-\frac{7}{8}
\right)
\right]\nn\\
&&+\frac{\nu^2 \tilde{\vec{L}} \cdot \tilde{\vec{S}}_1 (\tilde{\vec{S}}_1 \cdot \vec{n})^2}{\tilde{r}^6} \left[
\frac{15 \nu }{2}+\frac{15}{2}
+ \frac{\tilde{L}^2}{\tilde{r}}\left(
\frac{69}{16}-\frac{9 \nu }{8}
\right)
+\tilde{p}_r^2 \tilde{r} \left(
\frac{39 \nu }{8}+\frac{111}{16}
\right) \right.\nn\\
&&+\frac{1}{\nu q} \left(
\frac{15 \nu ^2}{2}+\frac{15 \nu }{2}-\frac{15}{2}
+ \frac{\tilde{L}^2}{\tilde{r}} \left(
-\frac{15 \nu ^2}{16}+\frac{39 \nu }{4}-\frac{69}{16}
\right)\right.\nn\\
&&\left. \left. 
+\tilde{p}_r^2 \tilde{r} \left(
\frac{75 \nu ^2}{16}+9 \nu -\frac{111}{16}
\right) \right) \right]\nn\\
&&+  \frac{\nu^2 \tilde{p}_r \tilde{\vec{L}} \cdot \tilde{\vec{S}}_1  \tilde{\vec{S}}_1 \cdot \vec{n}  \tilde{\vec{S}}_1 \times \tilde{\vec{L}} \cdot \vec{n} }{\tilde{r}^6} \left[ 
6 \nu +\frac{21}{8}
+\frac{1}{\nu q} \left(
\frac{45 \nu ^2}{8}-\frac{3 \nu }{4}-\frac{21}{8}
\right)
\right],
\eea
\bea
\tilde{H}^{\text{NLO}}_{(\text{ES}_1^2 ) \text{S}_1} &=&\frac{\nu^2 \tilde{\vec{L}} \cdot \tilde{\vec{S}}_1 \tilde{S}_1^2}{\tilde{r}^6} \left[
2 \nu -\frac{3}{2}
+ \frac{\tilde{L}^2}{\tilde{r}}\left(
\frac{9 \nu }{8}+\frac{15}{16}
\right)
+\tilde{p}_r^2 \tilde{r} \left(
\frac{9 \nu }{8}+\frac{63}{16}
\right) \right.\nn\\
&&\left. +\frac{1}{\nu q} \left(
2 \nu ^2-\frac{17 \nu }{2}+\frac{3}{2}
+ \frac{\tilde{L}^2}{\tilde{r}} \left(
\frac{27 \nu ^2}{16}+\frac{3 \nu }{4}-\frac{15}{16}
\right)
+\tilde{p}_r^2 \tilde{r} \left(
\frac{57 \nu ^2}{16}+\frac{27 \nu }{4}-\frac{63}{16}
\right) \right) \right]\nn\\
&&+  \frac{\nu^2 (\tilde{\vec{L}} \cdot \tilde{\vec{S}}_1)^3}{\tilde{r}^7} \left[ 
-\frac{3}{4}
+\frac{1}{\nu q} \left(
\frac{3}{4}-\frac{3 \nu }{2}
\right)
\right]\nn\\
&&+\frac{\nu^2 \tilde{\vec{L}} \cdot \tilde{\vec{S}}_1 (\tilde{\vec{S}}_1 \cdot \vec{n})^2}{\tilde{r}^6} \left[
24-\frac{27 \nu }{4}
+ \frac{\tilde{L}^2}{\tilde{r}}\left(
-\frac{45 \nu }{8}-\frac{87}{16}
\right)
+\tilde{p}_r^2 \tilde{r} \left(
-\frac{45 \nu }{8}-\frac{123}{16}
\right) \right.\nn\\
&&+\frac{1}{\nu q} \left(
-\frac{27 \nu ^2}{4}+\frac{273 \nu }{4}-24
+ \frac{\tilde{L}^2}{\tilde{r}} \left(
-\frac{135 \nu ^2}{16}-\frac{21 \nu }{4}+\frac{87}{16}
\right)\right.\nn\\
&&\left. \left. 
+\tilde{p}_r^2 \tilde{r} \left(
-\frac{285 \nu ^2}{16}-\frac{39 \nu }{4}+\frac{123}{16}
\right) \right) \right]\nn\\
&&+  \frac{\nu^2 \tilde{p}_r \tilde{\vec{L}} \cdot \tilde{\vec{S}}_1  \tilde{\vec{S}}_1 \cdot \vec{n}  \tilde{\vec{S}}_1 \times \tilde{\vec{L}} \cdot \vec{n} }{\tilde{r}^6} \left[ 
-\frac{9}{4}
+\frac{1}{\nu q} \left(
\frac{15 \nu ^2}{4}-\frac{9 \nu }{2}+\frac{9}{4}
\right)
\right],
\eea
\bea
\label{hamses2sq}
\tilde{H}^{\text{NLO}}_{C_{\text{ES}_1^2}^2 \text{S}_1^3} &=& \frac{\nu^2 \tilde{\vec{L}} \cdot \tilde{\vec{S}}_1 (\tilde{\vec{S}}_1 \cdot \vec{n})^2}{\tilde{r}^6} \left[ -\frac{3 \nu }{2}-3 +\frac{1}{\nu q} \left(  -\frac{3 \nu ^2}{2}-\frac{9 \nu }{2}+3 \right) \right],
\eea
\bea
\tilde{H}^{\text{NLO}}_{\text{BS}_1^3 } &=& \frac{\nu^2 \tilde{\vec{L}} \cdot \tilde{\vec{S}}_1 \tilde{S}_1^2}{\tilde{r}^6} \left[
\nu +7
+ \frac{\tilde{L}^2}{\tilde{r}}\left(
1-\nu
\right)
-\frac{7 \nu }{2} \tilde{p}_r^2 \tilde{r}   \right.\nn\\
&&\left. +\frac{1}{\nu q} \left(
\nu ^2+\frac{21 \nu }{2}-7
+ \frac{\tilde{L}^2}{\tilde{r}} \left(
-\nu ^2+3 \nu -1
\right)
+\tilde{p}_r^2 \tilde{r} \left(
\frac{7 \nu }{2}-\frac{7 \nu ^2}{2}
\right) \right) \right]\nn\\
&&+  \frac{\nu^2 (\tilde{\vec{L}} \cdot \tilde{\vec{S}}_1)^3}{\tilde{r}^7} \left[ 
-1
+\frac{1}{\nu q} \left(
1-2 \nu
\right)
\right]\nn\\
&&+\frac{\nu^2 \tilde{\vec{L}} \cdot \tilde{\vec{S}}_1 (\tilde{\vec{S}}_1 \cdot \vec{n})^2}{\tilde{r}^6} \left[
-5 \nu -36
+ \frac{\tilde{L}^2}{\tilde{r}}\left(
5 \nu -1
\right)
+\tilde{p}_r^2 \tilde{r} \left(
\frac{35 \nu }{2}-4
\right) \right.\nn\\
&&\left.+\frac{1}{\nu q} \left(
-5 \nu ^2-\frac{109 \nu }{2}+36
+ \frac{\tilde{L}^2}{\tilde{r}} \left(
5 \nu ^2-7 \nu +1
\right) +\tilde{p}_r^2 \tilde{r} \left(
\frac{35 \nu ^2}{2}-\frac{51 \nu }{2}+4
\right) \right) \right]\nn\\
&&+  \frac{\nu^2 \tilde{p}_r \tilde{\vec{L}} \cdot \tilde{\vec{S}}_1  \tilde{\vec{S}}_1 \cdot \vec{n}  \tilde{\vec{S}}_1 \times \tilde{\vec{L}} \cdot \vec{n} }{\tilde{r}^6} \left[ 
-5 \nu -3
+\frac{1}{\nu q} \left(
-5 \nu ^2-\nu +3
\right)
\right],
\eea
\bea
\tilde{H}^{\text{NLO}}_{\text{S}_1^2 \text{S}_2}  &=& \frac{\nu^2 \tilde{\vec{L}} \cdot \tilde{\vec{S}}_1 \tilde{\vec{S}}_1 \cdot \tilde{\vec{S}}_2}{\tilde{r}^6} \left[
-\frac{\nu }{4}-\frac{21}{2}
-\frac{9 \nu }{8} \frac{\tilde{L}^2}{\tilde{r}} 
+\frac{9 \nu }{8} \tilde{p}_r^2 \tilde{r}   \right.\nn\\
&&\left. +\frac{1}{ q} \left(
5-\frac{\nu }{4}
+ \frac{\tilde{L}^2}{\tilde{r}} \left(
\frac{3}{8}-\frac{9 \nu }{4}
\right)
+\tilde{p}_r^2 \tilde{r} \left(
-\frac{33 \nu }{8}-\frac{45}{8}
\right) \right) \right]\nn\\
&&+  \frac{\nu^2 \tilde{\vec{L}} \cdot \tilde{\vec{S}}_2 \tilde{S}_1^2}{\tilde{r}^6} \left[
-\frac{5 \nu }{4}-\frac{21}{4}
-\frac{15 \nu }{4} \tilde{p}_r^2 \tilde{r}     +\frac{1}{ q} \left(
-\frac{5 \nu }{4}-\frac{7}{2}
+\tilde{p}_r^2 \tilde{r} \left(
\frac{3}{2}-\frac{45 \nu }{16}
\right) \right) \right]\nn\\
&&+  \frac{\nu^2 (\tilde{\vec{L}} \cdot \tilde{\vec{S}}_1)^2 \tilde{\vec{L}} \cdot \tilde{\vec{S}}_2}{\tilde{r}^7} \left[ 
\frac{9 \nu }{4}
+\frac{1}{ q} \left(
\frac{15 \nu }{8}-\frac{3}{4}
\right)
\right]\nn\\
&&+\frac{\nu^2 \tilde{\vec{L}} \cdot \tilde{\vec{S}}_1 \tilde{\vec{S}}_1 \cdot \vec{n} \tilde{\vec{S}}_2 \cdot \vec{n} }{\tilde{r}^6} \left[
\frac{45}{2}-\frac{15 \nu }{4}
+ 3 \nu \frac{\tilde{L}^2}{\tilde{r}}
-9 \nu \tilde{p}_r^2 \tilde{r}  \right.\nn\\
&& \left. +\frac{1}{ q} \left(
-\frac{15 \nu }{4}-\frac{71}{2}
+ \frac{\tilde{L}^2}{\tilde{r}} \left(
\frac{63 \nu }{8}+\frac{15}{4}
\right) 
+\tilde{p}_r^2 \tilde{r} \left(
\frac{93 \nu }{8}+\frac{9}{2}
\right) \right) \right]\nn\\
&&+\frac{\nu^2 \tilde{\vec{L}} \cdot \tilde{\vec{S}}_2 (\tilde{\vec{S}}_1 \cdot \vec{n} )^2 }{\tilde{r}^6} \left[
\frac{33}{4}-\frac{3 \nu }{4}
+ \frac{9 \nu }{8} \frac{\tilde{L}^2}{\tilde{r}}
+ \frac{9 \nu }{8} \tilde{p}_r^2 \tilde{r}  \right.\nn\\
&& \left. +\frac{1}{ q} \left(
-\frac{3 \nu }{4}-\frac{7}{2}
+ \frac{\tilde{L}^2}{\tilde{r}} \left(
\frac{39 \nu }{16}+\frac{9}{8}
\right) 
+\tilde{p}_r^2 \tilde{r} \left(
\frac{69 \nu }{16}-\frac{15}{8}
\right) \right) \right]\nn\\
&&+  \frac{\nu^2 \tilde{p}_r   \tilde{\vec{S}}_1 \cdot \vec{n}  \tilde{\vec{S}}_1 \times \tilde{\vec{S}}_2 \cdot \vec{n} }{\tilde{r}^5} \left[
\frac{15 \nu }{4}-3 
-\frac{51 \nu }{8} \frac{\tilde{L}^2}{\tilde{r}}
+ \frac{21 \nu }{8} \tilde{p}_r^2 \tilde{r}  \right.\nn\\ 
&& \left. +\frac{1}{ q} \left(
\frac{15 \nu }{4}+\frac{37}{2}
+ \frac{\tilde{L}^2}{\tilde{r}} \left(
-\frac{27 \nu }{8}-\frac{45}{8}
\right)
+\tilde{p}_r^2 \tilde{r} \left(
\frac{3}{8}-\frac{3 \nu }{2}
\right) \right) \right]\nn\\
&&+  \frac{\nu^2 \tilde{p}_r \tilde{\vec{L}} \cdot \tilde{\vec{S}}_1  \tilde{\vec{S}}_2 \cdot \vec{n}  \tilde{\vec{S}}_1 \times \tilde{\vec{L}} \cdot \vec{n} }{\tilde{r}^6} \left[ 
\frac{21 \nu }{4}
+\frac{1}{ q} \left(
\frac{15 \nu }{8}-\frac{21}{4}
\right)
\right]\nn\\
&&+  \frac{\nu^2 \tilde{p}_r \tilde{\vec{L}} \cdot \tilde{\vec{S}}_2  \tilde{\vec{S}}_1 \cdot \vec{n}  \tilde{\vec{S}}_1 \times \tilde{\vec{L}} \cdot \vec{n} }{\tilde{r}^6} \left[ 
9 \nu
+\frac{1}{ q} \left(
\frac{15 \nu }{4}+3
\right)
\right],
\eea
\bea
\tilde{H}^{\text{NLO}}_{(\text{ES}_1^2 ) \text{S}_2}&=&\frac{\nu^2 \tilde{\vec{L}} \cdot \tilde{\vec{S}}_1 \tilde{\vec{S}}_1 \cdot \tilde{\vec{S}}_2}{\tilde{r}^6} \left[
\frac{3 \nu }{4}+\frac{1}{2}
-3 \nu \tilde{p}_r^2 \tilde{r}    +\frac{1}{ q} \left(
\frac{3 \nu }{4}-8
+\tilde{p}_r^2 \tilde{r} \left(
-\frac{15 \nu }{4}-3
\right) \right) \right]\nn\\
&&+  \frac{\nu^2 \tilde{\vec{L}} \cdot \tilde{\vec{S}}_2 \tilde{S}_1^2}{\tilde{r}^6} \left[
-\frac{\nu }{4}-\frac{41}{4}
+\frac{\tilde{L}^2}{\tilde{r}} \left(
\frac{15 \nu }{8}-\frac{15}{16}
\right)
+ \tilde{p}_r^2 \tilde{r} \left(
\frac{99 \nu }{8}-\frac{15}{16}
\right)   \right.\nn\\
&&\left. +\frac{1}{ q} \left(
-\frac{\nu }{4}-16
+\frac{39 \nu }{16}\frac{\tilde{L}^2}{\tilde{r}}
+\tilde{p}_r^2 \tilde{r} \left(
\frac{249 \nu }{16}-9
\right) \right) \right] +  \frac{3\nu^2 (\tilde{\vec{L}} \cdot \tilde{\vec{S}}_1)^2 \tilde{\vec{L}} \cdot \tilde{\vec{S}}_2}{q\tilde{r}^7} \nn\\
&&+\frac{\nu^2 \tilde{\vec{L}} \cdot \tilde{\vec{S}}_1 \tilde{\vec{S}}_1 \cdot \vec{n} \tilde{\vec{S}}_2 \cdot \vec{n} }{\tilde{r}^6} \left[
14-\frac{15 \nu }{4}
+  \frac{\tilde{L}^2}{\tilde{r}}\left(
\frac{15}{8}-\frac{15 \nu }{4}
\right)
+ \tilde{p}_r^2 \tilde{r}\left(
\frac{15}{8}-\frac{51 \nu }{4}
\right)  \right.\nn\\
&& \left. +\frac{1}{ q} \left(
55-\frac{15 \nu }{4}
-\frac{39 \nu }{8}\frac{\tilde{L}^2}{\tilde{r}} 
+\tilde{p}_r^2 \tilde{r} \left(
6-\frac{129 \nu }{8}
\right) \right) \right]\nn\\
&&+\frac{\nu^2 \tilde{\vec{L}} \cdot \tilde{\vec{S}}_2 (\tilde{\vec{S}}_1 \cdot \vec{n} )^2 }{\tilde{r}^6} \left[
\frac{3 \nu }{2}+\frac{121}{4}
+  \frac{\tilde{L}^2}{\tilde{r}} \left(
\frac{45}{16}-\frac{45 \nu }{8}
\right)
+  \tilde{p}_r^2 \tilde{r} \left(
\frac{45}{16}-\frac{273 \nu }{8}
\right) \right.\nn\\
&& \left. +\frac{1}{ q} \left(
\frac{3 \nu }{2}+44
+ \frac{\tilde{L}^2}{\tilde{r}} \left(
3-\frac{117 \nu }{16}
\right) 
+\tilde{p}_r^2 \tilde{r} \left(
6-\frac{687 \nu }{16}
\right) \right) \right]\nn\\
&&+   \frac{\nu^2 \tilde{p}_r   \tilde{\vec{S}}_1 \cdot \vec{n}  \tilde{\vec{S}}_1 \times \tilde{\vec{S}}_2 \cdot \vec{n} }{\tilde{r}^5} \left[ 
\frac{31}{2}-\frac{3 \nu }{4}
+\frac{\tilde{L}^2}{\tilde{r}} \left(
\frac{15}{8}-\frac{15 \nu }{4}
\right)
+\tilde{p}_r^2 \tilde{r} \left(
\frac{15}{8}-\frac{63 \nu }{4}
\right)\right.\nn\\
&&\left. +\frac{1}{ q} \left(
-\frac{3 \nu }{4}-11
-\frac{39 \nu }{8} \frac{\tilde{L}^2}{\tilde{r}}
+\tilde{p}_r^2 \tilde{r} \left(
3-\frac{159 \nu }{8}
\right)
\right)
\right]\nn\\
&&+  \frac{\nu^2 \tilde{p}_r \tilde{\vec{L}} \cdot \tilde{\vec{S}}_1  \tilde{\vec{S}}_2 \cdot \vec{n}  \tilde{\vec{S}}_1 \times \tilde{\vec{L}} \cdot \vec{n} }{\tilde{r}^6} \left[ 
3 \nu
+\frac{1}{ q} \left(
\frac{15 \nu }{4}+3
\right)
\right]\nn\\
&&+  \frac{\nu^2 \tilde{p}_r \tilde{\vec{L}} \cdot \tilde{\vec{S}}_2  \tilde{\vec{S}}_1 \cdot \vec{n}  \tilde{\vec{S}}_1 \times \tilde{\vec{L}} \cdot \vec{n} }{\tilde{r}^6} \left[ 
12 \nu
+\frac{1}{ q} \left(
15 \nu +6
\right)
\right].
\eea
As can be seen, the COM specification already results significant simplification of the general 
Hamiltonians. Yet the COM simplification has been the most detailed form of results provided in 
recent popular work via scattering-amplitudes methods which treat the unbound scattering problem. 

\subsubsection*{Comparison of COM results from two scattering-amplitudes works.}
Chen et al.~in [40], and Bern et al.~in [41], presented COM Hamiltonians for BHs, and for generic 
compact objects, respectively. To start with, their results are discrepant with each other in several 
ways, and already at the initial level of scattering amplitudes, from which their COM Hamiltonians 
are derived.
As we note below discrepancies of the two works with our results, we also touch on the discrepancies 
between them.
\begin{enumerate}
\item \textbf{Chen et al.}
In \cite{Chen:2021qkk} COM Hamiltonians for the case of BHs were presented in eqs.~(3.96), (3.97), 
for the S$_1^2$S$_2$, and S$_1^3$ parts of the potential, respectively. 
Both of these parts are discrepant with our results.  
As can be seen in their section $3.4$, \cite{Chen:2021qkk} could not succeed in relating their 
results to ours via canonical transformations, beyond the quadratic-in-spin sectors 
\cite{Levi:2015msa}, which were already well-confirmed then.
As noted in version $2$ of \cite{Chen:2021qkk}, ref.~\cite{Morales:2021} with our final results 
here was fully shared on summer $2021$ with Chen et al.~upon their request, and this discrepancy 
with our results has been known prior to version $1$ of \cite{Chen:2021qkk}. 

Moreover, \cite{Chen:2021qkk} presented in appendix B their preliminary scattering amplitudes 
for generic compact objects, in which they were notably missing the new contribution proportional 
to $C_{\text{ES}^2}^2$, that would correspond to our eq.~\eqref{hamses2sq} in the Hamiltonian. 
As this new sector contributes even when results are specified to the case of BHs, it is also 
clearly missing from their eq.~(3.97) in \cite{Chen:2021qkk} for the Hamiltonian. 
As we noted at the end of section \ref{finalactions}, and as we shall see in section 
\ref{selfchecks3} below, this new sector is also imposed by Poincar\'e invariance, 
and therefore the Hamiltonian of \cite{Chen:2021qkk} is generally not Poincar\`e-invariant, 
and thus is not only canonically inequivalent with our Hamiltonian, but is also inequivalent 
with any physically correct Hamiltonian. 

More generally, it should also be noted that the results in \cite{Chen:2021qkk} are discrepant 
already at the level of scattering amplitudes, even when restricted to the case of BHs, 
with various corresponding amplitudes, as in \cite{Bern:2022kto}, including the absence in 
\cite{Chen:2021qkk} of a part that is proportional to $C_{\text{ES}^2}^2$. 

\item \textbf{Bern et al.} The COM Hamiltonians for generic compact objects found in the ancillary 
files of \cite{Bern:2022kto} are discrepant with ours, even at the LO, and even when specified 
to the simpler case of BHs. 
As noted in \cite{Bern:2022kto}, their Hamiltonians contain, as of the cubic order in spin, new 
unfamiliar singularities in the COM momenta, $1/p^2$, where it was claimed there, that these 
singularities drop out for the case of BHs. 
However, it is easy to verify already at the LO cubic-in-spin sectors in \cite{Bern:2022kto}, that 
these singularities remain, even after restricting to the case of BHs, namely when the $2$ Wilson 
coefficients in these leading sectors are specified to unity, 
$C_{\text{ES}^2} = C_{\text{BS}^3} = 1$, as stipulated in our \cite{Levi:2015msa}. Thus the results 
in \cite{Bern:2022kto} are discrepant with our LO results in \cite{Levi:2014gsa} from $2014$, where 
those have since been well-confirmed via numerous independent methodologies, including in traditional 
GR methods. 

At the NLO the COM Hamiltonians in \cite{Bern:2022kto} for the cubic-in-spin sectors display similar 
singularities, even for BHs. 
Moreover, at the NLO the work in \cite{Bern:2022kto} included contributions with a claimed new 
Wilson coefficient, $H_2$, that they stipulated in their formulation. Such an extra free parameter 
violates spin-gauge invariance \cite{Levi:2015msa}, and is also discrepant and absent in other 
corresponding scattering amplitudes, as in \cite{Chen:2021qkk}. 

Finally, the singularities that appear in \cite{Bern:2022kto} as of the LO, as well as the above 
noted extra independent piece at the NLO in \cite{Bern:2022kto}, cannot be generated by canonical 
transformations, and therefore the Hamiltonians in \cite{Bern:2022kto} are not canonically 
equivalent to our well-verified Hamiltonians at the LO \cite{Levi:2014gsa}, nor to our new present 
Hamiltonians at the NLO, which we verified via Poincar\'e invariance, as shall be seen in 
section \ref{selfchecks3} below. 

\end{enumerate}

Next, we can further restrict the Hamiltonians to the aligned-spins configuration, in which the 
spins satisfy $\vec{S}_a \cdot \vec{n} = \vec{S}_a \cdot \vec{p}=0$, namely they are both aligned 
with the orbital angular momentum. It should be highlighted that the higher in spin the sectors are, 
the more dramatically they are affected by this simplification, with a greater loss of physical 
information as a consequence. This is in contrast to the simple spin-orbit sector, in which the single 
spin and the angular momentum are still trivially coupled \cite{Antonelli:2020ybz}. In the present higher-spin sectors, applying the 
aligned-spins constraints to our COM Hamiltonians yields:
\bea
\label{alignedham}
\tilde{H}^{\text{NLO}}_{\text{S}^3} &=&\frac{\nu^2 \tilde{L} \tilde{S}_1^3  }{\tilde{r}^6} \left[ -3 \nu -\frac{9}{4} + \frac{\tilde{L}^2}{\tilde{r}} \left( 2 \nu -\frac{11}{8} \right) + \tilde{p}_r^2 \tilde{r} \left( \frac{33}{16}-\frac{39 \nu }{8} \right) \right.\nn\\
&& + \frac{1}{\nu q} \left(-3 \nu ^2-\frac{3 \nu }{2}+\frac{9}{4}+ \frac{\tilde{L}^2}{\tilde{r}} \left( \frac{15 \nu ^2}{8}-\frac{19 \nu }{4}+\frac{11}{8} \right) + \tilde{p}_r^2 \tilde{r} \left( -\frac{75 \nu ^2}{16}+9 \nu -\frac{33}{16} \right) \right)\nn\\
&&+ C_{1 (\text{ES}^2)} \left( 2 \nu -\frac{3}{2} + \frac{\tilde{L}^2}{\tilde{r}} \left( \frac{9 \nu }{8}+\frac{3}{16} \right) + \tilde{p}_r^2 \tilde{r} \left( \frac{9 \nu }{8}+\frac{63}{16} \right) \right.\nn\\
&&\left. + \frac{1}{\nu q} \left( 2 \nu ^2-\frac{17 \nu }{2}+\frac{3}{2}+ \frac{\tilde{L}^2}{\tilde{r}} \left( \frac{27 \nu ^2}{16}-\frac{3 \nu }{4}-\frac{3}{16} \right) + \tilde{p}_r^2 \tilde{r} \left( \frac{57 \nu ^2}{16}+\frac{27 \nu }{4}-\frac{63}{16} \right) \right) \right)\nn\\
&&+ C_{1 (\text{BS}^3)} \left( \nu +7 -\nu \frac{\tilde{L}^2}{\tilde{r}}  -\frac{7 \nu }{2} \tilde{p}_r^2 \tilde{r}  + \frac{1}{\nu q} \left( \nu ^2+\frac{21 \nu }{2}-7 + \frac{\tilde{L}^2}{\tilde{r}} \left( \nu -\nu ^2 \right)\right. \right.\nn\\
&&\left.\left.\left. + \tilde{p}_r^2 \tilde{r} \left( \frac{7 \nu }{2}-\frac{7 \nu ^2}{2} \right) \right) \right) \right]
+ \frac{\nu^2 \tilde{L} \tilde{S}_1^2 \tilde{S}_2 }{\tilde{r}^6} \left[ -\frac{3 \nu }{2}-\frac{63}{4} + \frac{9 \nu }{8} \frac{\tilde{L}^2}{\tilde{r}}   -\frac{21 \nu }{8} \tilde{p}_r^2 \tilde{r} \right.\nn\\
&&+ \frac{1}{q} \left( \frac{3}{2}-\frac{3 \nu }{2} + \frac{\tilde{L}^2}{\tilde{r}} \left( -\frac{3 \nu }{8}-\frac{3}{8} \right) + \tilde{p}_r^2 \tilde{r} \left( -\frac{111 \nu }{16}-\frac{33}{8} \right)  \right)  \nn\\
&&+ C_{1 (\text{ES}^2)} \left( \frac{\nu }{2}-\frac{39}{4} + \frac{\tilde{L}^2}{\tilde{r}} \left( \frac{15 \nu }{8}-\frac{15}{16} \right) + \tilde{p}_r^2 \tilde{r} \left( \frac{75 \nu }{8}-\frac{15}{16} \right) \right.\nn\\
&&\left.\left. + \frac{1}{ q} \left( \frac{\nu }{2}-24+ \frac{\tilde{L}^2}{\tilde{r}} \left( \frac{39 \nu }{16}+3 \right) + \tilde{p}_r^2 \tilde{r} \left( \frac{189 \nu }{16}-12 \right) \right) \right) \right] + (1 \leftrightarrow 2),
\eea
where as noted above, the significant loss of physical information in higher-spin sectors, due to the 
aligned-spins simplification, even in comparison with the already simplified COM Hamiltonian, is 
evident. Moreover, 
notably one of the $6$ sectors of the potential drops upon this simplification -- the new distinct  
sector that appears in eqs.~\eqref{sqCES2}, \eqref{hamces2sq}, \eqref{hamses2sq}, with the 
$C_{\text{ES}^2}^2$ prefactor. Accordingly, we see now that this new unique feature will not show up in any of 
the familiar observables, be it for GWs, or in the scattering problem, which are all in the aligned-spins simplified kinematic configuration.

A final simplification appropriate for the inspiral phase, where the orbit is quasi-circular, is 
that the necessary circular-orbit condition, $p_r \equiv \vec{p} \cdot \vec{n}=0 \Rightarrow p^2=p_r^2+L^2/r^2\to L^2/r^2$, is satisfied. Further applying this condition to our aligned-spins Hamiltonians yields:
\bea
\tilde{H}^{\text{NLO}}_{\text{S}^3} &=&\frac{\nu^2 \tilde{L} \tilde{S}_1^3  }{\tilde{r}^6} \left[ -3 \nu -\frac{9}{4} + \frac{\tilde{L}^2}{\tilde{r}} \left( 2 \nu -\frac{11}{8} \right)   + \frac{1}{\nu q} \left(-3 \nu ^2-\frac{3 \nu }{2}+\frac{9}{4}\right.\right.\nn\\
&&\left.+ \frac{\tilde{L}^2}{\tilde{r}} \left( \frac{15 \nu ^2}{8}-\frac{19 \nu }{4}+\frac{11}{8} \right)  \right)
+ C_{1 (\text{ES}^2)} \left( 2 \nu -\frac{3}{2} + \frac{\tilde{L}^2}{\tilde{r}} \left( \frac{9 \nu }{8}+\frac{3}{16} \right)  \right.\nn\\
&&\left. + \frac{1}{\nu q} \left( 2 \nu ^2-\frac{17 \nu }{2}+\frac{3}{2}+ \frac{\tilde{L}^2}{\tilde{r}} \left( \frac{27 \nu ^2}{16}-\frac{3 \nu }{4}-\frac{3}{16} \right)  \right) \right)\nn\\
&&\left.+ C_{1 (\text{BS}^3)} \left( \nu +7 -\nu \frac{\tilde{L}^2}{\tilde{r}}    + \frac{1}{\nu q} \left( \nu ^2+\frac{21 \nu }{2}-7 + \frac{\tilde{L}^2}{\tilde{r}} \left( \nu -\nu ^2 \right) \right) \right) \right]\nn\\
&&+ \frac{\nu^2 \tilde{L} \tilde{S}_1^2 \tilde{S}_2 }{\tilde{r}^6} \left[ -\frac{3 \nu }{2}-\frac{63}{4} + \frac{9 \nu }{8} \frac{\tilde{L}^2}{\tilde{r}}   + \frac{1}{q} \left( \frac{3}{2}-\frac{3 \nu }{2} + \frac{\tilde{L}^2}{\tilde{r}} \left( -\frac{3 \nu }{8}-\frac{3}{8} \right)   \right) \right. \nn\\
&&\left.+ C_{1 (\text{ES}^2)} \left( \frac{\nu }{2}-\frac{39}{4} + \frac{\tilde{L}^2}{\tilde{r}} \left( \frac{15 \nu }{8}-\frac{15}{16} \right)   + \frac{1}{ q} \left( \frac{\nu }{2}-24+ \frac{\tilde{L}^2}{\tilde{r}} \left( \frac{39 \nu }{16}+3 \right)  \right) \right) \right]\nn\\
&& + (1 \leftrightarrow 2).
\eea

\section{Poincar\'e Algebra}
\label{Poincare}

The global Poincar\'e symmetry of isolated $N$-body systems in GR provides a powerful 
self-consistency check for the validity of general PN Hamiltonians in an arbitrary reference frame. 
From Noether's theorem, this global symmetry implies the existence of conserved integrals of motion, 
which form a representation of the Poincar\'e algebra in phase space. That is, the generators of 
Poincar\'e transformations satisfy the algebra that reads:
\begin{align}
\label{auto}
\{P_i, P_j\} = \{P_i, H\} = \{J_i, H\}=0, 
\quad \{J_i, J_j\} & = & \epsilon_{ijk}J_k, 
\quad \{J_i, P_j\} & = & \epsilon_{ijk}P_k, \\
\label{com}
\{G_i, P_j\} = \delta_{ij}H,
\quad \{G_i, H\} = P_i,
\quad \{G_i, G_j\} & = & -\epsilon_{ijk}J_k, 
\quad \{J_i, G_j\} & = & \epsilon_{ijk}G_k,
\end{align}
with $\vec{P}$ the total linear momentum, $H$ the Hamiltonian, $\vec{J}$ the total angular 
momentum, and $\vec{K}$ the boost generator, which is traded here for $\vec{G}$, the generalized 
relativistic ``center-of-mass'' (henceforth center-of-mass), using $\vec{K}\equiv 
\vec{G}-t\vec{P}$. Thus, $G_i/H$ is the center-of-mass that forms a canonical pair with the total 
linear momentum, $P_i$, but note that this center-of-mass does not satisfy the vanishing Poisson 
brackets of Newtonian center-of-mass vectors, but rather the relativistic Wigner rotation, 
$\{G_i, G_j\} = -\epsilon_{ijk}J_k$. 
Notice that the Poisson brackets in the sectors with spins are extended to include spin variables 
via the generalization \cite{Hanson:1974qy,Bel:1980}:
\bea
\{ f, g \} \equiv \{ f, g \}_x + \{ f, g \}_S,
\eea
with
\bea
\{ f, g \}_x &=& \sum_{I=1}^2 \left(\frac{\partial f}{\partial x_I} \cdot \frac{\partial g}{\partial p_I} - \frac{\partial f}{\partial p_I} \cdot \frac{\partial g}{\partial x_I} \right),\\
\{ f, g \}_S &=& \sum_{I=1}^2 S_I \times \frac{\partial f}{\partial S_I} \cdot \frac{\partial g}{\partial S_I},
\eea
where all terms on the RHS are understood to be vectors in scalar or triple products. 
By construction $H$ satisfies translation and rotation invariance, and thus the Poisson brackets in eq.~\eqref{auto} are trivially satisfied.
However, it is  far from trivial to solve for the center-of-mass, $\vec{G}$, which 
should satisfy the Poisson brackets in eq.~\eqref{com} with:
\bea
\vec{P} = \vec{p}_1 + \vec{p}_2, \quad \vec{J} = \sum_{I=1}^2 \left( \vec{x}_I \times \vec{p}_I +  \vec{S}_I \right),
\eea
for our binary system, and thus complete the full Poincar\'e algebra. 
Let us then turn to accomplish this ambitious task.

First, for $\vec{G}$ to satisfy: 
\be
\{G_i, P_j\} = \delta_{ij}H,
\ee
it should have the following form:
\bea
\vec{G} = h_1 \vec{x}_1 + h_2 \vec{x}_2 + \vec{Y},
\quad h_1 + h_2 = H,
\label{eq:padhY}
\eea
where $h_I$ and $\vec{Y}$ are translation-invariant, namely:
\bea
\{ h_I, P_i \} = \{ Y_i, P_j \} = 0,
\eea
which is equivalent to requiring that the dependence of $h_I$ and $\vec{Y}$ on the position 
variables should be only through $\vec{x}_1 - \vec{x}_2$, i.e.~in terms of $\vec{n}$ and $r$. 
$\vec{G}$ can then be uniquely solved from the constraint:
\bea
\label{tosolvefor1}
\{G_i, H\} & = & P_i.
\eea
As $\vec{P}$ and $H$ are symmetric under the exchange of worldline labels, $1 \leftrightarrow 2$, 
$\vec{G}$ should be symmetric under this exchange as well. This means that $\vec{Y}$ needs to be 
symmetric under $1 \leftrightarrow 2$, and $h_I$ can be written as:
\bea
h_I = \frac{H}{2} + h_I^{\text{AS}},
\label{eq:padec}
\eea
where $h_1^{AS} = -h_2^{AS}$ needs to be antisymmetric under $1 \leftrightarrow 2$. Since this 
just means that $h_1\vec{x}_1+h_2\vec{x}_2=H(\vec{x}_1+\vec{x}_2)/2+h_1^{AS}r\vec{n}$, we can just 
recast $\vec{G}$ as:
\be
\vec{G} = H (\vec{x}_1 + \vec{x}_2)/2 + \vec{Y},
\ee
where $h_1^{AS}r\vec{n}$ is just contained in $\vec{Y}$. Our task thus boils down to constructing 
$\vec{Y}$ from the most general constrained ansatz, and finding the unique solution for it, using 
eq.~\eqref{tosolvefor1}.

Let us then list the considerations for the construction of $\vec{G}$. First, the solution of 
$\vec{G}$ is decomposed into different sectors classified according to their PN order, spin 
order in S$_1$ and S$_2$, as well as possible factors of Wilson coefficients. The building blocks for 
$\vec{Y}$ are the vectors: $\vec{n}$, $\vec{p}_I$, $\vec{S}_I$, and we use dimensional analysis 
and Euclidean covariance, including parity invariance and time-reversal. 
The constraint in eq.~\eqref{com}, $\{J_i, G_j\} = \epsilon_{ijk}G_k$,
is automatically satisfied as long as $\vec{G}$ is constructed from vectors, $\vec{x}_I$, 
$\vec{p}_I$ and $\vec{S}_I$, to satisfy Euclidean covariance, such that $\vec{G}$ behaves as a 
vector under rotation.
One subtle point though is that at $3$ dimensions every $4$ vectors are dependent through the 
general identity:
\bea
\vec{A} (\vec{B} \cdot \vec{C} \times \vec{D}) 
- \vec{B} (\vec{C} \cdot \vec{D} \times \vec{A}) 
+ \vec{C} (\vec{D} \cdot \vec{A} \times \vec{B}) 
- \vec{D} (\vec{A} \cdot \vec{B} \times \vec{C}) 
= 0,
\label{eq:cyclid}
\eea
which can hide a certain redundancy in a general ansatz for sectors that contain more than $3$ 
vectors, as in any of the sectors with spins. 

Finally, in flat spacetime, i.e.~where $G_N \to 0$, 
the relativistic COM generator can be written in the following closed form \cite{Bel:1980}:
\bea
\vec{G}_{\text{flat}} = \sum_{I=1}^2 \left( \gamma_I m_I \vec{x}_I 
- \frac{\vec{S}_I \times \vec{p}_I}{m_I (1+ \gamma_I)}\right),
\label{comflat}
\eea
where $\gamma_I = \sqrt{1+ p_I^2/m_I^2}$. Eq.~\eqref{comflat} is then used to fix the
$\mathcal{O} (G_N^0)$ terms in $\vec{G}$ to agree with the special-relativistic limit. If the latter 
is used to constrain $\vec{G}$, then the remainder critical Poisson brackets involving the Wigner 
rotation, $\{G_i, G_j\} = -\epsilon_{ijk}J_k$, are also automatically satisfied in the solution for 
$\vec{G}$.

Following the various considerations above, we proceeded to solve for the full Poincar\'e algebra 
of the new complete precision-frontier at the $4.5$PN order, which includes only sectors with spins: 
The NLO cubic-in-spin sectors from our \cite{Levi:2019kgk} with the full general Hamiltonian first 
presented in section \ref{allhams} above, and the N$^3$LO spin-orbit sector with the general 
Hamiltonian provided first in our \cite{Kim:2022pou}. It should be highlighted 
that for the latter sector the general ansatz to solve for, contains an order of $\sim10^3$ free 
dimensionless coefficients. Therefore we needed to scale the solution of this problem, even 
compared to the most advanced Poincar\'e algebra at the $4$PN order, that we provided in 
\cite{Levi:2016ofk}. Note that the comprehensive construction of the Poincar\'e algebra is 
particularly strong as a consistency check of all sectors at the $4.5$PN order: Since there are no new 
Wilson coefficients introduced in any of the sectors at this order, the fulfilment of the Poincar\'e 
algebra is non-trivial for each of the relevant subsectors, so that they are all tested by the 
requirement of Poincar\'e invariance. This comprehensive check will thus establish the $4.5$PN order 
as the new precision frontier.

\subsection{NLO Cubic-in-Spin Sectors}
\label{selfchecks3}

Let us then enumerate all the sectors in $H$ and $\vec{G}$ relevant to the solution of $\vec{G}$ 
at the present NLO cubic-in-spin sectors. For the Hamiltonian we have:
\bea
H = H_{\text{N}} + H_{1\text{PN}} + H^{\text{LO}}_{\text{SO}} + H^{\text{LO}}_{\text{S}^2} 
+ H^{\text{NLO}}_{\text{SO}} + H^{\text{NLO}}_{\text{S}^2} 
+ H^{\text{LO}}_{\text{S}^3} + H^{\text{NLO}}_{\text{S}^3}, 
\eea
with
\bea
H^{\text{LO}}_{\text{SO}} &=& H^{\text{LO}}_{\text{S}_1} + (1 \leftrightarrow 2), \qquad H^{\text{NLO}}_{\text{SO}} = H^{\text{NLO}}_{\text{S}_1} + (1 \leftrightarrow 2),\\
H^{\text{LO}}_{\text{S}^2} &=& C_{1 \text{ES}^2} H^{\text{LO}}_{\text{ES}_1^2} + H^{\text{LO}}_{\text{S}_1 \text{S}_2} + (1 \leftrightarrow 2),\\
H^{\text{NLO}}_{\text{S}^2} &=& H^{\text{NLO}}_{\text{S}_1^2} + C_{1 \text{ES}^2} H^{\text{NLO}}_{\text{ES}_1^2} + H^{\text{NLO}}_{\text{S}_1 \text{S}_2} + (1 \leftrightarrow 2),\\
H^{\text{LO}}_{\text{S}^3} &=& C_{1 \text{ES}^2} H^{\text{LO}}_{(\text{ES}_1^2) \text{S}_1} 
+ C_{1 \text{BS}^3} H^{\text{LO}}_{\text{BS}_1^3}  + H^{\text{LO}}_{\text{S}_1^2 \text{S}_2} 
+ C_{1 \text{ES}^2} H^{\text{LO}}_{(\text{ES}_1^2) \text{S}_2} + (1 \leftrightarrow 2),\
\eea
while $H^{\text{NLO}}_{\text{S}^3}$ is given in eq.~\eqref{eq:hamcns3}. 
For the generalized COM, $\vec{G}$, we have:
\bea
\vec{G} = \vec{G}_{\text{N}} + \vec{G}_{1\text{PN}} + \vec{G}^{\text{LO}}_{\text{SO}} + \vec{G}^{\text{NLO}}_{\text{SO}} + \vec{G}^{\text{NLO}}_{\text{S}^2} + \vec{G}^{\text{NLO}}_{\text{S}^3},
\eea
with 
\bea
\vec{G}^{\text{LO}}_{\text{SO}} &=& \vec{G}^{\text{LO}}_{\text{S}_1} + (1 \leftrightarrow 2), \qquad \vec{G}^{\text{NLO}}_{\text{SO}} = \vec{G}^{\text{NLO}}_{\text{S}_1} + (1 \leftrightarrow 2),\\
\vec{G}^{\text{LO}}_{\text{S}^2} & = & \vec{G}^{\text{LO}}_{\text{S}^3}=\vec{0},\\
\vec{G}^{\text{NLO}}_{\text{S}^2} &=& \vec{G}^{\text{NLO}}_{\text{S}_1^2} + C_{1 \text{ES}^2} \vec{G}^{\text{NLO}}_{\text{ES}_1^2} + \vec{G}^{\text{NLO}}_{\text{S}_1 \text{S}_2} + (1 \leftrightarrow 2),\ 
\eea
and we need to solve for:
\begin{align}
\vec{G}^{\text{NLO}}_{\text{S}^3} = &
\vec{G}^{\text{NLO}}_{\text{S}_1^3 } 
+C_{1\text{ES}^2} \vec{G}^{\text{NLO}}_{(\text{ES}_1^2 ) \text{S}_1} 
 +  C_{1\text{ES}^2}^2 \vec{G}^{\text{NLO}}_{C_{\text{ES}_1^2}^2 \text{S}_1^3 }
+ C_{1\text{BS}^3} \vec{G}^{\text{NLO}}_{\text{BS}_1^3 } 
+\vec{G}^{\text{NLO}}_{\text{S}_1^2 \text{S}_2}  
+ C_{1\text{ES}^2} \vec{G}^{\text{NLO}}_{(\text{ES}_1^2 ) \text{S}_2} \nn \\
& +  (1 \leftrightarrow 2).
\end{align}
Note that all generators for LO sectors with spins (beyond the spin-orbit sector) are vanishing.

The decomposition above determines which Poisson brackets contribute in eq.~\eqref{tosolvefor1} to 
solve for a given sector. 
$\vec{G}^{\text{NLO}}_{\text{S}_1^3 }$ is solved by:
\bea
0&=&\{ \vec{G}^{\text{NLO}}_{\text{S}_1^3 },H_{\text{N}} \}_x + \{ \vec{G}_{\text{N}},  H^{\text{NLO}}_{\text{S}_1^3 } \}_x + \{ \vec{G}_{\text{S}_1}^{\text{LO}},  H^{\text{NLO}}_{\text{S}_1^2  } \}_x   +  \{ \vec{G}_{\text{S}_1^2}^{\text{NLO}},  H^{\text{LO}}_{\text{S}_1 } \}_x.
\eea
$\vec{G}^{\text{NLO}}_{(\text{ES}_1^2 ) \text{S}_1}$ is solved by:
\bea
0&=&\{ \vec{G}^{\text{NLO}}_{(\text{ES}_1^2 ) \text{S}_1},H_{\text{N}} \}_x+ \{ \vec{G}_{\text{N}},  H^{\text{NLO}}_{(\text{ES}_1^2 ) \text{S}_1} \}_x  + \{ \vec{G}_{\text{S}_1}^{\text{LO}},  H^{\text{NLO}}_{\text{ES}_1^2  } \}_x + \{ \vec{G}_{1\text{PN}},  H^{\text{LO}}_{(\text{ES}_1^2 ) \text{S}_1} \}_x \nn\\
&&+ \{ \vec{G}_{\text{S}_1}^{\text{NLO}},  H^{\text{LO}}_{\text{ES}_1^2  } \}_x +  \{ \vec{G}_{\text{ES}_1^2}^{\text{NLO}},  H^{\text{LO}}_{\text{S}_1 } \}_x +  \{ \vec{G}_{\text{S}_1}^{\text{LO}},  H^{\text{LO}}_{(\text{ES}_1^2 ) \text{S}_1 } \}_S +   \{ \vec{G}_{\text{S}_1^2}^{\text{NLO}},  H^{\text{LO}}_{\text{ES}_1^2 } \}_S.
\eea
$\vec{G}^{\text{NLO}}_{C_{\text{ES}_1^2}^2 \text{S}_1^3 } $ is solved by:
\bea
\label{nontrivialpoincare}
0 = \{ \vec{G}^{\text{NLO}}_{C_{\text{ES}_1^2}^2 \text{S}_1^3} , H_{\text{N}} \}_x   + \{  \vec{G}_{\text{N}} , H^{\text{NLO}}_{C_{\text{ES}_1^2}^2 \text{S}_1^3 } \}_x  + \{ \vec{G}_{\text{ES}_1^2}^{\text{NLO}},  H^{\text{LO}}_{\text{ES}_1^2 } \}_S.
\eea
$\vec{G}^{\text{NLO}}_{\text{BS}_1^3 }$ is solved by:
\bea
0&=&\{ \vec{G}^{\text{NLO}}_{\text{BS}_1^3 },H_{\text{N}} \}_x  + \{ \vec{G}_{\text{N} },H^{\text{NLO}}_{\text{BS}_1^3 } \}_x + \{ \vec{G}_{1\text{PN} },H^{\text{LO}}_{\text{BS}_1^3 } \}_x + \{ \vec{G}_{\text{S}_1 }^{\text{LO}},H^{\text{LO}}_{\text{BS}_1^3 } \}_S.
\eea
$\vec{G}^{\text{NLO}}_{\text{S}_1^2 \text{S}_2}$ is solved by:
\bea
0&=& \{ \vec{G}^{\text{NLO}}_{\text{S}_1^2 \text{S}_2},H_{\text{N}} \}_x  + \{ \vec{G}_{\text{N}},  H^{\text{NLO}}_{\text{S}_1^2 \text{S}_2 } \}_x + \{ \vec{G}_{\text{S}_2}^{\text{LO}},  H^{\text{NLO}}_{\text{S}_1^2  } \}_x + \{ \vec{G}_{\text{S}_1}^{\text{LO}}, 2 H^{\text{NLO}}_{\text{S}_1 \text{S}_2  } \}_x \nn\\
&&+ \{ \vec{G}_{1\text{PN}},  H^{\text{LO}}_{\text{S}_1^2 \text{S}_2 } \}_x   + \{ \vec{G}_{\text{S}_1}^{\text{NLO}},  2H^{\text{LO}}_{\text{S}_1 \text{S}_2 } \}_x + \{ 2\vec{G}_{\text{S}_1 \text{S}_2}^{\text{NLO}},  H^{\text{LO}}_{\text{S}_1  } \}_x +  \{ \vec{G}_{\text{S}_1^2}^{\text{NLO}},  H^{\text{LO}}_{\text{S}_2 } \}_x\nn\\
&&+ \{ \vec{G}_{\text{S}_2}^{\text{LO}},  H^{\text{LO}}_{\text{S}_1^2  \text{S}_2 } \}_S +  \{ \vec{G}_{\text{S}_1}^{\text{LO}},  H^{\text{LO}}_{\text{S}_1^2  \text{S}_2 } \}_S +\left[  \{ 2\vec{G}_{\text{S}_1 \text{S}_2}^{\text{NLO}}, 2 H^{\text{LO}}_{\text{S}_1 \text{S}_2  } \}_S  \right]_{ \text{S}_1^2 \text{S}_2}  +   \{ \vec{G}_{\text{S}_1^2}^{\text{NLO}},  2H^{\text{LO}}_{\text{S}_1 \text{S}_2 } \}_S ,\nn\\
\eea
where $[f]_{ \text{S}_1^{2} \text{S}_2}$ means extracting the part of $f$ that is quadratic in $S_1$ and linear in $S_2$.
Finally, $ \vec{G}^{\text{NLO}}_{(\text{ES}_1^2 ) \text{S}_2}$ is solved by:
\bea
0&=& \{  \vec{G}^{\text{NLO}}_{(\text{ES}_1^2 ) \text{S}_2},H_{\text{N}} \}_x  + \{ \vec{G}_{\text{N}},  H^{\text{NLO}}_{(\text{ES}_1^2) \text{S}_2 } \}_x + \{ \vec{G}_{\text{S}_2}^{\text{LO}},  H^{\text{NLO}}_{\text{ES}_1^2  } \}_x  + \{ \vec{G}_{1\text{PN}},  H^{\text{LO}}_{(\text{ES}_1^2) \text{S}_2 } \}_x \nn\\
&&+ \{ \vec{G}_{\text{S}_2}^{\text{NLO}},  H^{\text{LO}}_{\text{ES}_1^2 } \}_x   +  \{ \vec{G}_{\text{ES}_1^2}^{\text{NLO}},  H^{\text{LO}}_{\text{S}_2 } \}_x + \{ \vec{G}_{\text{S}_2}^{\text{LO}},  H^{\text{LO}}_{(\text{ES}_1^2)  \text{S}_2 } \}_S +  \{ \vec{G}_{\text{S}_1}^{\text{LO}},  H^{\text{LO}}_{(\text{ES}_1^2)  \text{S}_2 } \}_S  \nn\\
&& +  \{ 2\vec{G}_{\text{S}_1 \text{S}_2}^{\text{NLO}},  H^{\text{LO}}_{ \text{ES}_1^2  } \}_S +   \{ \vec{G}_{\text{ES}_1^2}^{\text{NLO}},  2H^{\text{LO}}_{\text{S}_1 \text{S}_2 } \}_S .
\eea
Each of the above $6$ equations corresponds to each of the $6$ subsectors that we saw in the final 
actions of section \ref{finalactions}. Notice that none of these equations is trivial, 
since the NLO cubic-in-spin sectors do not contain any new Wilson coefficients. Even the new 
subsector that is proportional to $C_{\text{ES}_1^2}^2$ depends on the already-encountered Wilson 
coefficient $C_{\text{ES}_1^2}$, and therefore, as can be seen, eq.~\eqref{nontrivialpoincare} is 
non-trivial, since it contains a third term from the spin Poisson brackets of two lower-order sectors. 
Thus even though, as we shall see below in eq.~\eqref{sol1}, the solution for the COM generator of 
$C_{\text{ES}_1^2}^2$ does not contribute to the general COM generator of the present NLO 
cubic-in-spin sectors, eq.~\eqref{nontrivialpoincare} is only fulfilled in a non-trivial manner, which 
actually proves that the new subsector is inevitable due to the requirement of Poincar\'e invariance. 

We recall that the solution for $\vec{G}^{\text{NLO}}_{\text{S}^3}$ is written as:
\bea 
\vec{G}^{\text{NLO}}_{\text{S}^3} &=& H^{\text{LO}}_{\text{S}^3} 
\frac{\left( \vec{x}_1 + \vec{x}_2 \right)}{2} +  \left( \vec{Y}^{\text{NLO}}_{\text{S}^3_1} + \vec{Y}^{\text{NLO}}_{\text{S}^2_1 \text{S}_2} + \left(1 \leftrightarrow 2 \right) \right),
\eea
and we find:
\bea
\label{sol1}
\vec{Y}^{\text{NLO}}_{\text{S}_1^3 } &=& - 	\frac{G C_{1\text{ES}^2} m_{2}}{8 m_{1}{}^3 r{}^3} \Big[ 15 \vec{S}_{1}\times\vec{n}\cdot\vec{p}_{1} S_{1}^2 \vec{n} - 15 \vec{S}_{1}\times\vec{n}\cdot\vec{p}_{1} \vec{n} ( \vec{S}_{1}\cdot\vec{n})^{2} \nn\\ 
&& - 12 \vec{S}_{1}\times\vec{n}\cdot\vec{p}_{1} \vec{S}_{1}\cdot\vec{n} \vec{S}_{1} + 2 S_{1}^2 \vec{S}_{1}\times\vec{p}_{1} + 6 ( \vec{S}_{1}\cdot\vec{n})^{2} \vec{S}_{1}\times\vec{p}_{1} \Big] \nn\\ 
&& + 	\frac{G C_{1\text{BS}^3} m_{2}}{6 m_{1}{}^3 r{}^3} \Big[ 6 \vec{S}_{1}\times\vec{n}\cdot\vec{p}_{1} \vec{S}_{1}\cdot\vec{n} \vec{S}_{1} + S_{1}^2 \vec{S}_{1}\times\vec{p}_{1} - 3 ( \vec{S}_{1}\cdot\vec{n})^{2} \vec{S}_{1}\times\vec{p}_{1} \Big] \nn\\ 
&& + 	\frac{G m_{2}}{4 m_{1}{}^3 r{}^3} \Big[ 3 \vec{S}_{1}\times\vec{n}\cdot\vec{p}_{1} S_{1}^2 \vec{n} - 9 S_{1}^2 \vec{S}_{1}\times\vec{n} \vec{p}_{1}\cdot\vec{n} - 3 \vec{S}_{1}\times\vec{n}\cdot\vec{p}_{1} \vec{S}_{1}\cdot\vec{n} \vec{S}_{1} \nn\\ 
&& + 2 S_{1}^2 \vec{S}_{1}\times\vec{p}_{1} + 6 ( \vec{S}_{1}\cdot\vec{n})^{2} \vec{S}_{1}\times\vec{p}_{1} \Big] + 	\frac{3 G C_{1\text{ES}^2}}{8 m_{1}{}^2 r{}^3} \Big[ S_{1}^2 \vec{S}_{1}\times\vec{n} \vec{p}_{2}\cdot\vec{n} \nn\\ 
&& + 5 \vec{S}_{1}\times\vec{n} \vec{p}_{2}\cdot\vec{n} ( \vec{S}_{1}\cdot\vec{n})^{2} - 2 \vec{S}_{1}\cdot\vec{n} \vec{p}_{2}\cdot\vec{S}_{1} \vec{S}_{1}\times\vec{n} - 3 S_{1}^2 \vec{S}_{1}\times\vec{p}_{2} \nn\\ 
&& + 7 ( \vec{S}_{1}\cdot\vec{n})^{2} \vec{S}_{1}\times\vec{p}_{2} \Big] - 	\frac{G C_{1\text{BS}^3}}{6 m_{1}{}^2 r{}^3} \Big[ 6 \vec{S}_{1}\times\vec{n}\cdot\vec{p}_{2} \vec{S}_{1}\cdot\vec{n} \vec{S}_{1} + S_{1}^2 \vec{S}_{1}\times\vec{p}_{2} \nn\\ 
&& - 3 ( \vec{S}_{1}\cdot\vec{n})^{2} \vec{S}_{1}\times\vec{p}_{2} \Big] - 	\frac{G}{4 m_{1}{}^2 r{}^3} \Big[ 3 S_{1}^2 \vec{S}_{1}\times\vec{n} \vec{p}_{2}\cdot\vec{n} - 12 \vec{S}_{1}\cdot\vec{n} \vec{p}_{2}\cdot\vec{S}_{1} \vec{S}_{1}\times\vec{n} \nn\\ 
&& - 5 S_{1}^2 \vec{S}_{1}\times\vec{p}_{2} + 12 ( \vec{S}_{1}\cdot\vec{n})^{2} \vec{S}_{1}\times\vec{p}_{2} \Big]  ,
\eea
\bea
\vec{Y}^{\text{NLO}}_{\text{S}_1^2 \text{S}_2 } &=& - 	\frac{G}{4 m_{1}{}^2 r{}^3} \Big[ 15 \vec{S}_{1}\times\vec{n}\cdot\vec{p}_{1} \vec{S}_{1}\cdot\vec{S}_{2} \vec{n} - 9 \vec{S}_{1}\cdot\vec{n} \vec{S}_{1}\times\vec{p}_{1}\cdot\vec{S}_{2} \vec{n} \nn\\ 
&& - 15 \vec{S}_{1}\times\vec{n}\cdot\vec{p}_{1} \vec{S}_{1}\cdot\vec{n} \vec{n} \vec{S}_{2}\cdot\vec{n} + 6 \vec{S}_{1}\times\vec{n}\cdot\vec{p}_{1} \vec{S}_{2}\cdot\vec{n} \vec{S}_{1} + 4 \vec{S}_{1}\times\vec{p}_{1}\cdot\vec{S}_{2} \vec{S}_{1} \nn\\ 
&& - 18 \vec{S}_{1}\times\vec{n}\cdot\vec{p}_{1} \vec{S}_{1}\cdot\vec{n} \vec{S}_{2} - 6 \vec{S}_{1}\cdot\vec{n} \vec{S}_{2}\cdot\vec{n} \vec{S}_{1}\times\vec{p}_{1} + 6 \vec{S}_{1}\cdot\vec{S}_{2} \vec{S}_{1}\times\vec{p}_{1} \Big] \nn\\ 
&& - 	\frac{G C_{1\text{ES}^2}}{8 m_{1}{}^2 r{}^3} \Big[ 24 \vec{S}_{2}\times\vec{n}\cdot\vec{p}_{1} S_{1}^2 \vec{n} - 6 \vec{S}_{1}\times\vec{n}\cdot\vec{p}_{1} \vec{S}_{1}\cdot\vec{S}_{2} \vec{n} + 27 \vec{S}_{1}\cdot\vec{n} \vec{S}_{1}\times\vec{p}_{1}\cdot\vec{S}_{2} \vec{n} \nn\\ 
&& + 21 \vec{S}_{1}\times\vec{n}\cdot\vec{S}_{2} \vec{S}_{1} \vec{p}_{1}\cdot\vec{n} + 3 \vec{S}_{1}\cdot\vec{S}_{2} \vec{S}_{1}\times\vec{n} \vec{p}_{1}\cdot\vec{n} + 24 \vec{S}_{1}\cdot\vec{n} \vec{S}_{1}\times\vec{S}_{2} \vec{p}_{1}\cdot\vec{n} \nn\\ 
&& - 15 \vec{S}_{2}\times\vec{n} \vec{p}_{1}\cdot\vec{n} ( \vec{S}_{1}\cdot\vec{n})^{2} - 18 \vec{S}_{1}\times\vec{n}\cdot\vec{p}_{1} \vec{S}_{2}\cdot\vec{n} \vec{S}_{1} - 31 \vec{S}_{1}\times\vec{p}_{1}\cdot\vec{S}_{2} \vec{S}_{1} \nn\\ 
&& - 27 \vec{S}_{1}\cdot\vec{n} \vec{S}_{2}\cdot\vec{n} \vec{S}_{1}\times\vec{p}_{1} + 15 \vec{S}_{1}\cdot\vec{S}_{2} \vec{S}_{1}\times\vec{p}_{1} - 19 \vec{p}_{1}\cdot\vec{S}_{1} \vec{S}_{1}\times\vec{S}_{2} \Big] \nn\\ 
&& + 	\frac{G C_{1\text{ES}^2}}{8 m_{1} m_{2} r{}^3} \Big[ 21 \vec{S}_{2}\times\vec{n}\cdot\vec{p}_{2} S_{1}^2 \vec{n} + 24 \vec{S}_{1}\cdot\vec{n} \vec{S}_{1}\times\vec{n}\cdot\vec{S}_{2} \vec{p}_{2}\cdot\vec{n} \vec{n} \nn\\ 
&& + 24 \vec{S}_{1}\times\vec{n}\cdot\vec{S}_{2} \vec{S}_{1} \vec{p}_{2}\cdot\vec{n} + 30 \vec{S}_{1}\cdot\vec{n} \vec{S}_{1}\times\vec{S}_{2} \vec{p}_{2}\cdot\vec{n} - 24 \vec{S}_{1}\times\vec{n}\cdot\vec{p}_{2} \vec{S}_{1}\cdot\vec{n} \vec{n} \vec{S}_{2}\cdot\vec{n} \nn\\ 
&& + 39 \vec{S}_{2}\times\vec{n}\cdot\vec{p}_{2} \vec{n} ( \vec{S}_{1}\cdot\vec{n})^{2} - 24 \vec{S}_{1}\times\vec{n}\cdot\vec{p}_{2} \vec{S}_{2}\cdot\vec{n} \vec{S}_{1} - 34 \vec{S}_{1}\times\vec{p}_{2}\cdot\vec{S}_{2} \vec{S}_{1} \nn\\ 
&& - 30 \vec{S}_{1}\cdot\vec{n} \vec{S}_{2}\cdot\vec{n} \vec{S}_{1}\times\vec{p}_{2} + 18 \vec{S}_{1}\cdot\vec{S}_{2} \vec{S}_{1}\times\vec{p}_{2} - 18 \vec{p}_{2}\cdot\vec{S}_{1} \vec{S}_{1}\times\vec{S}_{2} \Big] \nn\\ 
&& + 	\frac{G}{4 m_{1} m_{2} r{}^3} \Big[ 12 \vec{S}_{2}\times\vec{n}\cdot\vec{p}_{2} S_{1}^2 \vec{n} + 9 \vec{S}_{1}\times\vec{n}\cdot\vec{p}_{2} \vec{S}_{1}\cdot\vec{S}_{2} \vec{n} \nn\\ 
&& + 15 \vec{S}_{1}\cdot\vec{n} \vec{S}_{1}\times\vec{n}\cdot\vec{S}_{2} \vec{p}_{2}\cdot\vec{n} \vec{n} + 12 \vec{S}_{1}\times\vec{n}\cdot\vec{S}_{2} \vec{S}_{1} \vec{p}_{2}\cdot\vec{n} - 3 \vec{S}_{1}\cdot\vec{S}_{2} \vec{S}_{1}\times\vec{n} \vec{p}_{2}\cdot\vec{n} \nn\\ 
&& + 21 \vec{S}_{1}\cdot\vec{n} \vec{S}_{1}\times\vec{S}_{2} \vec{p}_{2}\cdot\vec{n} + 15 \vec{S}_{2}\times\vec{n} \vec{p}_{2}\cdot\vec{n} ( \vec{S}_{1}\cdot\vec{n})^{2} - 12 \vec{S}_{1}\times\vec{n}\cdot\vec{p}_{2} \vec{S}_{2}\cdot\vec{n} \vec{S}_{1} \nn\\ 
&& - 12 \vec{S}_{1}\times\vec{p}_{2}\cdot\vec{S}_{2} \vec{S}_{1} - 9 \vec{S}_{1}\times\vec{n}\cdot\vec{p}_{2} \vec{S}_{1}\cdot\vec{n} \vec{S}_{2} - 9 \vec{S}_{1}\cdot\vec{n} \vec{S}_{2}\cdot\vec{n} \vec{S}_{1}\times\vec{p}_{2} \nn\\ 
&& + 10 \vec{S}_{1}\cdot\vec{S}_{2} \vec{S}_{1}\times\vec{p}_{2} - 7 \vec{p}_{2}\cdot\vec{S}_{1} \vec{S}_{1}\times\vec{S}_{2} \Big] .
\eea
To recap, we solved for the complete Poincar\'e algebra of the present NLO cubic-in-spin sectors, 
which provides a strong confirmation for the validity of our new general Hamiltonian presented in 
eq.~\eqref{eq:hamcns3}.

\subsection{N$^3$LO Spin-Orbit Sector}
\label{selfchecks1}

To complete the Poincar\'e algebra at the $4.5$PN order we proceed to solve for the N$^3$LO 
spin-orbit sector.
Again, we enumerate all the sectors in $H$ and $\vec{G}$ relevant to the solution of $\vec{G}$ 
at the N$^3$LO spin-orbit sector. For the Hamiltonian we have:
\bea
H & = & H_{\text{N}} + H_{1\text{PN}} + H^{\text{LO}}_{\text{SO}} + H_{2\text{PN}} 
+ H^{\text{NLO}}_{\text{SO}} + H_{3\text{PN}} + H^{\text{N}^2\text{LO}}_{\text{SO}} 
+ H^{\text{N}^3\text{LO}}_{\text{SO}}, 
\eea
with
\bea
H^{\text{LO}}_{\text{SO}} &=& H^{\text{LO}}_{\text{S}_1} + (1 \leftrightarrow 2), \qquad H^{\text{NLO}}_{\text{SO}} = H^{\text{NLO}}_{\text{S}_1} + (1 \leftrightarrow 2),\\
H^{\text{N}^2\text{LO}}_{\text{SO}} &=& H^{\text{N}^2\text{LO}}_{\text{S}_1} + (1 \leftrightarrow 2), \qquad H^{\text{N}^3\text{LO}}_{\text{SO}} = H^{\text{N}^3\text{LO}}_{\text{S}_1} + (1 \leftrightarrow 2),
\eea
where $H^{\text{N$^3$LO}}_{\text{SO}}$ is given in our \cite{Kim:2022pou}. 
For the generalized COM, $\vec{G}$, we have:
\bea
\vec{G}&=& \vec{G}_{\text{N}} + \vec{G}_{1\text{PN}} + \vec{G}^{\text{LO}}_{\text{SO}}  
+ \vec{G}_{2\text{PN}} + \vec{G}^{\text{NLO}}_{\text{SO}}  + \vec{G}_{3\text{PN}} 
+ \vec{G}^{\text{N}^2\text{LO}}_{\text{SO}} + \vec{G}^{\text{N}^3\text{LO}}_{\text{SO}},
\eea
with
\bea
\vec{G}^{\text{LO}}_{\text{SO}} &=& \vec{G}^{\text{LO}}_{\text{S}_1} + (1 \leftrightarrow 2), \qquad \vec{G}^{\text{NLO}}_{\text{SO}} = \vec{G}^{\text{NLO}}_{\text{S}_1} + (1 \leftrightarrow 2),\\ 
\vec{G}^{\text{N}^2\text{LO}}_{\text{SO}} &=& \vec{G}^{\text{N}^2\text{LO}}_{\text{S}_1} + (1 \leftrightarrow 2),
\eea
and we need to solve for:
\bea
\vec{G}^{\text{N}^3\text{LO}}_{\text{SO}} &=& \vec{G}^{\text{N}^3\text{LO}}_{\text{S}_1} + (1 \leftrightarrow 2).
\eea

$ \vec{G}^{\text{N}^3\text{LO}}_{\text{S}_1}$ is solved by:
\bea
0&=& \{  \vec{G}^{\text{N}^3\text{LO}}_{\text{S}_1} , H_{\text{N}} \}_x + \{ \vec{G}_{\text{N}}  ,  H^{\text{N}^3\text{LO}}_{\text{S}_1}\}_x + \{ \vec{G}^{\text{LO}}_{\text{S}_1}  ,  H^{\text{N}^3\text{LO}}_{3\text{PN}}\}_x + \{ \vec{G}_{1\text{PN}}  ,  H^{\text{N}^2\text{LO}}_{\text{S}_1}\}_x \nn\\
&&+ \{ \vec{G}^{\text{NLO}}_{\text{S}_1}  ,  H^{\text{N}^2\text{LO}}_{2\text{PN}}\}_x + \{ \vec{G}^{\text{N}^2\text{LO}}_{2\text{PN}}  ,  H^{\text{NLO}}_{\text{S}_1}\}_x + \{ \vec{G}^{\text{N}^2\text{LO}}_{\text{S}_1}  ,  H^{}_{1\text{PN}}\}_x + \{ \vec{G}^{\text{N}^3\text{LO}}_{3\text{PN}}  ,  H^{\text{LO}}_{\text{S}_1}\}_x \nn\\
&&+ \{ \vec{G}^{\text{LO}}_{\text{S}_1}  ,  H^{\text{N}^2\text{LO}}_{\text{S}_1}\}_S + \{ \vec{G}^{\text{NLO}}_{\text{S}_1}  ,  H^{\text{NLO}}_{\text{S}_1}\}_S + \{ \vec{G}^{\text{N}^2\text{LO}}_{\text{S}_1}  ,  H^{\text{LO}}_{\text{S}_1}\}_S.
\eea
We recall that the solution for $\vec{G}^{\text{N$^3$LO}}_{\text{SO}}$ is written as:
\bea
\vec{G}^{\text{N}^3\text{LO}}_{\text{SO}} &=& 
H^{\text{N}^2\text{LO}}_{\text{SO}} \frac{\left( \vec{x}_1 + \vec{x}_2 \right)}{2} 
+  \left( \vec{Y}^{\text{N}^3\text{LO}}_{\text{S}_1}  + \left(1 \leftrightarrow 2 \right) \right),
\eea
and we find $\vec{Y}^{\text{N}^3\text{LO}}_{\text{S}_1 }$, which we provide in appendix 
\ref{thirdsubleadinggenwspin} due to its large volume.
Thus, we found the complete Poincar\'e algebra of the N$^3$LO spin-orbit sector, which provides a 
strong confirmation for the validity of the full general Hamiltonian first presented in our
\cite{Kim:2022pou}. 

To illustrate how stringent the consistency check of the Poincar\'e algebra is for the general 
Hamiltonian in an arbitrary reference frame, an important note on the present N$^3$LO spin-orbit 
sector should be made. 
We recall that in this sector we found in \cite{Kim:2022pou} that redefinition of the 
rotational variables was needed to be implemented for the first time -- beyond linear order. 
This gave rise to a unique novel addition to the spin potentials, originating from the 
rotational kinetic term, see section $4.1$ in \cite{Kim:2022pou}, and the final result 
there in eq.~(4.8), and also eq.~(4.25). 
In version $1$ of Mandal et al.~\cite{Mandal:2022nty} (which was posted a day after our 
\cite{Kim:2022pou}) the redefinitions in their section $4.1$ were treated incorrectly via an 
insertion of the EOM instead. This incorrect treatment was actually in contrast with 
\cite{Levi:2014sba} from $2014$, where we showed in section $5$, that such an insertion of 
the EOM can only be equivalent to linear order in the redefinitions, but not beyond linear order. 
Indeed, such a treatment is no longer correct in the N$^3$LO spin-orbit sector, as we showed 
in full rigour in section $4$ of \cite{Kim:2022pou}, and contributed to the incorrect result of 
\cite{Mandal:2022nty}.
Therefore, section $4.1$ of \cite{Mandal:2022nty} v$1$, and all the equations therein, 
i.e.~eqs.~(4.1)--(4.3), were corrected to eqs.~(4.1)--(4.4) in their later version $2$. 

Yet, the revised version $2$ of [24] still did not provide details or results, e.g.~for their 
redefinitions of variables, or their action, and their generic Hamiltonian was only provided in an 
ancillary file (Hamiltonian.m), which was also corrected between their v$1$ and v$2$. 
Thus their original result was also missing the above noted novel contribution, which we 
discovered and highlighted in \cite{Kim:2022pou}. It should be noted that \cite{Mandal:2022nty}, 
which also used our EFT framework and our work in \cite{Levi:2020kvb}, did not acknowledge the 
corrections they made, following our correct treatment, including this critical novelty related with 
the redefinitions, that was pointed out in our earlier \cite{Kim:2022pou}. 
As we verified, even if only this unique novel contribution would be missing from the general 
Hamiltonian, the Poincar\'e algebra would not have a solution, and the Hamiltonian would fail to pass
the Poincar\'e consistency check. Thus even such a contribution, which drops out already at the 
specific COM frame \cite{Kim:2022pou}, is essential for the Poincar\'e invariance to be satisfied, as 
we proved here. 
It should thus be highlighted that the Poincar\'e construction, which serves as a 
strong consistency-check, was especially critical in this case, since a valid general 
Hamiltonian of the N$^3$LO spin-orbit sector was obtained only in our \cite{Kim:2022pou}:
Even the corrected Hamiltonian in the ancillary file of the present revised version $2$ of 
\cite{Mandal:2022nty} failed to pass the Poincar\'e consistency-check, and thus clearly, is also not 
canonically related to ours in \cite{Kim:2022pou}. Apart from this discrepant result, the present 
revised version $2$ of \cite{Mandal:2022nty}, which followed our work, repeated on known results, 
that were also already published in \cite{Antonelli:2020ybz}.

\section{GW and Scattering Observables}
\label{obs}

The full Lagrangians and general Hamiltonians are both useful to derive the EOMs or the Poincar\'e 
invariance, and to construct EOB models essential for GW templates. Despite their wealth in 
general physical information, they are all gauge-dependent. Accordingly they are bulky and leave 
some room for possible ambiguities, which dramatically multiply when going to higher-order 
sectors, such as the present ones. For these reasons, it is crucial to also obtain some handy 
observables in various restricted kinematic configurations, which can be readily compared in GW 
measurements carried out in LIGO, Virgo, or KAGRA. With kinematic constraints, as outlined in 
section \ref{shams} above for simplified Hamiltonians, one can define the associated binding 
energies, $e$, using $e\equiv\tilde{H}$, and relate them to observables and gauge-invariant 
quantities. In this section we provide such meaningful gauge-invariant relations expressed via the 
measured frequency of GWs, which have been critical in the construction of GW templates. 

We also derive extrapolated scattering angles that are specific to aligned spins for the guidance of 
recent popular studies of the scattering problem in the weak-field approximation.

\subsection{Binding Energies and Gauge-Invariants}
\label{circular}

The gauge-invariant relations in this section are all derived under the condition of circular 
orbit, which is very fitting for the inspiral phase of the binary. Using 
$\dot{p}_r=-\partial\tilde{H}(\tilde{r},\tilde{L})/\partial\tilde{r}=0$, on top of the constraints 
listed in section \ref{shams}, enables to eliminate the coordinate dependence from the specialized 
Hamiltonians, and obtain the binding energy for the present sectors as a function of the total 
angular momentum:
\bea
(e)^{\text{NLO}}_{\text{S}^3} (\tilde{L}) &=& \frac{\nu^2 \tilde{S}_1^3}{\tilde{L}^{11}}\left[ \frac{167 \nu }{4}-\frac{389  }{8} +\left(11 \nu -\frac{927 }{8}\right)  C_{1\text{ES}^2} -13  C_{1 \text{BS}^3} \right.\nn\\
&&+ \frac{1}{\nu q} \left( \frac{291 \nu ^2}{8}+29 \nu +\frac{389}{8} + \left(\frac{103 \nu ^2}{8}-\frac{169 \nu }{4}+\frac{927}{8} \right) C_{1\text{ES}^2} \right.\nn\\
&&\left.\left.+\left(13-\frac{17 \nu }{2}\right) C_{1 \text{BS}^3}  \right)  \right] +\frac{\nu^2 \tilde{S}_1^2 \tilde{S}_2 }{\tilde{L}^{11}}\left[ \frac{201 \nu }{4}+\frac{2913}{4} +\left(5 \nu +\frac{1437}{8}\right)  C_{1\text{ES}^2}  \right.\nn\\
&&\left.+ \frac{1}{ q} \left( \frac{243 \nu }{8}+\frac{1917}{4} + \left(\frac{55 \nu }{8}+222 \right) C_{1\text{ES}^2} \right)  \right] + (1 \leftrightarrow 2).
\eea
Using the PN parameter, $x \equiv \tilde{\omega}^{2/3}$, for the gauge-invariant frequency, in Hamilton's equation for the orbital phase, 
$d\phi/d\tilde{t}\equiv\tilde{\omega}
=\partial\tilde{H}(\tilde{r},\tilde{L})/\partial\tilde{L}=0$, provides the relation of total angular momentum to the GW frequency:
\bea
\frac{1}{\tilde{L}^2 } & \supset & \nu^2 x^{11/2} \tilde{S}_1^3 \left[  \frac{247 \nu }{81}-\frac{757}{6} +\left(-\frac{443 \nu }{9}-\frac{361}{4} \right)  C_{1\text{ES}^2} +\left( 14 \nu +\frac{92}{3} \right) C_{1 \text{BS}^3} \right.\nn\\
&&+ \frac{1}{\nu q} \left( \frac{59 \nu ^2}{18}-\frac{202 \nu }{3}+\frac{757}{6}  + \left( -\frac{617 \nu ^2}{12}-\frac{293 \nu }{2}+\frac{361}{4} \right) C_{1\text{ES}^2} \right.\nn\\
&&\left.\left.+\left( 14 \nu ^2+\frac{149 \nu }{3}-\frac{92}{3} \right) C_{1 \text{BS}^3}  \right)  \right] + \nu^2 x^{11/2} \tilde{S}_1^2 \tilde{S}_2 \left[ \frac{4835 \nu }{54}+\frac{9329}{18} \right.\nn\\
&&\left.+\left( \frac{\nu }{9}-\frac{83}{12} \right)  C_{1\text{ES}^2}  + \frac{1}{ q} \left( \frac{1703 \nu }{18}+\frac{3083}{6} + \left( \frac{40}{3}-\frac{25 \nu }{12} \right) C_{1\text{ES}^2} \right)  \right] \nn\\
&& + (1 \leftrightarrow 2).
\eea

Then, by using the two previous relations, the binding energy can also be expressed in terms of 
the frequency:
\bea
(e)^{\text{NLO}}_{\text{S}^3} (x) &=& \nu^2 x^{11/2} \tilde{S}_1^3 \left[  \frac{4}{3}-\frac{128 \nu }{81} +\left( 2-\frac{20 \nu }{9} \right)  C_{1\text{ES}^2} +\left(-4 \nu -\frac{4}{3}\right) C_{1 \text{BS}^3} \right.\nn\\
&&+ \frac{1}{\nu q} \left(-\frac{8 \nu ^2}{9}+\frac{8 \nu }{3}-\frac{4}{3}  + \left( \frac{2 \nu ^2}{3}+16 \nu -2 \right) C_{1\text{ES}^2} \right.\nn\\
&&\left.\left.+\left( -4 \nu ^2-\frac{28 \nu }{3}+\frac{4}{3} \right) C_{1 \text{BS}^3}  \right)  \right] + \nu^2 x^{11/2} \tilde{S}_1^2 \tilde{S}_2 \left[ \frac{4 \nu }{27}+\frac{82}{9} \right.\nn\\
&&\left.+\left( \frac{64 \nu }{9}-\frac{32}{3} \right)  C_{1\text{ES}^2}  + \frac{1}{ q} \left( \frac{28}{3}-\frac{32 \nu }{9} + \left( 10 \nu -\frac{32}{3} \right) C_{1\text{ES}^2} \right)  \right] \nn\\&& + (1 \leftrightarrow 2).
\eea
These results for the binding energy which we provide here for the first time match those which were initially derived in \cite{Morales:2021}.

\subsection{Extrapolated Scattering Angles}
\label{angles}

In the so-called post-Minkowskian (PM) approximation for a weak gravitational field, where scattering 
events are studied in a perturbative expansion in $G$, the common observable is the scattering 
angle, defined in the COM frame for the simplified case of aligned spins. The extrapolated 
scattering angle can be computed at low perturbative orders, namely where logarithms do not 
show up yet, in the overlap of PN and PM approximations. For the PM approximation this link is not 
feasible as of the third subleading order, which amounts to reaching results only up to the $2$PN order. 
Yet recently a unique novel approach was put forward in \cite{Edison:2022cdu}, 
which capitalizes on amplitudes methods directly in the bound problem of the binary inspiral. Thus the 
approach in \cite{Edison:2022cdu} faces none of the obstructions that are common to all other 
amplitudes-based approaches that are set on the scattering problem.

At the present NLO sectors thus, the link with scattering can be achieved starting the computation from 
our aligned-spins Hamiltonians in eq.~\eqref{alignedham} by extending the binding energy 
of our PN Hamiltonian of a binary inspiral to the kinetic energy of scattering. 
These Hamiltonians are not specified to the ``quasi-isotropic'' gauge, as in all other scattering-based 
works, and we simply use the integration considerations outlined in \cite{Damour:1988mr}. Our 
scattering angles are thus computed similarly to our \cite{Kim:2022pou,Kim:2022bwv}, where here we just 
need to truncate our final expansion in $G$ at ${\cal{O}}(G^2)$. 

We remind some conventional notation \cite{Kim:2022bwv}:
\bea
p_{\infty} = \frac{m_1 m_2}{E} \sqrt{\gamma^2 - 1},\quad E
= \sqrt{m_1^2 + m_2^2 + 2 m_1 m_2 \gamma}, \quad \gamma 
= \frac{1}{\sqrt{1- v_\infty^2}},
\eea
and:
\bea
\tilde{b} = \frac{v_\infty^2 }{ Gm }b, 
\qquad \tilde{v} = \frac{v_\infty}{c}, 
\qquad \tilde{a}_i = \frac{ S_i }{b m_i c},
\qquad \Gamma = \frac{E}{mc^2} = \sqrt{1+2\nu (\gamma - 1) },
\eea
and thus we find that our consequent scattering angles in the present sectors are given by:
\bea
\theta_{\text{S}^3}^{\text{NLO}} = \theta_{\text{S}_1^3}^{\text{NLO}} 
+ \theta_{\text{S}_1^2 \text{S}_2}^{\text{NLO}}  +  (1 \leftrightarrow 2),
\eea
where
\bea
\frac{\theta_{\text{S}_1^3 }^{\text{NLO}}}{\Gamma} &=& \tilde{v} \tilde{a}_1^3 \left[-\frac{4}{\tilde{b}} C_{1\text{BS}^3} + \frac{\pi}{\tilde{b}^2} \left( \frac{15 \nu }{4} \tilde{v}^2 + \left( 3 \nu -6 -\left(\frac{3 \nu }{2}+\frac{27}{4}\right) \tilde{v}^2 \right) C_{1\text{ES}^2}\right.\right.\nn\\
&&+ \left(-6 + \left(\frac{27 \nu }{4}-\frac{33}{4}\right) \tilde{v}^2 \right) C_{1\text{BS}^3} \nn\\
&&\left.\left.+ \frac{\nu}{q}  \left( \frac{15}{4}\tilde{v}^2+ \left(3-\frac{3 }{2}\tilde{v}^2 \right) C_{1\text{ES}^2} + \frac{27}{4} \tilde{v}^2 C_{1\text{BS}^3} \right)\right)
\right],
\eea
\bea
\frac{\theta_{\text{S}_1^2 \text{S}_2}^{\text{NLO}} }{\Gamma} &=& \tilde{v} \tilde{a}_1^2 \tilde{a}_2 \left[- \frac{12}{\tilde{b}}C_{1\text{ES}^2} +   \frac{\pi}{\tilde{b}^2} \left( 6 \nu -12 + \tilde{v}^2 \left(\frac{27 \nu }{8}-\frac{99}{8} \right) \right.\right.\nn\\
&&+ \left( -3 \nu +\left(\frac{39 \nu }{8}-\frac{207}{8}\right) \tilde{v}^2 -21 \right)C_{1\text{ES}^2}\nn\\
&&\left.\left.
+ \frac{\nu}{q} \left(6  +\frac{27}{8} \tilde{v}^2 + \left(-3+ \frac{39 }{8}\tilde{v}^2 \right) C_{1\text{ES}^2} \right)  \right) 
\right].
\eea
Our scattering angles agree with the NLO PM ones derived for the case of BHs in 
\cite{Guevara:2018wpp}, namely when the Wilson coefficients for both of the objects are specified to 
unity, $C_{\text{ES}^2} = C_{\text{BS}^3} = 1$, as we prescribed in \cite{Levi:2015msa}. 
Note that the derivations in \cite{Guevara:2018wpp} built on our higher-spin worldline theory, 
and results presented in \cite{Levi:2014gsa,Levi:2015msa}, as this dependence was omitted in 
\cite{Guevara:2018wpp}. 
Thus the limited results in \cite{Guevara:2018wpp} are inherently dependent on our self-contained 
worldline framework.
It should also be highlighted that our results in this work -- in contrast with those in 
\cite{Guevara:2018wpp} -- are also not limited to the aligned-spins constraint, which is a significant 
and ever-growing simplification when going to higher-spin sectors. 
The latter was already clearly demonstrated at the quadratic-in-spin sectors in 
\cite{Kim:2022bwv}, in section \ref{shams} above, and in \cite{Levi:2022rrq}. 
The aligned-spins constraint entails a growing loss of physical information that is always absent from 
the scattering-angle observable.

\section{Conclusions}
\label{myfriendtheend}

We confirmed the generalized actions of the complete NLO cubic-in-spin interactions 
for generic compact binaries which were tackled first in \cite{Levi:2019kgk} via an extension of the 
EFT of spinning gravitating objects \cite{Levi:2015msa} and the public \texttt{EFTofPNG} code 
\cite{Levi:2017kzq}. These higher-spin sectors enter at the $4.5$PN order, and are at the present 
precision frontier in PN theory. The interaction potentials are made up of $6$ independent sectors, 
including a new unique sector that is proportional to the square of the quadrupolar deformation 
parameter, $C_{\text{ES}^2}$. From these actions the EOMs of both the position and spin can be 
directly obtained via straightforward variation \cite{Levi:2015msa,Morales:2021}. We derived the full 
general Hamiltonians in an arbitrary reference frame and in generic kinematic configurations. Such 
general Hamiltonians uniquely enable to study the full global Poincar\'e algebra in phase space, 
which also provides a critical consistency check of state-of-the-art PN theory.

We carried out such a complete study of the Poincar\'e algebra for all of the sectors at 
the $4.5$PN precision frontier, including the N$^3$LO spin-orbit sector that we presented for the 
first time in \cite{Kim:2022pou}, in order to establish the new precision frontier at this order. 
We fully solved for the Poincar\'e algebra of both the NLO cubic-in-spin sectors from our
\cite{Levi:2019kgk}, and the N$^3$LO spin-orbit sector from \cite{Levi:2020kvb,Kim:2022pou}. 
We note that to accomplish the latter it was crucial in particular to extend the formal procedure of 
redefinitions of rotational variables, which was first introduced in \cite{Levi:2014sba}. This 
extension was indeed carried out in \cite{Kim:2022pou} beyond linear order, but was critically missed 
in version $1$ of \cite{Mandal:2022nty}, and contributed to their incorrect result.  
It should thus be highlighted that the Poincar\'e construction, which serves as a strong 
consistency-check, was especially critical in this case, since a valid general Hamiltonian of the 
N$^3$LO spin-orbit sector was obtained only in \cite{Kim:2022pou}:
Even the corrected Hamiltonian in the present revised version $2$ of \cite{Mandal:2022nty} failed to 
pass the Poincar\'e consistency-check, and thus clearly, is also not canonically related to ours
in \cite{Kim:2022pou}. 

We note that similar to the N$^3$LO spin-orbit sector, the NLO cubic-in-spin sectors do not involve 
new operators/Wilson coefficients with respect to the LO ones at the $3.5$PN order (though the 
EFT evaluation of the NLO is obviously much more intricate). 
This is why the construction of the Poincar\'e algebra is particularly strong as a consistency check 
of all sectors at the $4.5$PN order: Since there are no new sectors or Wilson coefficients introduced 
in any of the sectors at this order, the fulfilment of the Poincar\'e algebra is non-trivial for each 
of the relevant subsectors, so that they are all tested by Poincar\'e invariance.
Nevertheless, the Wilson coefficients that appear only in such higher-order higher-spin sectors 
are critical to learn on strong gravity and QCD theories. 

Subsequently we derived simplified Hamiltonians under restricted kinematic constraints, where it is 
seen that the COM aligned-spins Hamiltonians get significantly less informative at higher-spin 
sectors. In particular the new potential proportional to $C_{\text{ES}^2}^2$ vanishes in the 
aligned-spins simplification. From these simplified Hamiltonians we derived the observable binding 
energies in terms of their gauge-invariant relations to the angular momentum and the frequency, which 
are critical for GW applications. We also derived the extrapolated scattering angles defined for the 
aligned-spins configuration in the scattering problem. We found agreement with the angles derived 
for the scattering of BHs via scattering-amplitudes methods, that built on our higher-spin theory, 
and are thus also dependent on our self-contained framework. Finally, our completion of the 
Poincar\'e algebra at the $4.5$PN order provides strong confidence that this new precision frontier 
for GW measurements has now been established.

\acknowledgments
We thank Fei Teng for pleasant discussions.  
ML has been supported by the Science and Technology Facilities Council (STFC) 
Rutherford Grant ST/V003895/2 ``\textit{Harnessing QFT for Gravity}'', 
and by the Mathematical Institute University of Oxford. 
ZY is supported by the Knut and Alice Wallenberg Foundation under grants 
KAW 2018.0116 and KAW 2018.0162.

\appendix

\section{Generalized Actions from EFT Evaluation}
\label{wherewerewe}

We collect below typos from manually printing results from our computer files for 
$5$ graphs in the journal version of \cite{Levi:2019kgk}, where the correct values 
are noted in boldface (after an arrow):
\begin{align}
&\text{Figure 2(a1)} \supset  \nn\\
&-C_{1(BS^3)}\frac{G}{r^4}\frac{m_2}{m_1^2} 
\Big[
\vec{S}_1 \cdot\vec{v}_1 \times \vec{v}_2 \big(
(2 \mathbf{\to 3})\vec{S}_1 \cdot\vec{v}_2\ \vec{S}_1 \cdot \vec{n}
+ \vec{v}_2 \cdot \vec{n}\big(S_1^2 - 5\big(S_1 \cdot \vec{n}\big)^2\big) 
- \vec{S}_1 \cdot \vec{v}_1\ \vec{S}_1 \cdot \vec{n}\big)\nn   \\
& +\vec{S}_1 \cdot\vec{v}_1 \times \vec{n}\Big(
\big(S^2_1 - 5\big(\vec{S}_1 \cdot\vec{n}\big)^2\big)  
\vec{v}_1 \cdot\vec{v}_2 + \vec{S}_1 \cdot \vec{v}_2 
\big(\vec{S}_1 \cdot \vec{v}_2 - 5\vec{S}_1 \cdot \vec{n}\ \vec{v}_2 \cdot \vec{n}\big)\nn   \\
&- \vec{S}_1 \cdot \vec{v}_2 \big(\vec{S}_1 \cdot \vec{v}_1 
- 5\vec{S}_1 \cdot\vec{n}\ \vec{v}_1 \cdot \vec{n}\big)\Big)
+\vec{S}_1 \cdot\vec{v}_2 \times \vec{n}
\Big(\frac{1}{2}S^2_1\big(v^2_1 + v^2_2\big)\nn   \\
& - \vec{S}_1 \cdot\vec{v}_1\big(\vec{S}_1 \cdot\vec{v}_2
- 5\vec{S}_1 \cdot\vec{n}\ \vec{v}_2 \cdot\vec{n}\big)
- \frac{5}{2}\big(\vec{S}_1 \cdot\vec{n}\big)^2\big(v^2_1 + v^2_2\big)\Big) \nn   \\
& -\frac{1}{2}\vec{v}_1 \cdot\vec{v}_2 \times\vec{n}\ \vec{S}_1 \cdot\vec{v}_1
\big(S^2_1 - 5\big(\vec{S}_1 \cdot\vec{n}\big)^2 \big)\Big],\\ 
&\text{Figure 2(a2)} \supset \nn\\
&\frac{1}{2}C_{1(BS^3)}\frac{G}{r^4}\frac{m_2}{m_1^2}\
\Big[\vec{S}_1\cdot \vec{v}_1\times \vec{n}\Big(
S_1^2\big(v_1^2+3v_2^2\big)
-2\vec{S}_1\cdot (\vec{n} \mathbf{\to \vec{v}_1} )
\big(\vec{S}_1\cdot\vec{v}_2-5\ 
\vec{S}_1\cdot\vec{n}\ \vec{v}_2\cdot\vec{n}\big) \nn \\
& -5(\vec{S}_1\cdot\vec{n})^2\big(v_1^2+3v_2^2\big)\Big)
-2\vec{S}_1\cdot \vec{v}_1\times \vec{v}_2\ \vec{S}_1 \cdot\vec{v}_1\ \vec{S}_1 \cdot \vec{n}
\Big],\\
&\text{Figure 2(a4)} \supset \nn\\
& +C_{1(ES^2)}\frac{G}{r^3}\frac{1}{m_1} \Big[
2\ \vec{S}_1 \cdot \vec{S}_2 \times \vec{v}_2\
\big( \vec{S}_1 \cdot \vec{a}_1 +\dot{\vec{S}}_1 \cdot \vec{v}_1 \big)
+2\dot{\vec{S}}_1 \cdot \vec{S}_2 \times \vec{v}_2\ 
\vec{S}_1 \cdot \vec{v}_1  \nn  \\
& +4\vec{S}_2 \cdot \vec{v}_1 \times \vec{v}_2\ 
\dot{\vec{S}}_1 \cdot \vec{S}_1\
+ (2\vec{S}_2 \cdot \vec{v}_2 \times \vec{a}_1\ S^2_1 \mathbf{\to 0})\
+ \vec{S}_2 \cdot \vec{v}_2 \times \vec{a}_2 \big( S^2_1
-3 \big( \vec{S}_1 \cdot \vec{n} \big)^2 \big) \nn \\
& -6\vec{S}_2 \cdot \vec{v}_2 \times \vec{n} 
\Big((S^2_1\ \vec{a}_1 \cdot \vec{n} \mathbf{\to 0})
-2\dot{\vec{S}}_1 \cdot \vec{S}_1\ \vec{v}_1 \cdot \vec{n}
+ \vec{S}_1 \cdot \vec{v}_1\ \dot{\vec{S}}_1 \cdot \vec{n}
+ \dot{\vec{S}}_1 \cdot \vec{v}_1\ \vec{S}_1 \cdot \vec{n}  \nn  \\
& +\vec{S}_1 \cdot \vec{a}_1\ \vec{S}_1 \cdot \vec{n}  \Big)\Big], \\
&\text{Figure 2(a9)} \supset \nn\\
&
-\frac{3}{2}C_{1(ES^2)}\frac{G}{r^4}\frac{1}{m_1} \Big[
2 \vec{S}_1 \cdot \vec{S}_2 \times \vec{v}_2 \Big(
\vec{S}_1 \cdot \vec{v}_1\ \vec{v}_2 \cdot \vec{n}\
+\vec{S}_1 \cdot \vec{v}_2\ \vec{v}_1 \cdot \vec{n}\
\nn \\
&+ \vec{S}_1 \cdot \vec{n}\ \vec{v}_1 \cdot \vec{v}_2\
- 5\vec{S}_1 \cdot \vec{n}\ \vec{v}_1 \cdot \vec{n}\ \vec{v}_2 \cdot \vec{n}\ \Big)
- \vec{S}_2 \cdot \vec{v}_1 \times \vec{v}_2 
\Big( (+ \mathbf{\to -}) S^2_1 \vec{v}_2 \cdot \vec{n}
+2 \vec{S}_1 \cdot \vec{v}_2\ \vec{S}_1 \cdot \vec{n}\ \nn \\
&
-5\big( \vec{S}_1 \cdot \vec{n} \big)^2 \vec{v}_2 \cdot \vec{n}\Big)
- \vec{S}_2 \cdot \vec{v}_2 \times \vec{n}\
\Big(S_1^2 \big(\vec{v}_1 \cdot \vec{v}_2\
- 5\ \vec{v}_1 \cdot \vec{n}\  \vec{v}_2 \cdot \vec{n}\big)
-2\vec{S}_1 \cdot \vec{v}_1\ \vec{S}_1 \cdot \vec{v}_2
\nn \\
&
+ 10 \vec{S}_1 \cdot \vec{v}_1\ \vec{S}_1 \cdot \vec{n}\ \vec{v}_2 \cdot \vec{n}
+ 10 \vec{S}_1 \cdot \vec{v}_2\ \vec{S}_1 \cdot \vec{n}\ \vec{v}_1 \cdot \vec{n} \Big)\nn \\
&
+ 5\big(\vec{S}_1 \cdot \vec{n} \big)^2 \big(\vec{v}_1 \cdot \vec{v}_2\
- 7\ \vec{v}_1 \cdot \vec{n}\  \vec{v}_2 \cdot \vec{n}  \big)\Big)\Big],\\
&\text{Figure 3(b6)} \supset \nn\\
& C_{1(ES^2)}\frac{G^2}{r^5}\frac{m_2}{m_1}
\left[-(23 \mathbf{\to 39})\vec{S}_1\cdot \vec{S}_2\times \vec{v}_1\,\vec{S}_1\cdot\vec{n}
+13\vec{S}_1\cdot \vec{S}_2\times \vec{v}_2\,\vec{S}_1\cdot\vec{n}\right.\nn \\
&-\vec{S}_2\cdot \vec{v}_1\times \vec{n}\big((31 \mathbf{\to 15})S_1^2 
- 66\big(\vec{S}_1 \cdot\vec{n}\big)^2\big)-
\vec{S}_1\cdot\vec{S}_2\times\vec{n}\left(10\vec{S}_1\cdot\vec{v}_1
- (51 \mathbf{\to 63})\vec{S}_1\cdot\vec{n}\,\vec{v}_1\cdot\vec{n}\right)\nn \\
&+\left.\vec{S}_1\cdot\vec{S}_2\times\vec{n}\left(11\vec{S}_1\cdot\vec{v}_2
-(54 \mathbf{\to 66})\vec{S}_1\cdot\vec{n}\,\vec{v}_2\cdot\vec{n}\right)\right].
\end{align}
As we noted in \cite{Levi:2019kgk} and section \ref{actions} above the generalized actions of the 
NLO cubic-in-spin interactions are confirmed (eqs.~(5.2) and (5.13) there). There were $2$ 
independent typos in print (correction marked boldface after an arrow) compared with our computer 
files. 
In the last term of $L_{(2)}$ (eq.~(5.4) in 
\cite{Levi:2019kgk}):
\bea
L_{(2)}& \supset & + 3\ \vec{S}_1 \cdot \vec{v}_2 \times\vec{n}
\Big( \vec{S}_1 \cdot \vec{S}_2\ \big(v_1^2 
- 5\ \vec{v}_1 \cdot \vec{n}\ \vec{v}_2 \cdot \vec{n} \big)
- \vec{S}_1 \cdot \vec{v}_1\ \vec{S}_2 \cdot \vec{v}_1 
\nn \\
&&
+ (15 \mathbf{\to 5})\ 
\vec{S}_1 \cdot \vec{v}_2\ \vec{S}_2 \cdot \vec{n}\ \vec{v}_1 \cdot \vec{n}\Big),
\eea
and in $L_{(6)}$ (eq.~(5.8) in \cite{Levi:2019kgk}):
\begin{align}
L_{(6)} & \supset &
+\frac{1}{2}\ \dot{\vec{S}}_1 \cdot \vec{S}_2 \times \vec{v}_2 
\ \Big(\vec{S}_1 \cdot \vec{v}_1 
- (1 \mathbf{\to 2})\vec{S}_1 \cdot \vec{v}_2
-3\ \vec{S}_1 \cdot \vec{n}\ \vec{v}_1 \cdot \vec{n}
-3\ \vec{S}_1 \cdot \vec{n}\ \vec{v}_2 \cdot \vec{n} \Big).
\end{align}
Let us stress that the computer files of \cite{Levi:2019kgk}, also included in the 
ancillary files to the present publication, contain the correct results.

\section{New Redefinitions at the NLO Cubic-In-Spin Sectors}
\label{newredefs}

The new position shifts fixed in the present sectors can be written as:
\bea
\left(\Delta \vec{x}_1\right)^{\text{NLO}}_{\text{S}^3} &=&\left(\Delta \vec{x}_1\right)^{\text{NLO}}_{\text{S}_1^3}+\left(\Delta \vec{x}_1\right)^{\text{NLO}}_{\text{S}_1^2 \text{S}_2} +\left(\Delta \vec{x}_1\right)^{\text{NLO}}_{\text{S}_1 \text{S}_2^2} +\left(\Delta \vec{x}_1\right)^{\text{NLO}}_{\text{S}_2^3},
\eea
where 
\bea
\left(\Delta \vec{x}_1\right)^{\text{NLO}}_{\text{S}_1^3} &=&- 	\frac{3 G m_{2}}{16 m_{1}{}^3 r{}^3} \Big[ 17 S_{1}^2 \vec{S}_{1}\times\vec{n}\cdot\vec{v}_{1} \vec{n} + 4 S_{1}^2 \vec{S}_{1}\times\vec{n}\cdot\vec{v}_{2} \vec{n} - S_{1}^2 \vec{S}_{1}\times\vec{n} \big( 7 \vec{v}_{1}\cdot\vec{n} \nn\\ 
&& + 5 \vec{v}_{2}\cdot\vec{n} \big) - 5 \vec{S}_{1}\cdot\vec{n} \vec{S}_{1}\times\vec{n}\cdot\vec{v}_{1} \vec{S}_{1} - 4 \vec{S}_{1}\cdot\vec{n} \vec{S}_{1}\times\vec{n}\cdot\vec{v}_{2} \vec{S}_{1} \nn\\ 
&& + 12 \vec{S}_{1}\cdot\vec{n} \vec{S}_{1}\cdot\vec{v}_{1} \vec{S}_{1}\times\vec{n} + 8 S_{1}^2 \vec{S}_{1}\times\vec{v}_{1} + 3 S_{1}^2 \vec{S}_{1}\times\vec{v}_{2} \Big] \nn\\ 
&& - 	\frac{G C_{1\text{ES}^2} m_{2}}{4 m_{1}{}^3 r{}^3} \Big[ 6 S_{1}^2 \vec{S}_{1}\times\vec{n}\cdot\vec{v}_{1} \vec{n} + 12 S_{1}^2 \vec{S}_{1}\times\vec{n} \big( \vec{v}_{1}\cdot\vec{n} - \vec{v}_{2}\cdot\vec{n} \big) \nn\\ 
&& - 9 \vec{S}_{1}\cdot\vec{n} \vec{S}_{1}\times\vec{n}\cdot\vec{v}_{1} \vec{S}_{1} - 24 \vec{S}_{1}\cdot\vec{n} \vec{S}_{1}\cdot\vec{v}_{1} \vec{S}_{1}\times\vec{n} + 24 \vec{S}_{1}\cdot\vec{n} \vec{S}_{1}\cdot\vec{v}_{2} \vec{S}_{1}\times\vec{n} \nn\\ 
&& - 7 S_{1}^2 \vec{S}_{1}\times\vec{v}_{1} + 15 ( \vec{S}_{1}\cdot\vec{n})^{2} \vec{S}_{1}\times\vec{v}_{1} + 12 S_{1}^2 \vec{S}_{1}\times\vec{v}_{2} - 24 ( \vec{S}_{1}\cdot\vec{n})^{2} \vec{S}_{1}\times\vec{v}_{2} \Big] \nn\\ 
&& - 	\frac{G C_{1\text{BS}^3} m_{2}}{6 m_{1}{}^3 r{}^3} \Big[ 3 S_{1}^2 \vec{S}_{1}\times\vec{n} \vec{v}_{2}\cdot\vec{n} - 15 \vec{S}_{1}\times\vec{n} \vec{v}_{2}\cdot\vec{n} ( \vec{S}_{1}\cdot\vec{n})^{2} \nn\\ 
&& - 6 \vec{S}_{1}\cdot\vec{n} \vec{S}_{1}\times\vec{n}\cdot\vec{v}_{1} \vec{S}_{1} + 6 \vec{S}_{1}\cdot\vec{n} \vec{S}_{1}\times\vec{n}\cdot\vec{v}_{2} \vec{S}_{1} - 6 \vec{S}_{1}\cdot\vec{n} \vec{S}_{1}\cdot\vec{v}_{1} \vec{S}_{1}\times\vec{n} \nn\\ 
&& + S_{1}^2 \vec{S}_{1}\times\vec{v}_{2} - 3 ( \vec{S}_{1}\cdot\vec{n})^{2} \vec{S}_{1}\times\vec{v}_{2} \Big]\nn\\
&&+ 	\frac{G C_{1\text{ES}^2} m_{2}}{2 m_{1}{}^3 r{}^2} \Big[ \vec{S}_{1}\times\vec{n}\cdot\dot{\vec{S}}_{1} \vec{S}_{1} - \vec{S}_{1}\cdot\dot{\vec{S}}_{1} \vec{S}_{1}\times\vec{n} \Big] + 	\frac{G m_{2}}{16 m_{1}{}^3 r{}^2} \Big[ 2 \vec{S}_{1}\times\vec{n}\cdot\dot{\vec{S}}_{1} \vec{S}_{1} \nn\\ 
&& + 3 \vec{S}_{1}\cdot\dot{\vec{S}}_{1} \vec{S}_{1}\times\vec{n} - 5 \vec{S}_{1}\cdot\vec{n} \vec{S}_{1}\times\dot{\vec{S}}_{1} \Big]\nn\\
&& 	+\frac{1}{8 m_{1}{}^3} \Big[ (2 \vec{S}_{1}\cdot\vec{v}_{1} \vec{S}_{1}\times\ddot{\vec{S}}_{1} - 2 \vec{S}_{1}\cdot\ddot{\vec{S}}_{1} \vec{S}_{1}\times\vec{v}_{1}) + (2 \vec{S}_{1}\times\dot{\vec{S}}_{1}\cdot\vec{v}_{1} \dot{\vec{S}}_{1} \nn\\ 
&& - \dot{\vec{S}}_{1}\cdot\vec{v}_{1} \vec{S}_{1}\times\dot{\vec{S}}_{1} + \dot{S}_{1}^2 \vec{S}_{1}\times\vec{v}_{1}) - S_{1}^2 \vec{S}_{1}\times\dot{\vec{a}}_{1} + ( \vec{S}_{1}\times\dot{\vec{S}}_{1}\cdot\vec{a}_{1} \vec{S}_{1} \nn\\ 
&& + 2 \vec{S}_{1}\cdot\vec{a}_{1} \vec{S}_{1}\times\dot{\vec{S}}_{1} - 3 \vec{S}_{1}\cdot\dot{\vec{S}}_{1} \vec{S}_{1}\times\vec{a}_{1}) - \vec{S}_{1}\cdot\dot{\vec{S}}_{1} \dot{\vec{S}}_{1}\times\vec{v}_{1} - S_{1}^2 \dot{\vec{S}}_{1}\times\vec{a}_{1} \Big],
\eea
\bea
\left(\Delta \vec{x}_1\right)^{\text{NLO}}_{\text{S}_1^2 \text{S}_2} &=& - 	\frac{G}{16 m_{1}{}^2 r{}^3} \Big[ 12 \vec{S}_{1}\times\vec{n}\cdot\vec{S}_{2} \vec{S}_{1}\cdot\vec{v}_{1} \vec{n} - 12 \vec{S}_{1}\cdot\vec{S}_{2} \vec{S}_{1}\times\vec{n}\cdot\vec{v}_{1} \vec{n} \nn\\ 
&& - 15 S_{1}^2 \vec{S}_{2}\times\vec{n}\cdot\vec{v}_{2} \vec{n} + 12 \vec{S}_{1}\cdot\vec{S}_{2} \vec{S}_{1}\times\vec{n} \big( 4 \vec{v}_{1}\cdot\vec{n} - \vec{v}_{2}\cdot\vec{n} \big) - 12 S_{1}^2 \vec{S}_{2}\times\vec{n} \vec{v}_{1}\cdot\vec{n} \nn\\ 
&& - 12 \vec{S}_{1}\cdot\vec{n} \vec{S}_{1}\times\vec{S}_{2} \big( \vec{v}_{1}\cdot\vec{n} - \vec{v}_{2}\cdot\vec{n} \big) + 12 \vec{S}_{1}\cdot\vec{n} \vec{S}_{2}\times\vec{n}\cdot\vec{v}_{1} \vec{S}_{1} \nn\\ 
&& + 24 \vec{S}_{1}\times\vec{S}_{2}\cdot\vec{v}_{1} \vec{S}_{1} + 15 \vec{S}_{1}\cdot\vec{n} \vec{S}_{2}\times\vec{n}\cdot\vec{v}_{2} \vec{S}_{1} + 12 \vec{S}_{1}\cdot\vec{n} \vec{S}_{1}\times\vec{n}\cdot\vec{v}_{1} \vec{S}_{2} \nn\\ 
&& - 36 \vec{S}_{1}\cdot\vec{n} \vec{S}_{2}\cdot\vec{v}_{1} \vec{S}_{1}\times\vec{n} + 12 \vec{S}_{1}\cdot\vec{n} \vec{S}_{1}\cdot\vec{v}_{1} \vec{S}_{2}\times\vec{n} - 8 \vec{S}_{1}\cdot\vec{v}_{1} \vec{S}_{1}\times\vec{S}_{2} \nn\\ 
&& + \vec{S}_{1}\cdot\vec{v}_{2} \vec{S}_{1}\times\vec{S}_{2} - 16 S_{1}^2 \vec{S}_{2}\times\vec{v}_{1} - \vec{S}_{1}\cdot\vec{S}_{2} \vec{S}_{1}\times\vec{v}_{2} \Big] \nn\\ 
&& + 	\frac{G C_{1\text{ES}^2}}{8 m_{1}{}^2 r{}^3} \Big[ 24 \vec{S}_{1}\cdot\vec{n} \vec{S}_{1}\times\vec{n}\cdot\vec{S}_{2} \big( \vec{v}_{1} - \vec{v}_{2} \big) - 12 \vec{S}_{1}\times\vec{n}\cdot\vec{S}_{2} \vec{S}_{1}\cdot\vec{v}_{1} \vec{n} \nn\\ 
&& - 24 S_{1}^2 \vec{S}_{2}\times\vec{n}\cdot\vec{v}_{1} \vec{n} + 24 \vec{S}_{1}\times\vec{n}\cdot\vec{S}_{2} \vec{S}_{1}\cdot\vec{v}_{2} \vec{n} + 27 S_{1}^2 \vec{S}_{2}\times\vec{n}\cdot\vec{v}_{2} \vec{n} \nn\\ 
&& + 6 \vec{S}_{1}\cdot\vec{n} \vec{S}_{1}\times\vec{S}_{2}\cdot\vec{v}_{2} \vec{n} - 24 \vec{S}_{1}\times\vec{n}\cdot\vec{S}_{2} \vec{S}_{1} \vec{v}_{2}\cdot\vec{n} - 12 \vec{S}_{1}\cdot\vec{S}_{2} \vec{S}_{1}\times\vec{n} \big( \vec{v}_{1}\cdot\vec{n} \nn\\ 
&& - 2 \vec{v}_{2}\cdot\vec{n} \big) - 12 S_{1}^2 \vec{S}_{2}\times\vec{n} \big( \vec{v}_{1}\cdot\vec{n} + 3 \vec{v}_{2}\cdot\vec{n} \big) + 12 \vec{S}_{1}\cdot\vec{n} \vec{S}_{1}\times\vec{S}_{2} \vec{v}_{1}\cdot\vec{n} \nn\\ 
&& + 15 \vec{S}_{2}\times\vec{n}\cdot\vec{v}_{2} \vec{n} ( \vec{S}_{1}\cdot\vec{n})^{2} + 60 \vec{S}_{2}\times\vec{n} \vec{v}_{2}\cdot\vec{n} ( \vec{S}_{1}\cdot\vec{n})^{2} + 12 \vec{S}_{1}\cdot\vec{n} \vec{S}_{2}\times\vec{n}\cdot\vec{v}_{1} \vec{S}_{1} \nn\\ 
&& - 18 \vec{S}_{1}\cdot\vec{n} \vec{S}_{2}\times\vec{n}\cdot\vec{v}_{2} \vec{S}_{1} + 8 \vec{S}_{1}\times\vec{S}_{2}\cdot\vec{v}_{2} \vec{S}_{1} + 12 \vec{S}_{1}\cdot\vec{n} \vec{S}_{1}\cdot\vec{v}_{1} \vec{S}_{2}\times\vec{n} \nn\\ 
&& - 6 \vec{S}_{1}\cdot\vec{v}_{2} \vec{S}_{1}\times\vec{S}_{2} + 14 \vec{S}_{1}\cdot\vec{S}_{2} \vec{S}_{1}\times\vec{v}_{2} + 7 S_{1}^2 \vec{S}_{2}\times\vec{v}_{2} - 33 ( \vec{S}_{1}\cdot\vec{n})^{2} \vec{S}_{2}\times\vec{v}_{2} \Big]\nn\\
&&- 	\frac{G C_{1\text{ES}^2}}{2 m_{1}{}^2 r{}^2} \Big[ 2 \vec{S}_{1}\times\vec{n}\cdot\dot{\vec{S}}_{2} \vec{S}_{1} - 2 \vec{S}_{1}\cdot\dot{\vec{S}}_{2} \vec{S}_{1}\times\vec{n} + S_{1}^2 \dot{\vec{S}}_{2}\times\vec{n} - 3 ( \vec{S}_{1}\cdot\vec{n})^{2} \dot{\vec{S}}_{2}\times\vec{n} \Big] \nn\\ 
&& + 	\frac{G}{4 m_{1}{}^2 r{}^2} \Big[ \vec{S}_{1}\cdot\dot{\vec{S}}_{2} \vec{S}_{1}\times\vec{n} - \vec{S}_{1}\cdot\vec{n} \vec{S}_{1}\times\dot{\vec{S}}_{2} \Big],
\eea
\bea
\left(\Delta \vec{x}_1\right)^{\text{NLO}}_{\text{S}_1 \text{S}_2^2} &=& 	\frac{G C_{2\text{ES}^2}}{4 m_{1} m_{2} r{}^3} \Big[ 12 \vec{S}_{1}\times\vec{n}\cdot\vec{S}_{2} \vec{S}_{2} \vec{v}_{2}\cdot\vec{n} - 6 S_{2}^2 \vec{S}_{1}\times\vec{n} \vec{v}_{2}\cdot\vec{n} + 12 \vec{S}_{1}\cdot\vec{S}_{2} \vec{S}_{2}\times\vec{n} \vec{v}_{2}\cdot\vec{n} \nn\\ 
&& + 30 \vec{S}_{1}\times\vec{n} \vec{v}_{2}\cdot\vec{n} ( \vec{S}_{2}\cdot\vec{n})^{2} + 4 \vec{S}_{1}\times\vec{S}_{2}\cdot\vec{v}_{2} \vec{S}_{2} - 12 \vec{S}_{2}\cdot\vec{n} \vec{S}_{2}\cdot\vec{v}_{2} \vec{S}_{1}\times\vec{n} \nn\\ 
&& + S_{2}^2 \vec{S}_{1}\times\vec{v}_{1} - 3 ( \vec{S}_{2}\cdot\vec{n})^{2} \vec{S}_{1}\times\vec{v}_{1} - 6 S_{2}^2 \vec{S}_{1}\times\vec{v}_{2} + 18 ( \vec{S}_{2}\cdot\vec{n})^{2} \vec{S}_{1}\times\vec{v}_{2} \nn\\ 
&& - 4 \vec{S}_{1}\cdot\vec{S}_{2} \vec{S}_{2}\times\vec{v}_{2} \Big] - 	\frac{G}{4 m_{1} m_{2} r{}^3} \Big[ 3 \vec{S}_{1}\cdot\vec{S}_{2} \vec{S}_{2}\times\vec{n}\cdot\vec{v}_{2} \vec{n} - 3 \vec{S}_{2}\cdot\vec{n} \vec{S}_{1}\times\vec{S}_{2}\cdot\vec{v}_{2} \vec{n} \nn\\ 
&& - 3 S_{2}^2 \vec{S}_{1}\times\vec{n} \vec{v}_{2}\cdot\vec{n} - 6 \vec{S}_{1}\cdot\vec{S}_{2} \vec{S}_{2}\times\vec{n} \vec{v}_{2}\cdot\vec{n} - 15 \vec{S}_{1}\cdot\vec{n} \vec{S}_{2}\cdot\vec{n} \vec{n} \vec{S}_{2}\times\vec{n}\cdot\vec{v}_{2} \nn\\ 
&& + 3 \vec{S}_{2}\cdot\vec{n} \vec{S}_{2}\times\vec{n}\cdot\vec{v}_{2} \vec{S}_{1} - 3 \vec{S}_{1}\cdot\vec{n} \vec{S}_{2}\times\vec{n}\cdot\vec{v}_{2} \vec{S}_{2} - \vec{S}_{1}\times\vec{S}_{2}\cdot\vec{v}_{2} \vec{S}_{2} \nn\\ 
&& + 3 \vec{S}_{2}\cdot\vec{n} \vec{S}_{2}\cdot\vec{v}_{2} \vec{S}_{1}\times\vec{n} + 6 \vec{S}_{2}\cdot\vec{n} \vec{S}_{1}\cdot\vec{v}_{2} \vec{S}_{2}\times\vec{n} - \vec{S}_{2}\cdot\vec{v}_{2} \vec{S}_{1}\times\vec{S}_{2} + S_{2}^2 \vec{S}_{1}\times\vec{v}_{2} \nn\\ 
&& + 3 \vec{S}_{1}\cdot\vec{n} \vec{S}_{2}\cdot\vec{n} \vec{S}_{2}\times\vec{v}_{2} + \vec{S}_{1}\cdot\vec{S}_{2} \vec{S}_{2}\times\vec{v}_{2} \Big]\nn\\
&&- 	\frac{G C_{2\text{ES}^2}}{m_{1} m_{2} r{}^2} \Big[ \vec{S}_{1}\times\vec{n}\cdot\dot{\vec{S}}_{2} \vec{S}_{2} + \vec{S}_{1}\times\vec{n}\cdot\vec{S}_{2} \dot{\vec{S}}_{2} - 3 \vec{S}_{2}\cdot\vec{n} \dot{\vec{S}}_{2}\cdot\vec{n} \vec{S}_{1}\times\vec{n} \nn\\ 
&& + \vec{S}_{2}\cdot\dot{\vec{S}}_{2} \vec{S}_{1}\times\vec{n} + \vec{S}_{1}\cdot\dot{\vec{S}}_{2} \vec{S}_{2}\times\vec{n} + \vec{S}_{1}\cdot\vec{S}_{2} \dot{\vec{S}}_{2}\times\vec{n} \Big] ,
\eea
\bea
\left(\Delta \vec{x}_1\right)^{\text{NLO}}_{\text{S}_2^3} &=& 	\frac{G}{4 m_{2}{}^2 r{}^3} \Big[ 9 S_{2}^2 \vec{S}_{2}\times\vec{n}\cdot\vec{v}_{2} \vec{n} - 12 S_{2}^2 \vec{S}_{2}\times\vec{n} \vec{v}_{2}\cdot\vec{n} + 12 \vec{S}_{2}\cdot\vec{n} \vec{S}_{2}\cdot\vec{v}_{2} \vec{S}_{2}\times\vec{n} \nn\\ 
&& + 7 S_{2}^2 \vec{S}_{2}\times\vec{v}_{2} \Big] - 	\frac{G C_{2\text{BS}^3}}{6 m_{2}{}^2 r{}^3} \Big[ 3 S_{2}^2 \vec{S}_{2}\times\vec{n} \vec{v}_{2}\cdot\vec{n} - 15 \vec{S}_{2}\times\vec{n} \vec{v}_{2}\cdot\vec{n} ( \vec{S}_{2}\cdot\vec{n})^{2} \nn\\ 
&& + 6 \vec{S}_{2}\cdot\vec{n} \vec{S}_{2}\cdot\vec{v}_{2} \vec{S}_{2}\times\vec{n} - S_{2}^2 \vec{S}_{2}\times\vec{v}_{2} + 3 ( \vec{S}_{2}\cdot\vec{n})^{2} \vec{S}_{2}\times\vec{v}_{2} \Big] \nn\\ 
&& + 	\frac{3 G C_{2\text{ES}^2}}{8 m_{2}{}^2 r{}^3} \Big[ S_{2}^2 \vec{S}_{2}\times\vec{n}\cdot\vec{v}_{2} \vec{n} + 5 \vec{S}_{2}\times\vec{n}\cdot\vec{v}_{2} \vec{n} ( \vec{S}_{2}\cdot\vec{n})^{2} - 2 \vec{S}_{2}\cdot\vec{n} \vec{S}_{2}\times\vec{n}\cdot\vec{v}_{2} \vec{S}_{2} \nn\\ 
&& + 3 S_{2}^2 \vec{S}_{2}\times\vec{v}_{2} - 7 ( \vec{S}_{2}\cdot\vec{n})^{2} \vec{S}_{2}\times\vec{v}_{2} \Big]\nn\\
&&+ 	\frac{G C_{2\text{BS}^3}}{6 m_{2}{}^2 r{}^2} \Big[ 6 \vec{S}_{2}\cdot\vec{n} \dot{\vec{S}}_{2}\cdot\vec{n} \vec{S}_{2}\times\vec{n} - 2 \vec{S}_{2}\cdot\dot{\vec{S}}_{2} \vec{S}_{2}\times\vec{n} - S_{2}^2 \dot{\vec{S}}_{2}\times\vec{n} \nn\\ 
&& + 3 ( \vec{S}_{2}\cdot\vec{n})^{2} \dot{\vec{S}}_{2}\times\vec{n} \Big].
\eea

The new redefinitions of rotational variables fixed in the present sectors can be written as:
\bea
\left(\omega^{ij}_1\right)^{\text{NLO}}_{\text{S}^3} =\left(\omega^{ij}_1\right)^{\text{NLO}}_{\text{S}_1^3}+\left(\omega^{ij}_1\right)^{\text{NLO}}_{\text{S}_1^2 \text{S}_2} +\left(\omega^{ij}_1\right)^{\text{NLO}}_{\text{S}_1 \text{S}_2^2}  - (i \leftrightarrow j),
\eea
where 
\bea
\left(\omega^{ij}_1\right)^{\text{NLO}}_{\text{S}_1^3}&=&- 	\frac{G C_{1\text{BS}^3} m_{2}}{6 m_{1}{}^2 r{}^3} \Big[ 3 S_{1}^2 \big( v_{2}^i v_{1}^j -3 \vec{v}_{2}\cdot\vec{n} v_{1}^i n^j + 3 \vec{v}_{2}\cdot\vec{n} v_{2}^i n^j \big) + 18 \vec{S}_{1}\cdot\vec{n} \vec{S}_{1}\cdot\vec{v}_{1} \big( v_{1}^i n^j \nn\\ 
&& - v_{2}^i n^j \big) - 6 \vec{S}_{1}\cdot\vec{n} S_{1}^i \big( v_{1}^2 n^j - \vec{v}_{1}\cdot\vec{v}_{2} n^j + 5 \vec{v}_{1}\cdot\vec{n} \vec{v}_{2}\cdot\vec{n} n^j - 5 ( \vec{v}_{2}\cdot\vec{n})^{2} n^j \big) \nn\\ 
&& - 6 \vec{S}_{1}\cdot\vec{v}_{1} S_{1}^i \big( \vec{v}_{1}\cdot\vec{n} n^j - 2 \vec{v}_{2}\cdot\vec{n} n^j \big) - 6 \vec{S}_{1}\cdot\vec{v}_{2} S_{1}^i \vec{v}_{2}\cdot\vec{n} n^j - 6 \vec{S}_{1}\cdot\vec{n} S_{1}^j \big( \vec{v}_{1}\cdot\vec{n} v_{1}^i \nn\\ 
&& + 5 \vec{v}_{2}\cdot\vec{n} v_{1}^i - 2 \vec{v}_{1}\cdot\vec{n} v_{2}^i - 4 \vec{v}_{2}\cdot\vec{n} v_{2}^i \big) - 2 \vec{S}_{1}\cdot\vec{v}_{1} S_{1}^j \big( 3 v_{1}^i - 2 v_{2}^i \big) + 2 \vec{S}_{1}\cdot\vec{v}_{2} S_{1}^j v_{1}^i \nn\\ 
&& - 9 \big( v_{2}^i v_{1}^j -5 \vec{v}_{2}\cdot\vec{n} v_{1}^i n^j + 5 \vec{v}_{2}\cdot\vec{n} v_{2}^i n^j \big) ( \vec{S}_{1}\cdot\vec{n})^{2} \Big] \nn\\ 
&& - 	\frac{G m_{2}}{8 m_{1}{}^2 r{}^3} \Big[ 2 S_{1}^2 \big( 16 v_{2}^i v_{1}^j -9 \vec{v}_{1}\cdot\vec{n} v_{1}^i n^j + 24 \vec{v}_{2}\cdot\vec{n} v_{1}^i n^j - 30 \vec{v}_{1}\cdot\vec{n} v_{2}^i n^j \big) \nn\\ 
&& - 6 \vec{S}_{1}\cdot\vec{n} \vec{S}_{1}\cdot\vec{v}_{1} \big( 7 v_{1}^i n^j - 10 v_{2}^i n^j \big) + 12 \vec{S}_{1}\cdot\vec{n} \vec{S}_{1}\cdot\vec{v}_{2} v_{1}^i n^j + 9 \vec{S}_{1}\cdot\vec{n} S_{1}^i \big( 3 v_{1}^2 n^j \nn\\ 
&& - 4 \vec{v}_{1}\cdot\vec{v}_{2} n^j \big) + 9 \vec{S}_{1}\cdot\vec{v}_{1} S_{1}^i \big( 2 \vec{v}_{1}\cdot\vec{n} n^j - 5 \vec{v}_{2}\cdot\vec{n} n^j \big) + 36 \vec{S}_{1}\cdot\vec{v}_{2} S_{1}^i \vec{v}_{1}\cdot\vec{n} n^j \nn\\ 
&& + 15 \vec{S}_{1}\cdot\vec{n} S_{1}^j \big( 3 \vec{v}_{1}\cdot\vec{n} v_{1}^i - 2 \vec{v}_{2}\cdot\vec{n} v_{1}^i \big) + 5 \vec{S}_{1}\cdot\vec{v}_{1} S_{1}^j \big( v_{1}^i - 7 v_{2}^i \big) + 18 \vec{S}_{1}\cdot\vec{v}_{2} S_{1}^j v_{1}^i \nn\\ 
&& + 12 v_{2}^i v_{1}^j ( \vec{S}_{1}\cdot\vec{n})^{2} \Big] + 	\frac{3 G C_{1\text{ES}^2} m_{2}}{8 m_{1}{}^2 r{}^3} \Big[ 4 S_{1}^2 \big( 7 v_{2}^i v_{1}^j + 3 \vec{v}_{1}\cdot\vec{n} v_{1}^i n^j + 7 \vec{v}_{2}\cdot\vec{n} v_{1}^i n^j \nn\\ 
&& - 10 \vec{v}_{1}\cdot\vec{n} v_{2}^i n^j \big) - 5 \vec{S}_{1}\cdot\vec{n} \vec{S}_{1}\cdot\vec{v}_{1} \big( 5 v_{1}^i n^j - 8 v_{2}^i n^j \big) - 16 \vec{S}_{1}\cdot\vec{n} \vec{S}_{1}\cdot\vec{v}_{2} v_{1}^i n^j \nn\\ 
&& + 8 \vec{S}_{1}\cdot\vec{n} S_{1}^i \big( v_{1}^2 n^j - \vec{v}_{1}\cdot\vec{v}_{2} n^j \big) - 32 \vec{S}_{1}\cdot\vec{v}_{1} S_{1}^i \vec{v}_{2}\cdot\vec{n} n^j + 32 \vec{S}_{1}\cdot\vec{v}_{2} S_{1}^i \vec{v}_{1}\cdot\vec{n} n^j \nn\\ 
&& + \vec{S}_{1}\cdot\vec{n} S_{1}^j \big( \vec{v}_{1}\cdot\vec{n} v_{1}^i - 24 \vec{v}_{2}\cdot\vec{n} v_{1}^i + 24 \vec{v}_{1}\cdot\vec{n} v_{2}^i \big) + 8 \vec{S}_{1}\cdot\vec{v}_{1} S_{1}^j \big( v_{1}^i - 4 v_{2}^i \big) \nn\\ 
&& + 24 \vec{S}_{1}\cdot\vec{v}_{2} S_{1}^j v_{1}^i - 16 v_{2}^i v_{1}^j ( \vec{S}_{1}\cdot\vec{n})^{2} \Big]\nn\\ && - 	\frac{6 G^2 C_{1\text{ES}^2} m_{2}}{m_{1} r{}^4} \vec{S}_{1}\cdot\vec{n} S_{1}^i n^j - 	\frac{5 G^2 m_{2}{}^2}{8 m_{1}{}^2 r{}^4} \vec{S}_{1}\cdot\vec{n} S_{1}^i n^j - 	\frac{2 G^2 C_{1\text{ES}^2} m_{2}{}^2}{m_{1}{}^2 r{}^4} \vec{S}_{1}\cdot\vec{n} S_{1}^i n^j \nn\\
&&- 	\frac{G C_{1\text{ES}^2} m_{2}}{2 m_{1}{}^2 r{}^2} \Big[ S_{1}^2 a_{1}^i n^j - \vec{S}_{1}\cdot\vec{n} S_{1}^j a_{1}^i \Big] + 	\frac{G m_{2}}{8 m_{1}{}^2 r{}^2} \Big[ 5 S_{1}^2 a_{1}^i n^j - 6 \vec{S}_{1}\cdot\vec{a}_{1} S_{1}^i n^j \nn\\ 
&& + \vec{S}_{1}\cdot\vec{n} S_{1}^j a_{1}^i \Big] - 	\frac{G m_{2}}{16 m_{1}{}^2 r{}^2} \Big[ 8 \vec{S}_{1}\cdot\dot{\vec{S}}_{1} v_{1}^i n^j - 6 \dot{\vec{S}}_{1}\cdot\vec{v}_{1} S_{1}^i n^j + 7 \vec{S}_{1}\cdot\vec{v}_{1} \dot{S}_{1}^i n^j \nn\\ 
&& - 2 \dot{\vec{S}}_{1}\cdot\vec{n} S_{1}^j v_{1}^i - 9 \dot{S}_{1}^i S_{1}^j \vec{v}_{1}\cdot\vec{n} - 16 \vec{S}_{1}\cdot\vec{n} \dot{S}_{1}^j v_{1}^i \Big] - 	\frac{G C_{1\text{ES}^2} m_{2}}{4 m_{1}{}^2 r{}^2} \Big[ 2 \vec{S}_{1}\cdot\vec{v}_{1} \dot{S}_{1}^i n^j \nn\\ 
&& - \dot{S}_{1}^i S_{1}^j \vec{v}_{1}\cdot\vec{n} - \vec{S}_{1}\cdot\vec{n} \dot{S}_{1}^j v_{1}^i \Big]\nn\\
&&- 	\frac{1}{8 m_{1}{}^2} \Big[ ( \vec{S}_{1}\cdot\vec{v}_{1} \dot{S}_{1}^j a_{1}^i + \vec{S}_{1}\cdot\vec{a}_{1} \dot{S}_{1}^j v_{1}^i) + \vec{S}_{1}\cdot\vec{v}_{1} \ddot{S}_{1}^j v_{1}^i + \dot{\vec{S}}_{1}\cdot\vec{v}_{1} \dot{S}_{1}^j v_{1}^i \Big],
\eea
\bea
\left(\omega^{ij}_1\right)^{\text{NLO}}_{\text{S}_1^2 \text{S}_2}&=& 	\frac{G}{8 m_{1} r{}^3} \Big[ 3 \vec{S}_{1}\cdot\vec{n} \vec{S}_{2}\cdot\vec{n} v_{2}^i v_{1}^j + \vec{S}_{1}\cdot\vec{S}_{2} \big( v_{2}^i v_{1}^j + 12 \vec{v}_{1}\cdot\vec{n} v_{1}^i n^j - 15 \vec{v}_{2}\cdot\vec{n} v_{1}^i n^j \nn\\ 
&& - 21 \vec{v}_{1}\cdot\vec{n} v_{2}^i n^j \big) + 9 \vec{S}_{2}\cdot\vec{n} \vec{S}_{1}\cdot\vec{v}_{1} v_{1}^i n^j + 21 \vec{S}_{1}\cdot\vec{n} \vec{S}_{2}\cdot\vec{v}_{1} v_{2}^i n^j \nn\\ 
&& + 3 \vec{S}_{2}\cdot\vec{n} \vec{S}_{1}\cdot\vec{v}_{2} v_{1}^i n^j - 12 \vec{S}_{1}\cdot\vec{n} \vec{S}_{2}\cdot\vec{v}_{2} v_{1}^i n^j - 12 \vec{S}_{2}\cdot\vec{v}_{1} S_{1}^i \vec{v}_{1}\cdot\vec{n} n^j \nn\\ 
&& + 12 \vec{S}_{2}\cdot\vec{v}_{2} S_{1}^i \vec{v}_{1}\cdot\vec{n} n^j - 3 \vec{S}_{1}\cdot\vec{n} S_{2}^i \big( 4 v_{1}^2 n^j - \vec{v}_{1}\cdot\vec{v}_{2} n^j \big) + 12 \vec{S}_{1}\cdot\vec{v}_{1} S_{2}^i \vec{v}_{1}\cdot\vec{n} n^j \nn\\ 
&& - 3 \vec{S}_{1}\cdot\vec{v}_{2} S_{2}^i \vec{v}_{1}\cdot\vec{n} n^j + 3 \vec{S}_{2}\cdot\vec{n} S_{1}^j \vec{v}_{1}\cdot\vec{n} v_{1}^i - \vec{S}_{2}\cdot\vec{v}_{1} S_{1}^j \big( 5 v_{1}^i + 7 v_{2}^i \big) \nn\\ 
&& + 8 \vec{S}_{2}\cdot\vec{v}_{2} S_{1}^j v_{1}^i + S_{2}^i S_{1}^j \big( 8 v_{1}^2 - 5 \vec{v}_{1}\cdot\vec{v}_{2} + 12 \vec{v}_{1}\cdot\vec{n} \vec{v}_{2}\cdot\vec{n} - 12 ( \vec{v}_{1}\cdot\vec{n})^{2} \big) \nn\\ 
&& - 3 \vec{S}_{1}\cdot\vec{n} S_{2}^j \big( 4 \vec{v}_{1}\cdot\vec{n} v_{1}^i - 9 \vec{v}_{2}\cdot\vec{n} v_{1}^i \big) + 5 \vec{S}_{1}\cdot\vec{v}_{1} S_{2}^j v_{1}^i - 11 \vec{S}_{1}\cdot\vec{v}_{2} S_{2}^j v_{1}^i \Big] \nn\\ 
&& + 	\frac{G C_{1\text{ES}^2}}{4 m_{1} r{}^3} \Big[ 3 \vec{S}_{1}\cdot\vec{n} \vec{S}_{2}\cdot\vec{n} \big( 3 v_{2}^i v_{1}^j -20 \vec{v}_{2}\cdot\vec{n} v_{1}^i n^j - 5 \vec{v}_{1}\cdot\vec{n} v_{2}^i n^j + 15 \vec{v}_{2}\cdot\vec{n} v_{2}^i n^j \big) \nn\\ 
&& - 3 \vec{S}_{1}\cdot\vec{S}_{2} \big( 7 v_{2}^i v_{1}^j -8 \vec{v}_{1}\cdot\vec{n} v_{1}^i n^j + 10 \vec{v}_{2}\cdot\vec{n} v_{1}^i n^j - 5 \vec{v}_{1}\cdot\vec{n} v_{2}^i n^j + \vec{v}_{2}\cdot\vec{n} v_{2}^i n^j \big) \nn\\ 
&& - 3 \vec{S}_{2}\cdot\vec{n} \vec{S}_{1}\cdot\vec{v}_{1} \big( 2 v_{1}^i n^j + 11 v_{2}^i n^j \big) - 3 \vec{S}_{1}\cdot\vec{n} \vec{S}_{2}\cdot\vec{v}_{1} \big( 2 v_{1}^i n^j - 3 v_{2}^i n^j \big) \nn\\ 
&& + 3 \vec{S}_{2}\cdot\vec{n} \vec{S}_{1}\cdot\vec{v}_{2} \big( 14 v_{1}^i n^j - v_{2}^i n^j \big) - 3 \vec{S}_{1}\cdot\vec{n} \vec{S}_{2}\cdot\vec{v}_{2} v_{2}^i n^j + 6 \vec{S}_{2}\cdot\vec{n} S_{1}^i \big( v_{1}^2 n^j \nn\\ 
&& - 2 \vec{v}_{1}\cdot\vec{v}_{2} n^j + v_{2}^2 n^j -5 ( \vec{v}_{2}\cdot\vec{n})^{2} n^j \big) - 6 \vec{S}_{2}\cdot\vec{v}_{1} S_{1}^i \vec{v}_{1}\cdot\vec{n} n^j \nn\\ 
&& + 12 \vec{S}_{2}\cdot\vec{v}_{2} S_{1}^i \vec{v}_{1}\cdot\vec{n} n^j - 3 \vec{S}_{1}\cdot\vec{n} S_{2}^i \big( 2 v_{1}^2 n^j - 5 \vec{v}_{1}\cdot\vec{v}_{2} n^j \nn\\ 
&& + 3 v_{2}^2 n^j -25 \vec{v}_{1}\cdot\vec{n} \vec{v}_{2}\cdot\vec{n} n^j - 5 ( \vec{v}_{2}\cdot\vec{n})^{2} n^j \big) - 3 \vec{S}_{1}\cdot\vec{v}_{1} S_{2}^i \big( 2 \vec{v}_{1}\cdot\vec{n} n^j \nn\\ 
&& - 7 \vec{v}_{2}\cdot\vec{n} n^j \big) - 3 \vec{S}_{1}\cdot\vec{v}_{2} S_{2}^i \big( 13 \vec{v}_{1}\cdot\vec{n} n^j - 2 \vec{v}_{2}\cdot\vec{n} n^j \big) + 3 \vec{S}_{2}\cdot\vec{n} S_{1}^j \big( 4 \vec{v}_{2}\cdot\vec{n} v_{1}^i \nn\\ 
&& + \vec{v}_{1}\cdot\vec{n} v_{2}^i - 5 \vec{v}_{2}\cdot\vec{n} v_{2}^i \big) - 3 \vec{S}_{2}\cdot\vec{v}_{1} S_{1}^j v_{2}^i + \vec{S}_{2}\cdot\vec{v}_{2} S_{1}^j \big( 4 v_{1}^i - v_{2}^i \big) + S_{2}^i S_{1}^j \big( 6 v_{1}^2 \nn\\ 
&& - 13 \vec{v}_{1}\cdot\vec{v}_{2} + 7 v_{2}^2 -3 \vec{v}_{1}\cdot\vec{n} \vec{v}_{2}\cdot\vec{n} - 6 ( \vec{v}_{1}\cdot\vec{n})^{2} - 15 ( \vec{v}_{2}\cdot\vec{n})^{2} \big) \nn\\ 
&& - 3 \vec{S}_{1}\cdot\vec{n} S_{2}^j \big( 6 \vec{v}_{1}\cdot\vec{n} v_{1}^i - 11 \vec{v}_{2}\cdot\vec{n} v_{1}^i - 6 \vec{v}_{1}\cdot\vec{n} v_{2}^i + 3 \vec{v}_{2}\cdot\vec{n} v_{2}^i \big) + 6 \vec{S}_{1}\cdot\vec{v}_{1} S_{2}^j \big( v_{1}^i \nn\\ 
&& + 3 v_{2}^i \big) - \vec{S}_{1}\cdot\vec{v}_{2} S_{2}^j \big( 29 v_{1}^i - 5 v_{2}^i \big) \Big]\nn\\ && - 	\frac{2 G^2 C_{1\text{ES}^2}}{r{}^4} \Big[ \vec{S}_{2}\cdot\vec{n} S_{1}^i n^j - 2 \vec{S}_{1}\cdot\vec{n} S_{2}^i n^j + S_{2}^i S_{1}^j \Big] + 	\frac{4 G^2 C_{1\text{ES}^2} m_{2}}{m_{1} r{}^4} \Big[ \vec{S}_{2}\cdot\vec{n} S_{1}^i n^j \nn\\ 
&& - 2 \vec{S}_{1}\cdot\vec{n} S_{2}^i n^j + S_{2}^i S_{1}^j \Big] - 	\frac{3 G^2}{4 r{}^4} \Big[ 3 \vec{S}_{2}\cdot\vec{n} S_{1}^i n^j + 2 S_{2}^i S_{1}^j \Big] + 	\frac{3 G^2 m_{2}}{m_{1} r{}^4} \vec{S}_{1}\cdot\vec{n} S_{2}^i n^j \nn\\
&&- 	\frac{G C_{1\text{ES}^2}}{4 m_{1} r{}^2} \Big[ 3 \vec{S}_{1}\cdot\vec{n} \vec{S}_{2}\cdot\vec{n} a_{2}^i n^j - 11 \vec{S}_{1}\cdot\vec{S}_{2} a_{2}^i n^j + 6 \vec{S}_{2}\cdot\vec{n} S_{1}^i \vec{a}_{2}\cdot\vec{n} n^j \nn\\ 
&& + 2 \vec{S}_{2}\cdot\vec{a}_{2} S_{1}^i n^j - 15 \vec{S}_{1}\cdot\vec{n} S_{2}^i \vec{a}_{2}\cdot\vec{n} n^j + 7 \vec{S}_{1}\cdot\vec{a}_{2} S_{2}^i n^j + 5 \vec{S}_{2}\cdot\vec{n} S_{1}^j a_{2}^i \nn\\ 
&& + 3 S_{2}^i S_{1}^j \vec{a}_{2}\cdot\vec{n} - 2 \vec{S}_{1}\cdot\vec{n} S_{2}^j a_{2}^i \Big] - 	\frac{G}{4 m_{1} r{}^2} \Big[ 2 \vec{S}_{1}\cdot\dot{\vec{S}}_{2} v_{1}^i n^j - \dot{\vec{S}}_{2}\cdot\vec{v}_{1} S_{1}^i n^j \nn\\ 
&& - \dot{S}_{2}^i S_{1}^j \vec{v}_{1}\cdot\vec{n} - 2 \vec{S}_{1}\cdot\vec{n} \dot{S}_{2}^j v_{1}^i \Big] - 	\frac{G C_{1\text{ES}^2}}{4 m_{1} r{}^2} \Big[ 3 \vec{S}_{1}\cdot\vec{n} \dot{\vec{S}}_{2}\cdot\vec{n} \big( 2 v_{1}^i n^j - v_{2}^i n^j \big) \nn\\ 
&& + \vec{S}_{1}\cdot\dot{\vec{S}}_{2} \big( 2 v_{1}^i n^j - 5 v_{2}^i n^j \big) - 3 \dot{\vec{S}}_{2}\cdot\vec{n} S_{1}^i \big( \vec{v}_{1}\cdot\vec{n} n^j - 4 \vec{v}_{2}\cdot\vec{n} n^j \big) + \dot{\vec{S}}_{2}\cdot\vec{v}_{1} S_{1}^i n^j \nn\\ 
&& - 21 \vec{S}_{1}\cdot\vec{n} \dot{S}_{2}^i \vec{v}_{2}\cdot\vec{n} n^j - 4 \vec{S}_{1}\cdot\vec{v}_{1} \dot{S}_{2}^i n^j + 5 \vec{S}_{1}\cdot\vec{v}_{2} \dot{S}_{2}^i n^j - \dot{\vec{S}}_{2}\cdot\vec{n} S_{1}^j \big( 3 v_{1}^i - 5 v_{2}^i \big) \nn\\ 
&& + \dot{S}_{2}^i S_{1}^j \big( \vec{v}_{1}\cdot\vec{n} + 7 \vec{v}_{2}\cdot\vec{n} \big) - 2 \vec{S}_{1}\cdot\vec{n} \dot{S}_{2}^j \big( v_{1}^i + v_{2}^i \big) \Big]\nn\\
&&- 	\frac{G C_{1\text{ES}^2}}{6 m_{1} r} \Big[ (3 \ddot{\vec{S}}_{2}\cdot\vec{n} S_{1}^i n^j - 6 \vec{S}_{1}\cdot\vec{n} \ddot{S}_{2}^i n^j + 3 \ddot{S}_{2}^i S_{1}^j) + ( \dot{\vec{S}}_{2}\cdot\vec{n} \dot{S}_{1}^i n^j \nn\\ 
&& - 2 \dot{\vec{S}}_{1}\cdot\vec{n} \dot{S}_{2}^i n^j + \dot{S}_{2}^i \dot{S}_{1}^j) \Big],
\eea
\bea
\left(\omega^{ij}_1\right)^{\text{NLO}}_{\text{S}_1 \text{S}_2^2}&=&- 	\frac{G C_{2\text{ES}^2}}{2 m_{2} r{}^3} \Big[ S_{2}^2 \big( 5 v_{2}^i v_{1}^j + 3 \vec{v}_{2}\cdot\vec{n} v_{1}^i n^j - 3 \vec{v}_{2}\cdot\vec{n} v_{2}^i n^j \big) - 6 \vec{S}_{2}\cdot\vec{n} \vec{S}_{2}\cdot\vec{v}_{2} \big( v_{1}^i n^j \nn\\ 
&& - v_{2}^i n^j \big) - 6 \vec{S}_{2}\cdot\vec{n} S_{2}^j \big( \vec{v}_{2}\cdot\vec{n} v_{1}^i - \vec{v}_{2}\cdot\vec{n} v_{2}^i \big) + 2 \vec{S}_{2}\cdot\vec{v}_{2} S_{2}^j \big( v_{1}^i - v_{2}^i \big) \nn\\ 
&& - 3 \big( 3 v_{2}^i v_{1}^j -5 \vec{v}_{2}\cdot\vec{n} v_{1}^i n^j + 5 \vec{v}_{2}\cdot\vec{n} v_{2}^i n^j \big) ( \vec{S}_{2}\cdot\vec{n})^{2} \Big] - 	\frac{G}{4 m_{2} r{}^3} \Big[ S_{2}^2 \big( 2 v_{2}^i v_{1}^j \nn\\ 
&& + 9 \vec{v}_{2}\cdot\vec{n} v_{1}^i n^j - 3 \vec{v}_{1}\cdot\vec{n} v_{2}^i n^j - 9 \vec{v}_{2}\cdot\vec{n} v_{2}^i n^j \big) - 3 \vec{S}_{2}\cdot\vec{n} \vec{S}_{2}\cdot\vec{v}_{2} \big( 3 v_{1}^i n^j - 2 v_{2}^i n^j \big) \nn\\ 
&& + 3 \vec{S}_{2}\cdot\vec{n} S_{2}^i \big( 3 \vec{v}_{1}\cdot\vec{v}_{2} n^j - 3 v_{2}^2 n^j -5 \vec{v}_{1}\cdot\vec{n} \vec{v}_{2}\cdot\vec{n} n^j - 5 ( \vec{v}_{2}\cdot\vec{n})^{2} n^j \big) \nn\\ 
&& - 9 \vec{S}_{2}\cdot\vec{v}_{1} S_{2}^i \vec{v}_{2}\cdot\vec{n} n^j + 3 \vec{S}_{2}\cdot\vec{v}_{2} S_{2}^i \big( \vec{v}_{1}\cdot\vec{n} n^j + 4 \vec{v}_{2}\cdot\vec{n} n^j \big) + 3 \vec{S}_{2}\cdot\vec{n} S_{2}^j \big( \vec{v}_{2}\cdot\vec{n} v_{1}^i \nn\\ 
&& - 3 \vec{v}_{1}\cdot\vec{n} v_{2}^i - 4 \vec{v}_{2}\cdot\vec{n} v_{2}^i \big) - 3 \vec{S}_{2}\cdot\vec{v}_{1} S_{2}^j v_{2}^i + 2 \vec{S}_{2}\cdot\vec{v}_{2} S_{2}^j \big( v_{1}^i + v_{2}^i \big) + 3 \big( v_{2}^i v_{1}^j \nn\\ 
&& + 5 \vec{v}_{1}\cdot\vec{n} v_{2}^i n^j + 5 \vec{v}_{2}\cdot\vec{n} v_{2}^i n^j \big) ( \vec{S}_{2}\cdot\vec{n})^{2} \Big]\nn\\ && +  	\frac{2 G^2}{r{}^4} \vec{S}_{2}\cdot\vec{n} S_{2}^i n^j - 	\frac{G^2 m_{1}}{2 m_{2} r{}^4} \vec{S}_{2}\cdot\vec{n} S_{2}^i n^j + 	\frac{39 G^2 C_{2\text{ES}^2} m_{1}}{4 m_{2} r{}^4} \vec{S}_{2}\cdot\vec{n} S_{2}^i n^j \nn\\
&&- 	\frac{G}{4 m_{2} r{}^2} \Big[ S_{2}^2 a_{2}^i n^j - 3 \vec{S}_{2}\cdot\vec{n} S_{2}^i \vec{a}_{2}\cdot\vec{n} n^j - \vec{S}_{2}\cdot\vec{a}_{2} S_{2}^i n^j - 3 \vec{S}_{2}\cdot\vec{n} S_{2}^j a_{2}^i \nn\\ 
&& + 3 a_{2}^i n^j ( \vec{S}_{2}\cdot\vec{n})^{2} \Big] - 	\frac{3 G}{4 m_{2} r{}^2} \Big[ 2 \vec{S}_{2}\cdot\vec{n} \dot{\vec{S}}_{2}\cdot\vec{n} v_{2}^i n^j - \dot{\vec{S}}_{2}\cdot\vec{n} S_{2}^i \vec{v}_{2}\cdot\vec{n} n^j \nn\\ 
&& - \vec{S}_{2}\cdot\vec{n} \dot{S}_{2}^i \vec{v}_{2}\cdot\vec{n} n^j - \dot{\vec{S}}_{2}\cdot\vec{n} S_{2}^j v_{2}^i - \vec{S}_{2}\cdot\vec{n} \dot{S}_{2}^j v_{2}^i \Big] - 	\frac{G C_{2\text{ES}^2}}{4 m_{2} r{}^2} \Big[ 6 \vec{S}_{2}\cdot\vec{n} \dot{\vec{S}}_{2}\cdot\vec{n} \big( v_{1}^i n^j \nn\\ 
&& - v_{2}^i n^j \big) - 2 \vec{S}_{2}\cdot\dot{\vec{S}}_{2} \big( 3 v_{1}^i n^j + v_{2}^i n^j \big) - 3 \dot{\vec{S}}_{2}\cdot\vec{n} S_{2}^i \vec{v}_{2}\cdot\vec{n} n^j + \dot{\vec{S}}_{2}\cdot\vec{v}_{2} S_{2}^i n^j \nn\\ 
&& - 3 \vec{S}_{2}\cdot\vec{n} \dot{S}_{2}^i \vec{v}_{2}\cdot\vec{n} n^j + \vec{S}_{2}\cdot\vec{v}_{2} \dot{S}_{2}^i n^j + \dot{\vec{S}}_{2}\cdot\vec{n} S_{2}^j \big( 2 v_{1}^i + v_{2}^i \big) + \vec{S}_{2}\cdot\vec{n} \dot{S}_{2}^j \big( 2 v_{1}^i \nn\\ 
&& + v_{2}^i \big) \Big]\nn\\
&&+ 	\frac{G C_{2\text{ES}^2}}{3 m_{2} r} \dot{\vec{S}}_{2}\cdot\vec{n} \dot{S}_{2}^i n^j .
\eea

\section{Final Actions}
\label{distilledaction}

The final potentials that we obtain for the NLO cubic-in-sectors, comprise the following $6$ 
distinct sectors:
\begin{align}
	V^{\text{NLO}}_{\text{S}^3} = &
	V^{\text{NLO}}_{\text{S}_1^3 } 
	+ C_{1\text{ES}^2} V^{\text{NLO}}_{(\text{ES}_1^2 ) \text{S}_1} 
	+ C_{1\text{ES}^2}^2 V^{\text{NLO}}_{C_{\text{ES}_1^2}^2 \text{S}_1^3 }
	+ C_{1\text{BS}^3} V^{\text{NLO}}_{\text{BS}_1^3 } 
	+ V^{\text{NLO}}_{\text{S}_1^2 \text{S}_2}  
	+ C_{1\text{ES}^2} V^{\text{NLO}}_{(\text{ES}_1^2 ) \text{S}_2} \nn\\
	& + (1 \leftrightarrow 2),
\end{align}
where 
\bea
V^{\text{NLO}}_{\text{S}_1^3 } &=&  	\frac{G m_{2}}{16 m_{1}{}^2 r{}^4} \Big[ S_{1}^2 \vec{S}_{1}\times\vec{n}\cdot\vec{v}_{1} \big( 22 v_{1}^2 - 12 \vec{v}_{1}\cdot\vec{v}_{2} + 36 v_{2}^2 + 180 \vec{v}_{1}\cdot\vec{n} \vec{v}_{2}\cdot\vec{n} - 55 ( \vec{v}_{1}\cdot\vec{n})^{2} \nn\\ 
&& - 180 ( \vec{v}_{2}\cdot\vec{n})^{2} \big) - 12 S_{1}^2 \vec{S}_{1}\times\vec{n}\cdot\vec{v}_{2} \big( 4 \vec{v}_{1}\cdot\vec{v}_{2} -20 \vec{v}_{1}\cdot\vec{n} \vec{v}_{2}\cdot\vec{n} + 15 ( \vec{v}_{1}\cdot\vec{n})^{2} \big) \nn\\ 
&& + 12 S_{1}^2 \vec{S}_{1}\times\vec{v}_{1}\cdot\vec{v}_{2} \big( 11 \vec{v}_{1}\cdot\vec{n} - 10 \vec{v}_{2}\cdot\vec{n} \big) - 10 \vec{S}_{1}\cdot\vec{n} \vec{S}_{1}\cdot\vec{v}_{1} \big( 7 \vec{v}_{1}\cdot\vec{n} \nn\\ 
&& - 18 \vec{v}_{2}\cdot\vec{n} \big) \vec{S}_{1}\times\vec{n}\cdot\vec{v}_{1} - 60 \vec{S}_{1}\cdot\vec{v}_{1} \vec{S}_{1}\times\vec{n}\cdot\vec{v}_{1} \vec{S}_{1}\cdot\vec{v}_{2} + 120 \vec{S}_{1}\cdot\vec{n} \vec{S}_{1}\cdot\vec{v}_{1} \big( \vec{v}_{1}\cdot\vec{n} \nn\\ 
&& - 2 \vec{v}_{2}\cdot\vec{n} \big) \vec{S}_{1}\times\vec{n}\cdot\vec{v}_{2} + 48 \vec{S}_{1}\cdot\vec{v}_{1} \vec{S}_{1}\cdot\vec{v}_{2} \vec{S}_{1}\times\vec{n}\cdot\vec{v}_{2} - 12 \vec{S}_{1}\cdot\vec{n} \vec{S}_{1}\cdot\vec{v}_{1} \vec{S}_{1}\times\vec{v}_{1}\cdot\vec{v}_{2} \nn\\ 
&& - 55 \vec{S}_{1}\times\vec{n}\cdot\vec{v}_{1} v_{1}^2 ( \vec{S}_{1}\cdot\vec{n})^{2} + 60 \vec{S}_{1}\times\vec{n}\cdot\vec{v}_{2} v_{1}^2 ( \vec{S}_{1}\cdot\vec{n})^{2} + 14 \vec{S}_{1}\times\vec{n}\cdot\vec{v}_{1} ( \vec{S}_{1}\cdot\vec{v}_{1})^{2} \Big]\nn\\ && - 	\frac{G^2 m_{2}}{2 m_{1} r{}^5} \Big[ S_{1}^2 \vec{S}_{1}\times\vec{n}\cdot\vec{v}_{1} + 6 \vec{S}_{1}\times\vec{n}\cdot\vec{v}_{1} ( \vec{S}_{1}\cdot\vec{n})^{2} \Big] - 	\frac{G^2 m_{2}{}^2}{8 m_{1}{}^2 r{}^5} \Big[ 12 S_{1}^2 \vec{S}_{1}\times\vec{n}\cdot\vec{v}_{1} \nn\\ 
&& + 4 S_{1}^2 \vec{S}_{1}\times\vec{n}\cdot\vec{v}_{2} - 3 \vec{S}_{1}\times\vec{n}\cdot\vec{v}_{1} ( \vec{S}_{1}\cdot\vec{n})^{2} - 12 \vec{S}_{1}\times\vec{n}\cdot\vec{v}_{2} ( \vec{S}_{1}\cdot\vec{n})^{2} \Big],
\eea
\bea
V^{\text{NLO}}_{(\text{ES}_1^2 ) \text{S}_1}&=&- 	\frac{3 G m_{2}}{16 m_{1}{}^2 r{}^4} \Big[ S_{1}^2 \vec{S}_{1}\times\vec{n}\cdot\vec{v}_{1} \big( 3 v_{1}^2 - 8 \vec{v}_{1}\cdot\vec{v}_{2} + 4 v_{2}^2 -40 \vec{v}_{1}\cdot\vec{n} \vec{v}_{2}\cdot\vec{n} + 20 ( \vec{v}_{1}\cdot\vec{n})^{2} \nn\\ 
&& + 10 ( \vec{v}_{2}\cdot\vec{n})^{2} \big) - 4 S_{1}^2 \vec{S}_{1}\times\vec{v}_{1}\cdot\vec{v}_{2} \big( 6 \vec{v}_{1}\cdot\vec{n} - 5 \vec{v}_{2}\cdot\vec{n} \big) - 20 \vec{S}_{1}\cdot\vec{n} \vec{S}_{1}\cdot\vec{v}_{1} \big( \vec{v}_{1}\cdot\vec{n} \nn\\ 
&& - 2 \vec{v}_{2}\cdot\vec{n} \big) \vec{S}_{1}\times\vec{n}\cdot\vec{v}_{1} - 40 \vec{S}_{1}\cdot\vec{n} \vec{S}_{1}\times\vec{n}\cdot\vec{v}_{1} \vec{v}_{2}\cdot\vec{n} \vec{S}_{1}\cdot\vec{v}_{2} - 8 \vec{S}_{1}\cdot\vec{v}_{1} \vec{S}_{1}\times\vec{n}\cdot\vec{v}_{1} \vec{S}_{1}\cdot\vec{v}_{2} \nn\\ 
&& - 24 \vec{S}_{1}\cdot\vec{n} \vec{S}_{1}\cdot\vec{v}_{1} \vec{S}_{1}\times\vec{v}_{1}\cdot\vec{v}_{2} + 24 \vec{S}_{1}\cdot\vec{n} \vec{S}_{1}\cdot\vec{v}_{2} \vec{S}_{1}\times\vec{v}_{1}\cdot\vec{v}_{2} - 5 \vec{S}_{1}\times\vec{n}\cdot\vec{v}_{1} \big( 7 v_{1}^2 \nn\\ 
&& - 16 \vec{v}_{1}\cdot\vec{v}_{2} + 8 v_{2}^2 -14 ( \vec{v}_{2}\cdot\vec{n})^{2} \big) ( \vec{S}_{1}\cdot\vec{n})^{2} + 20 \vec{S}_{1}\times\vec{v}_{1}\cdot\vec{v}_{2} \big( 4 \vec{v}_{1}\cdot\vec{n} \nn\\ 
&& - 3 \vec{v}_{2}\cdot\vec{n} \big) ( \vec{S}_{1}\cdot\vec{n})^{2} + 4 \vec{S}_{1}\times\vec{n}\cdot\vec{v}_{1} ( \vec{S}_{1}\cdot\vec{v}_{1})^{2} + 4 \vec{S}_{1}\times\vec{n}\cdot\vec{v}_{1} ( \vec{S}_{1}\cdot\vec{v}_{2})^{2} \Big]\nn\\ && +  	\frac{3 G^2 m_{2}{}^2}{2 m_{1}{}^2 r{}^5} \Big[ 2 S_{1}^2 \vec{S}_{1}\times\vec{n}\cdot\vec{v}_{1} + 3 S_{1}^2 \vec{S}_{1}\times\vec{n}\cdot\vec{v}_{2} - 13 \vec{S}_{1}\times\vec{n}\cdot\vec{v}_{1} ( \vec{S}_{1}\cdot\vec{n})^{2} \nn\\ 
&& - 11 \vec{S}_{1}\times\vec{n}\cdot\vec{v}_{2} ( \vec{S}_{1}\cdot\vec{n})^{2} \Big] + 	\frac{G^2 m_{2}}{4 m_{1} r{}^5} \Big[ 23 S_{1}^2 \vec{S}_{1}\times\vec{n}\cdot\vec{v}_{1} + 2 S_{1}^2 \vec{S}_{1}\times\vec{n}\cdot\vec{v}_{2} \nn\\ 
&& - 102 \vec{S}_{1}\times\vec{n}\cdot\vec{v}_{1} ( \vec{S}_{1}\cdot\vec{n})^{2} - 18 \vec{S}_{1}\times\vec{n}\cdot\vec{v}_{2} ( \vec{S}_{1}\cdot\vec{n})^{2} \Big],
\eea
\bea
\label{sqCES2}
V^{\text{NLO}}_{C_{\text{ES}_1^2}^2 \text{S}_1^3} &=&- 	\frac{3 G^2 m_{2}{}^2}{2 m_{1}{}^2 r{}^5} \Big[ 2 \vec{S}_{1}\times\vec{v}_{1}\cdot\vec{n} ( \vec{S}_{1}\cdot\vec{n})^{2} - 3 \vec{S}_{1}\times\vec{v}_{2}\cdot\vec{n} ( \vec{S}_{1}\cdot\vec{n})^{2} \Big] ,
\eea
\bea
V^{\text{NLO}}_{\text{BS}_1^3 }&=& 	\frac{G m_{2}}{2 m_{1}{}^2 r{}^4} \Big[ S_{1}^2 \vec{S}_{1}\times\vec{n}\cdot\vec{v}_{1} \big( v_{1}^2 - \vec{v}_{1}\cdot\vec{v}_{2} + v_{2}^2 -5 \vec{v}_{1}\cdot\vec{n} \vec{v}_{2}\cdot\vec{n} \big) - S_{1}^2 \vec{S}_{1}\times\vec{n}\cdot\vec{v}_{2} \big( v_{1}^2 \nn\\ 
&& - \vec{v}_{1}\cdot\vec{v}_{2} + v_{2}^2 -5 \vec{v}_{1}\cdot\vec{n} \vec{v}_{2}\cdot\vec{n} \big) + S_{1}^2 \vec{S}_{1}\times\vec{v}_{1}\cdot\vec{v}_{2} \big( \vec{v}_{1}\cdot\vec{n} - \vec{v}_{2}\cdot\vec{n} \big) \nn\\ 
&& + 5 \vec{S}_{1}\cdot\vec{n} \vec{S}_{1}\cdot\vec{v}_{1} \big( 2 \vec{v}_{1}\cdot\vec{n} - \vec{v}_{2}\cdot\vec{n} \big) \vec{S}_{1}\times\vec{n}\cdot\vec{v}_{1} - 10 \vec{S}_{1}\cdot\vec{n} \vec{S}_{1}\times\vec{n}\cdot\vec{v}_{1} \vec{v}_{1}\cdot\vec{n} \vec{S}_{1}\cdot\vec{v}_{2} \nn\\ 
&& + \vec{S}_{1}\cdot\vec{v}_{1} \vec{S}_{1}\times\vec{n}\cdot\vec{v}_{1} \vec{S}_{1}\cdot\vec{v}_{2} - 5 \vec{S}_{1}\cdot\vec{n} \vec{S}_{1}\cdot\vec{v}_{1} \big( 3 \vec{v}_{1}\cdot\vec{n} - 2 \vec{v}_{2}\cdot\vec{n} \big) \vec{S}_{1}\times\vec{n}\cdot\vec{v}_{2} \nn\\ 
&& + 10 \vec{S}_{1}\cdot\vec{n} \vec{S}_{1}\cdot\vec{v}_{2} \vec{v}_{1}\cdot\vec{n} \vec{S}_{1}\times\vec{n}\cdot\vec{v}_{2} - 2 \vec{S}_{1}\cdot\vec{v}_{1} \vec{S}_{1}\cdot\vec{v}_{2} \vec{S}_{1}\times\vec{n}\cdot\vec{v}_{2} \nn\\ 
&& + 6 \vec{S}_{1}\cdot\vec{n} \vec{S}_{1}\cdot\vec{v}_{1} \vec{S}_{1}\times\vec{v}_{1}\cdot\vec{v}_{2} - 2 \vec{S}_{1}\cdot\vec{n} \vec{S}_{1}\cdot\vec{v}_{2} \vec{S}_{1}\times\vec{v}_{1}\cdot\vec{v}_{2} - 5 \vec{S}_{1}\times\vec{n}\cdot\vec{v}_{1} \big( v_{1}^2 - \vec{v}_{1}\cdot\vec{v}_{2} \nn\\ 
&& + v_{2}^2 -7 \vec{v}_{1}\cdot\vec{n} \vec{v}_{2}\cdot\vec{n} \big) ( \vec{S}_{1}\cdot\vec{n})^{2} + 5 \vec{S}_{1}\times\vec{n}\cdot\vec{v}_{2} \big( v_{1}^2 - \vec{v}_{1}\cdot\vec{v}_{2} \nn\\ 
&& + v_{2}^2 -7 \vec{v}_{1}\cdot\vec{n} \vec{v}_{2}\cdot\vec{n} \big) ( \vec{S}_{1}\cdot\vec{n})^{2} - 5 \vec{S}_{1}\times\vec{v}_{1}\cdot\vec{v}_{2} \big( \vec{v}_{1}\cdot\vec{n} - \vec{v}_{2}\cdot\vec{n} \big) ( \vec{S}_{1}\cdot\vec{n})^{2} \nn\\ 
&& - 2 \vec{S}_{1}\times\vec{n}\cdot\vec{v}_{1} ( \vec{S}_{1}\cdot\vec{v}_{1})^{2} + 3 \vec{S}_{1}\times\vec{n}\cdot\vec{v}_{2} ( \vec{S}_{1}\cdot\vec{v}_{1})^{2} \Big]\nn\\ && +  	\frac{G^2 m_{2}}{6 m_{1} r{}^5} \Big[ 4 S_{1}^2 \vec{S}_{1}\times\vec{n}\cdot\vec{v}_{1} - 3 S_{1}^2 \vec{S}_{1}\times\vec{n}\cdot\vec{v}_{2} - 18 \vec{S}_{1}\times\vec{n}\cdot\vec{v}_{1} ( \vec{S}_{1}\cdot\vec{n})^{2} \nn\\ 
&& + 15 \vec{S}_{1}\times\vec{n}\cdot\vec{v}_{2} ( \vec{S}_{1}\cdot\vec{n})^{2} \Big] - 	\frac{G^2 m_{2}{}^2}{6 m_{1}{}^2 r{}^5} \Big[ 24 S_{1}^2 \vec{S}_{1}\times\vec{n}\cdot\vec{v}_{1} - 25 S_{1}^2 \vec{S}_{1}\times\vec{n}\cdot\vec{v}_{2} \nn\\ 
&& - 126 \vec{S}_{1}\times\vec{n}\cdot\vec{v}_{1} ( \vec{S}_{1}\cdot\vec{n})^{2} + 129 \vec{S}_{1}\times\vec{n}\cdot\vec{v}_{2} ( \vec{S}_{1}\cdot\vec{n})^{2} \Big],
\eea
\bea
V^{\text{NLO}}_{\text{S}_1^2 \text{S}_2 } &=&
\frac{3 G}{16 m_{1} r{}^4} \Big[ 4 \vec{S}_{1}\times\vec{n}\cdot\vec{S}_{2} \vec{S}_{1}\cdot\vec{v}_{1} \big( v_{1}^2 - 4 \vec{v}_{1}\cdot\vec{v}_{2} + 2 v_{2}^2 -10 ( \vec{v}_{2}\cdot\vec{n})^{2} \big) \nn\\ 
&& + 2 \vec{S}_{1}\cdot\vec{S}_{2} \vec{S}_{1}\times\vec{n}\cdot\vec{v}_{1} \big( 11 v_{1}^2 - 5 \vec{v}_{1}\cdot\vec{v}_{2} - 2 v_{2}^2 + 40 \vec{v}_{1}\cdot\vec{n} \vec{v}_{2}\cdot\vec{n} - 30 ( \vec{v}_{1}\cdot\vec{n})^{2} \nn\\ 
&& + 20 ( \vec{v}_{2}\cdot\vec{n})^{2} \big) - 4 S_{1}^2 \vec{S}_{2}\times\vec{n}\cdot\vec{v}_{1} \big( 7 v_{1}^2 + 6 \vec{v}_{1}\cdot\vec{v}_{2} - 2 v_{2}^2 -10 ( \vec{v}_{1}\cdot\vec{n})^{2} + 10 ( \vec{v}_{2}\cdot\vec{n})^{2} \big) \nn\\ 
&& + 2 \vec{S}_{1}\cdot\vec{n} \vec{S}_{1}\times\vec{S}_{2}\cdot\vec{v}_{1} \big( v_{1}^2 - 4 \vec{v}_{1}\cdot\vec{v}_{2} + 2 v_{2}^2 -10 ( \vec{v}_{2}\cdot\vec{n})^{2} \big) \nn\\ 
&& + 8 \vec{S}_{1}\times\vec{S}_{2}\cdot\vec{v}_{1} \vec{S}_{1}\cdot\vec{v}_{2} \vec{v}_{1}\cdot\vec{n} + 4 \vec{S}_{1}\cdot\vec{S}_{2} \vec{S}_{1}\times\vec{n}\cdot\vec{v}_{2} \big( 6 v_{1}^2 + 3 \vec{v}_{1}\cdot\vec{v}_{2} -10 \vec{v}_{1}\cdot\vec{n} \vec{v}_{2}\cdot\vec{n} \big) \nn\\ 
&& + S_{1}^2 \vec{S}_{2}\times\vec{n}\cdot\vec{v}_{2} \big( 16 v_{1}^2 - 12 \vec{v}_{1}\cdot\vec{v}_{2} -5 ( \vec{v}_{1}\cdot\vec{n})^{2} \big) - 12 \vec{S}_{1}\cdot\vec{v}_{1} \vec{S}_{1}\times\vec{S}_{2}\cdot\vec{v}_{2} \vec{v}_{1}\cdot\vec{n} \nn\\ 
&& + 4 \vec{S}_{1}\cdot\vec{S}_{2} \vec{S}_{1}\times\vec{v}_{1}\cdot\vec{v}_{2} \big( 4 \vec{v}_{1}\cdot\vec{n} + \vec{v}_{2}\cdot\vec{n} \big) - 16 S_{1}^2 \vec{S}_{2}\times\vec{v}_{1}\cdot\vec{v}_{2} \vec{v}_{1}\cdot\vec{n} \nn\\ 
&& + 10 \vec{S}_{1}\cdot\vec{n} \vec{S}_{2}\cdot\vec{n} \big( v_{1}^2 - 4 \vec{v}_{1}\cdot\vec{v}_{2} + 2 v_{2}^2 -14 ( \vec{v}_{2}\cdot\vec{n})^{2} \big) \vec{S}_{1}\times\vec{n}\cdot\vec{v}_{1} \nn\\ 
&& + 20 \vec{S}_{1}\cdot\vec{n} \vec{S}_{2}\cdot\vec{v}_{1} \big( 3 \vec{v}_{1}\cdot\vec{n} - 4 \vec{v}_{2}\cdot\vec{n} \big) \vec{S}_{1}\times\vec{n}\cdot\vec{v}_{1} - 28 \vec{S}_{1}\cdot\vec{v}_{1} \vec{S}_{2}\cdot\vec{v}_{1} \vec{S}_{1}\times\vec{n}\cdot\vec{v}_{1} \nn\\ 
&& - 80 \vec{S}_{1}\cdot\vec{n} \vec{S}_{1}\cdot\vec{v}_{1} \vec{v}_{1}\cdot\vec{n} \vec{S}_{2}\times\vec{n}\cdot\vec{v}_{1} + 20 \vec{S}_{2}\cdot\vec{n} \vec{S}_{1}\times\vec{n}\cdot\vec{v}_{1} \vec{v}_{2}\cdot\vec{n} \vec{S}_{1}\cdot\vec{v}_{2} \nn\\ 
&& + 26 \vec{S}_{2}\cdot\vec{v}_{1} \vec{S}_{1}\times\vec{n}\cdot\vec{v}_{1} \vec{S}_{1}\cdot\vec{v}_{2} + 8 \vec{S}_{1}\cdot\vec{v}_{1} \vec{S}_{2}\times\vec{n}\cdot\vec{v}_{1} \vec{S}_{1}\cdot\vec{v}_{2} \nn\\ 
&& + 40 \vec{S}_{1}\cdot\vec{n} \vec{S}_{1}\times\vec{n}\cdot\vec{v}_{1} \vec{v}_{2}\cdot\vec{n} \vec{S}_{2}\cdot\vec{v}_{2} + 8 \vec{S}_{1}\cdot\vec{v}_{1} \vec{S}_{1}\times\vec{n}\cdot\vec{v}_{1} \vec{S}_{2}\cdot\vec{v}_{2} \nn\\ 
&& - 8 \vec{S}_{1}\times\vec{n}\cdot\vec{v}_{1} \vec{S}_{1}\cdot\vec{v}_{2} \vec{S}_{2}\cdot\vec{v}_{2} - 40 \vec{S}_{1}\cdot\vec{n} \vec{S}_{2}\cdot\vec{n} v_{1}^2 \vec{S}_{1}\times\vec{n}\cdot\vec{v}_{2} \nn\\ 
&& + 40 \vec{S}_{2}\cdot\vec{n} \vec{S}_{1}\cdot\vec{v}_{1} \vec{v}_{1}\cdot\vec{n} \vec{S}_{1}\times\vec{n}\cdot\vec{v}_{2} + 40 \vec{S}_{1}\cdot\vec{n} \vec{S}_{2}\cdot\vec{v}_{1} \vec{v}_{2}\cdot\vec{n} \vec{S}_{1}\times\vec{n}\cdot\vec{v}_{2} \nn\\ 
&& - 24 \vec{S}_{1}\cdot\vec{v}_{1} \vec{S}_{2}\cdot\vec{v}_{1} \vec{S}_{1}\times\vec{n}\cdot\vec{v}_{2} - 12 \vec{S}_{2}\cdot\vec{v}_{1} \vec{S}_{1}\cdot\vec{v}_{2} \vec{S}_{1}\times\vec{n}\cdot\vec{v}_{2} \nn\\ 
&& + 30 \vec{S}_{1}\cdot\vec{n} \vec{S}_{1}\cdot\vec{v}_{1} \vec{v}_{1}\cdot\vec{n} \vec{S}_{2}\times\vec{n}\cdot\vec{v}_{2} - 20 \vec{S}_{1}\cdot\vec{n} \vec{S}_{1}\cdot\vec{v}_{2} \vec{v}_{1}\cdot\vec{n} \vec{S}_{2}\times\vec{n}\cdot\vec{v}_{2} \nn\\ 
&& + 12 \vec{S}_{1}\cdot\vec{v}_{1} \vec{S}_{1}\cdot\vec{v}_{2} \vec{S}_{2}\times\vec{n}\cdot\vec{v}_{2} - 40 \vec{S}_{1}\cdot\vec{n} \vec{S}_{2}\cdot\vec{n} \vec{v}_{1}\cdot\vec{n} \vec{S}_{1}\times\vec{v}_{1}\cdot\vec{v}_{2} \nn\\ 
&& + 8 \vec{S}_{2}\cdot\vec{n} \vec{S}_{1}\cdot\vec{v}_{1} \vec{S}_{1}\times\vec{v}_{1}\cdot\vec{v}_{2} - 4 \vec{S}_{2}\cdot\vec{n} \vec{S}_{1}\cdot\vec{v}_{2} \vec{S}_{1}\times\vec{v}_{1}\cdot\vec{v}_{2} \nn\\ 
&& + 4 \vec{S}_{1}\cdot\vec{n} \vec{S}_{1}\cdot\vec{v}_{1} \vec{S}_{2}\times\vec{v}_{1}\cdot\vec{v}_{2} + 40 \vec{S}_{2}\times\vec{n}\cdot\vec{v}_{1} v_{1}^2 ( \vec{S}_{1}\cdot\vec{n})^{2} - 5 \vec{S}_{2}\times\vec{n}\cdot\vec{v}_{2} \big( 5 v_{1}^2 \nn\\ 
&& - 4 \vec{v}_{1}\cdot\vec{v}_{2} \big) ( \vec{S}_{1}\cdot\vec{n})^{2} + 32 \vec{S}_{2}\times\vec{n}\cdot\vec{v}_{1} ( \vec{S}_{1}\cdot\vec{v}_{1})^{2} - 16 \vec{S}_{2}\times\vec{n}\cdot\vec{v}_{2} ( \vec{S}_{1}\cdot\vec{v}_{1})^{2} \Big]\nn\\ && - 	\frac{G^2 m_{2}}{8 m_{1} r{}^5} \Big[ 6 \vec{S}_{1}\cdot\vec{n} \vec{S}_{1}\times\vec{n}\cdot\vec{S}_{2} \big( 13 \vec{v}_{1}\cdot\vec{n} - 10 \vec{v}_{2}\cdot\vec{n} \big) + 12 \vec{S}_{1}\times\vec{n}\cdot\vec{S}_{2} \vec{S}_{1}\cdot\vec{v}_{1} \nn\\ 
&& + 78 \vec{S}_{1}\cdot\vec{n} \vec{S}_{2}\cdot\vec{n} \vec{S}_{1}\times\vec{n}\cdot\vec{v}_{1} - 6 \vec{S}_{1}\cdot\vec{S}_{2} \vec{S}_{1}\times\vec{n}\cdot\vec{v}_{1} + 10 S_{1}^2 \vec{S}_{2}\times\vec{n}\cdot\vec{v}_{1} \nn\\ 
&& + 40 \vec{S}_{1}\cdot\vec{n} \vec{S}_{1}\times\vec{S}_{2}\cdot\vec{v}_{1} + 200 \vec{S}_{1}\times\vec{n}\cdot\vec{S}_{2} \vec{S}_{1}\cdot\vec{v}_{2} + 456 \vec{S}_{1}\cdot\vec{n} \vec{S}_{2}\cdot\vec{n} \vec{S}_{1}\times\vec{n}\cdot\vec{v}_{2} \nn\\ 
&& - 312 \vec{S}_{1}\cdot\vec{S}_{2} \vec{S}_{1}\times\vec{n}\cdot\vec{v}_{2} + 194 S_{1}^2 \vec{S}_{2}\times\vec{n}\cdot\vec{v}_{2} + 69 \vec{S}_{1}\cdot\vec{n} \vec{S}_{1}\times\vec{S}_{2}\cdot\vec{v}_{2} \nn\\ 
&& + 6 \vec{S}_{2}\times\vec{n}\cdot\vec{v}_{1} ( \vec{S}_{1}\cdot\vec{n})^{2} + 3 \vec{S}_{2}\times\vec{n}\cdot\vec{v}_{2} ( \vec{S}_{1}\cdot\vec{n})^{2} \Big] + 	\frac{G^2}{4 r{}^5} \Big[ 6 \vec{S}_{1}\cdot\vec{n} \vec{S}_{1}\times\vec{n}\cdot\vec{S}_{2} \big( 5 \vec{v}_{1}\cdot\vec{n} \nn\\ 
&& - 3 \vec{v}_{2}\cdot\vec{n} \big) - 99 \vec{S}_{1}\times\vec{n}\cdot\vec{S}_{2} \vec{S}_{1}\cdot\vec{v}_{1} - 240 \vec{S}_{1}\cdot\vec{n} \vec{S}_{2}\cdot\vec{n} \vec{S}_{1}\times\vec{n}\cdot\vec{v}_{1} \nn\\ 
&& + 136 \vec{S}_{1}\cdot\vec{S}_{2} \vec{S}_{1}\times\vec{n}\cdot\vec{v}_{1} - 98 S_{1}^2 \vec{S}_{2}\times\vec{n}\cdot\vec{v}_{1} - 19 \vec{S}_{1}\cdot\vec{n} \vec{S}_{1}\times\vec{S}_{2}\cdot\vec{v}_{1} \nn\\ 
&& - 26 \vec{S}_{1}\times\vec{n}\cdot\vec{S}_{2} \vec{S}_{1}\cdot\vec{v}_{2} - 72 \vec{S}_{1}\cdot\vec{n} \vec{S}_{2}\cdot\vec{n} \vec{S}_{1}\times\vec{n}\cdot\vec{v}_{2} + 48 \vec{S}_{1}\cdot\vec{S}_{2} \vec{S}_{1}\times\vec{n}\cdot\vec{v}_{2} \nn\\ 
&& - 25 S_{1}^2 \vec{S}_{2}\times\vec{n}\cdot\vec{v}_{2} - 40 \vec{S}_{1}\cdot\vec{n} \vec{S}_{1}\times\vec{S}_{2}\cdot\vec{v}_{2} + 33 \vec{S}_{2}\times\vec{n}\cdot\vec{v}_{1} ( \vec{S}_{1}\cdot\vec{n})^{2} \nn\\ 
&& - 27 \vec{S}_{2}\times\vec{n}\cdot\vec{v}_{2} ( \vec{S}_{1}\cdot\vec{n})^{2} \Big],
\eea
\bea
V^{\text{NLO}}_{(\text{ES}_1^2 ) \text{S}_2}&=& 	\frac{3 G}{16 m_{1} r{}^4} \Big[ 8 S_{1}^2 \vec{S}_{2}\times\vec{n}\cdot\vec{v}_{1} \big( 3 v_{1}^2 - 3 \vec{v}_{1}\cdot\vec{v}_{2} + 5 \vec{v}_{1}\cdot\vec{n} \vec{v}_{2}\cdot\vec{n} - 10 ( \vec{v}_{1}\cdot\vec{n})^{2} \big) \nn\\ 
&& - 16 \vec{S}_{1}\cdot\vec{n} \vec{S}_{1}\times\vec{S}_{2}\cdot\vec{v}_{1} \big( v_{1}^2 - \vec{v}_{1}\cdot\vec{v}_{2} -5 \vec{v}_{1}\cdot\vec{n} \vec{v}_{2}\cdot\vec{n} \big) + 16 \vec{S}_{1}\cdot\vec{v}_{1} \vec{S}_{1}\times\vec{S}_{2}\cdot\vec{v}_{1} \big( \vec{v}_{1}\cdot\vec{n} \nn\\ 
&& - \vec{v}_{2}\cdot\vec{n} \big) - 16 \vec{S}_{1}\times\vec{S}_{2}\cdot\vec{v}_{1} \vec{S}_{1}\cdot\vec{v}_{2} \vec{v}_{1}\cdot\vec{n} - S_{1}^2 \vec{S}_{2}\times\vec{n}\cdot\vec{v}_{2} \big( 16 v_{1}^2 - 16 \vec{v}_{1}\cdot\vec{v}_{2} + v_{2}^2 \nn\\ 
&& + 40 \vec{v}_{1}\cdot\vec{n} \vec{v}_{2}\cdot\vec{n} - 70 ( \vec{v}_{1}\cdot\vec{n})^{2} \big) + 2 \vec{S}_{1}\cdot\vec{n} \vec{S}_{1}\times\vec{S}_{2}\cdot\vec{v}_{2} \big( v_{2}^2 -40 \vec{v}_{1}\cdot\vec{n} \vec{v}_{2}\cdot\vec{n} \nn\\ 
&& + 10 ( \vec{v}_{1}\cdot\vec{n})^{2} \big) - 4 \vec{S}_{1}\cdot\vec{v}_{1} \vec{S}_{1}\times\vec{S}_{2}\cdot\vec{v}_{2} \big( 5 \vec{v}_{1}\cdot\vec{n} - 4 \vec{v}_{2}\cdot\vec{n} \big) \nn\\ 
&& + 16 \vec{S}_{1}\cdot\vec{v}_{2} \vec{S}_{1}\times\vec{S}_{2}\cdot\vec{v}_{2} \vec{v}_{1}\cdot\vec{n} - 4 S_{1}^2 \vec{S}_{2}\times\vec{v}_{1}\cdot\vec{v}_{2} \big( 9 \vec{v}_{1}\cdot\vec{n} - 8 \vec{v}_{2}\cdot\vec{n} \big) \nn\\ 
&& + 80 \vec{S}_{1}\cdot\vec{n} \vec{S}_{1}\cdot\vec{v}_{1} \big( \vec{v}_{1}\cdot\vec{n} - \vec{v}_{2}\cdot\vec{n} \big) \vec{S}_{2}\times\vec{n}\cdot\vec{v}_{1} - 80 \vec{S}_{1}\cdot\vec{n} \vec{S}_{2}\times\vec{n}\cdot\vec{v}_{1} \vec{v}_{1}\cdot\vec{n} \vec{S}_{1}\cdot\vec{v}_{2} \nn\\ 
&& + 16 \vec{S}_{1}\cdot\vec{v}_{1} \vec{S}_{2}\times\vec{n}\cdot\vec{v}_{1} \vec{S}_{1}\cdot\vec{v}_{2} - 20 \vec{S}_{1}\cdot\vec{n} \vec{S}_{1}\cdot\vec{v}_{1} \big( 5 \vec{v}_{1}\cdot\vec{n} - 4 \vec{v}_{2}\cdot\vec{n} \big) \vec{S}_{2}\times\vec{n}\cdot\vec{v}_{2} \nn\\ 
&& + 80 \vec{S}_{1}\cdot\vec{n} \vec{S}_{1}\cdot\vec{v}_{2} \vec{v}_{1}\cdot\vec{n} \vec{S}_{2}\times\vec{n}\cdot\vec{v}_{2} - 16 \vec{S}_{1}\cdot\vec{v}_{1} \vec{S}_{1}\cdot\vec{v}_{2} \vec{S}_{2}\times\vec{n}\cdot\vec{v}_{2} \nn\\ 
&& - 28 \vec{S}_{1}\cdot\vec{n} \vec{S}_{1}\cdot\vec{v}_{1} \vec{S}_{2}\times\vec{v}_{1}\cdot\vec{v}_{2} + 32 \vec{S}_{1}\cdot\vec{n} \vec{S}_{1}\cdot\vec{v}_{2} \vec{S}_{2}\times\vec{v}_{1}\cdot\vec{v}_{2} - 40 \vec{S}_{2}\times\vec{n}\cdot\vec{v}_{1} \big( v_{1}^2 \nn\\ 
&& - \vec{v}_{1}\cdot\vec{v}_{2} -7 \vec{v}_{1}\cdot\vec{n} \vec{v}_{2}\cdot\vec{n} \big) ( \vec{S}_{1}\cdot\vec{n})^{2} + 5 \vec{S}_{2}\times\vec{n}\cdot\vec{v}_{2} \big( v_{2}^2 -56 \vec{v}_{1}\cdot\vec{n} \vec{v}_{2}\cdot\vec{n} \nn\\ 
&& + 14 ( \vec{v}_{1}\cdot\vec{n})^{2} \big) ( \vec{S}_{1}\cdot\vec{n})^{2} + 20 \vec{S}_{2}\times\vec{v}_{1}\cdot\vec{v}_{2} \big( 5 \vec{v}_{1}\cdot\vec{n} - 4 \vec{v}_{2}\cdot\vec{n} \big) ( \vec{S}_{1}\cdot\vec{n})^{2} \nn\\ 
&& - 16 \vec{S}_{2}\times\vec{n}\cdot\vec{v}_{1} ( \vec{S}_{1}\cdot\vec{v}_{1})^{2} + 16 \vec{S}_{2}\times\vec{n}\cdot\vec{v}_{2} ( \vec{S}_{1}\cdot\vec{v}_{1})^{2} \Big]\nn\\ && +  	\frac{G^2 m_{2}}{4 m_{1} r{}^5} \Big[ 252 \vec{S}_{1}\cdot\vec{n} \vec{S}_{1}\times\vec{n}\cdot\vec{S}_{2} \big( \vec{v}_{1}\cdot\vec{n} - \vec{v}_{2}\cdot\vec{n} \big) - 24 \vec{S}_{1}\times\vec{n}\cdot\vec{S}_{2} \vec{S}_{1}\cdot\vec{v}_{1} \nn\\ 
&& - 40 S_{1}^2 \vec{S}_{2}\times\vec{n}\cdot\vec{v}_{1} + 116 \vec{S}_{1}\cdot\vec{n} \vec{S}_{1}\times\vec{S}_{2}\cdot\vec{v}_{1} + 18 \vec{S}_{1}\times\vec{n}\cdot\vec{S}_{2} \vec{S}_{1}\cdot\vec{v}_{2} - 3 S_{1}^2 \vec{S}_{2}\times\vec{n}\cdot\vec{v}_{2} \nn\\ 
&& - 47 \vec{S}_{1}\cdot\vec{n} \vec{S}_{1}\times\vec{S}_{2}\cdot\vec{v}_{2} + 180 \vec{S}_{2}\times\vec{n}\cdot\vec{v}_{1} ( \vec{S}_{1}\cdot\vec{n})^{2} \Big] - 	\frac{G^2}{4 r{}^5} \Big[ 6 \vec{S}_{1}\cdot\vec{n} \vec{S}_{1}\times\vec{n}\cdot\vec{S}_{2} \big( \vec{v}_{1}\cdot\vec{n} \nn\\ 
&& - \vec{v}_{2}\cdot\vec{n} \big) + 2 \vec{S}_{1}\times\vec{n}\cdot\vec{S}_{2} \vec{S}_{1}\cdot\vec{v}_{1} + 31 S_{1}^2 \vec{S}_{2}\times\vec{n}\cdot\vec{v}_{1} - 40 \vec{S}_{1}\cdot\vec{n} \vec{S}_{1}\times\vec{S}_{2}\cdot\vec{v}_{1} \nn\\ 
&& + 2 \vec{S}_{1}\times\vec{n}\cdot\vec{S}_{2} \vec{S}_{1}\cdot\vec{v}_{2} - 3 S_{1}^2 \vec{S}_{2}\times\vec{n}\cdot\vec{v}_{2} + 4 \vec{S}_{1}\cdot\vec{n} \vec{S}_{1}\times\vec{S}_{2}\cdot\vec{v}_{2} \nn\\ 
&& - 123 \vec{S}_{2}\times\vec{n}\cdot\vec{v}_{1} ( \vec{S}_{1}\cdot\vec{n})^{2} + 15 \vec{S}_{2}\times\vec{n}\cdot\vec{v}_{2} ( \vec{S}_{1}\cdot\vec{n})^{2} \Big].
\eea

\section{General Hamiltonians}
\label{sixgenhams}

our full general Hamiltonian for the present NLO cubic-in-spin sectors is 
comprised of $6$ distinct sectors:
\begin{align}
	H^{\text{NLO}}_{\text{S}^3} = & 
	H^{\text{NLO}}_{\text{S}_1^3 } 
	+ C_{1\text{ES}^2} H^{\text{NLO}}_{(\text{ES}_1^2 ) \text{S}_1} 
	+ C_{1\text{ES}^2}^2 H^{\text{NLO}}_{C_{\text{ES}_1^2}^2 \text{S}_1^3 }
	+ C_{1\text{BS}^3} H^{\text{NLO}}_{\text{BS}_1^3 } 
	+ H^{\text{NLO}}_{\text{S}_1^2 \text{S}_2}  
	+ C_{1\text{ES}^2} H^{\text{NLO}}_{(\text{ES}_1^2 ) \text{S}_2} \nn \\
	& +  (1 \leftrightarrow 2),
	\label{eq:hamcns3}
\end{align}
where 
\bea
H^{\text{NLO}}_{\text{S}_1^3 } &=& 	\frac{3 G}{4 m_{1}{}^3 m_{2} r{}^4} \Big[ 3 \vec{S}_{1}\times\vec{n}\cdot\vec{p}_{1} S_{1}^2 \big( p_{2}^2 -5 ( \vec{p}_{2}\cdot\vec{n})^{2} \big) \nn\\ 
&& - 4 \vec{S}_{1}\times\vec{n}\cdot\vec{p}_{2} S_{1}^2 \big( \vec{p}_{1}\cdot\vec{p}_{2} -5 \vec{p}_{1}\cdot\vec{n} \vec{p}_{2}\cdot\vec{n} \big) - 10 \vec{S}_{1}\times\vec{p}_{1}\cdot\vec{p}_{2} S_{1}^2 \vec{p}_{2}\cdot\vec{n} \nn\\ 
&& - 20 \vec{S}_{1}\times\vec{n}\cdot\vec{p}_{2} \vec{S}_{1}\cdot\vec{n} \vec{p}_{2}\cdot\vec{n} \vec{p}_{1}\cdot\vec{S}_{1} + 4 \vec{S}_{1}\times\vec{n}\cdot\vec{p}_{2} \vec{p}_{1}\cdot\vec{S}_{1} \vec{p}_{2}\cdot\vec{S}_{1} \Big] \nn\\ 
&& + 	\frac{G m_{2}}{16 m_{1}{}^5 r{}^4} \Big[ 11 \vec{S}_{1}\times\vec{n}\cdot\vec{p}_{1} S_{1}^2 \big( 2 p_{1}^2 -5 ( \vec{p}_{1}\cdot\vec{n})^{2} \big) - 70 \vec{S}_{1}\times\vec{n}\cdot\vec{p}_{1} \vec{S}_{1}\cdot\vec{n} \vec{p}_{1}\cdot\vec{n} \vec{p}_{1}\cdot\vec{S}_{1} \nn\\ 
&& - 55 \vec{S}_{1}\times\vec{n}\cdot\vec{p}_{1} p_{1}^2 ( \vec{S}_{1}\cdot\vec{n})^{2} + 14 \vec{S}_{1}\times\vec{n}\cdot\vec{p}_{1} ( \vec{p}_{1}\cdot\vec{S}_{1})^{2} \Big] \nn\\ 
&& - 	\frac{3 G}{4 m_{1}{}^4 r{}^4} \Big[ \vec{S}_{1}\times\vec{n}\cdot\vec{p}_{1} S_{1}^2 \big( \vec{p}_{1}\cdot\vec{p}_{2} -15 \vec{p}_{1}\cdot\vec{n} \vec{p}_{2}\cdot\vec{n} \big) + 15 \vec{S}_{1}\times\vec{n}\cdot\vec{p}_{2} S_{1}^2 ( \vec{p}_{1}\cdot\vec{n})^{2} \nn\\ 
&& - 11 \vec{S}_{1}\times\vec{p}_{1}\cdot\vec{p}_{2} S_{1}^2 \vec{p}_{1}\cdot\vec{n} - 15 \vec{S}_{1}\times\vec{n}\cdot\vec{p}_{1} \vec{S}_{1}\cdot\vec{n} \vec{p}_{2}\cdot\vec{n} \vec{p}_{1}\cdot\vec{S}_{1} \nn\\ 
&& - 10 \vec{S}_{1}\times\vec{n}\cdot\vec{p}_{2} \vec{S}_{1}\cdot\vec{n} \vec{p}_{1}\cdot\vec{n} \vec{p}_{1}\cdot\vec{S}_{1} + \vec{S}_{1}\times\vec{p}_{1}\cdot\vec{p}_{2} \vec{S}_{1}\cdot\vec{n} \vec{p}_{1}\cdot\vec{S}_{1} \nn\\ 
&& + 5 \vec{S}_{1}\times\vec{n}\cdot\vec{p}_{1} \vec{p}_{1}\cdot\vec{S}_{1} \vec{p}_{2}\cdot\vec{S}_{1} - 5 \vec{S}_{1}\times\vec{n}\cdot\vec{p}_{2} p_{1}^2 ( \vec{S}_{1}\cdot\vec{n})^{2} \Big]\nn\\ && +  	\frac{3 G^2 m_{2}{}^2}{4 m_{1}{}^3 r{}^5} \Big[ 3 \vec{S}_{1}\times\vec{n}\cdot\vec{p}_{1} S_{1}^2 - 10 \vec{S}_{1}\times\vec{n}\cdot\vec{p}_{1} ( \vec{S}_{1}\cdot\vec{n})^{2} \Big] + 	\frac{G^2 m_{2}}{4 m_{1}{}^2 r{}^5} \Big[ 10 \vec{S}_{1}\times\vec{n}\cdot\vec{p}_{1} S_{1}^2 \nn\\ 
&& - 11 \vec{S}_{1}\times\vec{n}\cdot\vec{p}_{2} S_{1}^2 - 36 \vec{S}_{1}\times\vec{n}\cdot\vec{p}_{1} ( \vec{S}_{1}\cdot\vec{n})^{2} + 24 \vec{S}_{1}\times\vec{n}\cdot\vec{p}_{2} ( \vec{S}_{1}\cdot\vec{n})^{2} \Big],
\eea
\bea
H^{\text{NLO}}_{(\text{ES}_1^2 ) \text{S}_1} 
&=& - 	\frac{3 G m_{2}}{16 m_{1}{}^5 r{}^4} \Big[ \vec{S}_{1}\times\vec{n}\cdot\vec{p}_{1} S_{1}^2 \big( p_{1}^2 + 20 ( \vec{p}_{1}\cdot\vec{n})^{2} \big) - 20 \vec{S}_{1}\times\vec{n}\cdot\vec{p}_{1} \vec{S}_{1}\cdot\vec{n} \vec{p}_{1}\cdot\vec{n} \vec{p}_{1}\cdot\vec{S}_{1} \nn\\ 
&& - 25 \vec{S}_{1}\times\vec{n}\cdot\vec{p}_{1} p_{1}^2 ( \vec{S}_{1}\cdot\vec{n})^{2} + 4 \vec{S}_{1}\times\vec{n}\cdot\vec{p}_{1} ( \vec{p}_{1}\cdot\vec{S}_{1})^{2} \Big] \nn\\ 
&& + 	\frac{3 G}{2 m_{1}{}^4 r{}^4} \Big[ \vec{S}_{1}\times\vec{n}\cdot\vec{p}_{1} S_{1}^2 \big( \vec{p}_{1}\cdot\vec{p}_{2} + 5 \vec{p}_{1}\cdot\vec{n} \vec{p}_{2}\cdot\vec{n} \big) + 3 \vec{S}_{1}\times\vec{p}_{1}\cdot\vec{p}_{2} S_{1}^2 \vec{p}_{1}\cdot\vec{n} \nn\\ 
&& - 5 \vec{S}_{1}\times\vec{n}\cdot\vec{p}_{1} \vec{S}_{1}\cdot\vec{n} \vec{p}_{2}\cdot\vec{n} \vec{p}_{1}\cdot\vec{S}_{1} + 3 \vec{S}_{1}\times\vec{p}_{1}\cdot\vec{p}_{2} \vec{S}_{1}\cdot\vec{n} \vec{p}_{1}\cdot\vec{S}_{1} \nn\\ 
&& + \vec{S}_{1}\times\vec{n}\cdot\vec{p}_{1} \vec{p}_{1}\cdot\vec{S}_{1} \vec{p}_{2}\cdot\vec{S}_{1} - 10 \vec{S}_{1}\times\vec{n}\cdot\vec{p}_{1} \vec{p}_{1}\cdot\vec{p}_{2} ( \vec{S}_{1}\cdot\vec{n})^{2} \nn\\ 
&& - 10 \vec{S}_{1}\times\vec{p}_{1}\cdot\vec{p}_{2} \vec{p}_{1}\cdot\vec{n} ( \vec{S}_{1}\cdot\vec{n})^{2} \Big] - 	\frac{3 G}{8 m_{1}{}^3 m_{2} r{}^4} \Big[ \vec{S}_{1}\times\vec{n}\cdot\vec{p}_{1} S_{1}^2 \big( 2 p_{2}^2 + 5 ( \vec{p}_{2}\cdot\vec{n})^{2} \big) \nn\\ 
&& + 10 \vec{S}_{1}\times\vec{p}_{1}\cdot\vec{p}_{2} S_{1}^2 \vec{p}_{2}\cdot\vec{n} - 20 \vec{S}_{1}\times\vec{n}\cdot\vec{p}_{1} \vec{S}_{1}\cdot\vec{n} \vec{p}_{2}\cdot\vec{n} \vec{p}_{2}\cdot\vec{S}_{1} \nn\\ 
&& + 12 \vec{S}_{1}\times\vec{p}_{1}\cdot\vec{p}_{2} \vec{S}_{1}\cdot\vec{n} \vec{p}_{2}\cdot\vec{S}_{1} - 5 \vec{S}_{1}\times\vec{n}\cdot\vec{p}_{1} \big( 4 p_{2}^2 -7 ( \vec{p}_{2}\cdot\vec{n})^{2} \big) ( \vec{S}_{1}\cdot\vec{n})^{2} \nn\\ 
&& - 30 \vec{S}_{1}\times\vec{p}_{1}\cdot\vec{p}_{2} \vec{p}_{2}\cdot\vec{n} ( \vec{S}_{1}\cdot\vec{n})^{2} + 2 \vec{S}_{1}\times\vec{n}\cdot\vec{p}_{1} ( \vec{p}_{2}\cdot\vec{S}_{1})^{2} \Big]\nn\\ && +  	\frac{3 G^2 m_{2}{}^2}{2 m_{1}{}^3 r{}^5} \Big[ \vec{S}_{1}\times\vec{n}\cdot\vec{p}_{1} S_{1}^2 - 16 \vec{S}_{1}\times\vec{n}\cdot\vec{p}_{1} ( \vec{S}_{1}\cdot\vec{n})^{2} \Big] + 	\frac{G^2}{2 m_{1} r{}^5} \Big[ 7 \vec{S}_{1}\times\vec{n}\cdot\vec{p}_{2} S_{1}^2 \nn\\ 
&& - 27 \vec{S}_{1}\times\vec{n}\cdot\vec{p}_{2} ( \vec{S}_{1}\cdot\vec{n})^{2} \Big] + 	\frac{G^2 m_{2}}{4 m_{1}{}^2 r{}^5} \Big[ 5 \vec{S}_{1}\times\vec{n}\cdot\vec{p}_{1} S_{1}^2 + 21 \vec{S}_{1}\times\vec{n}\cdot\vec{p}_{2} S_{1}^2 \nn\\ 
&& - 60 \vec{S}_{1}\times\vec{n}\cdot\vec{p}_{1} ( \vec{S}_{1}\cdot\vec{n})^{2} - 45 \vec{S}_{1}\times\vec{n}\cdot\vec{p}_{2} ( \vec{S}_{1}\cdot\vec{n})^{2} \Big],
\eea
\bea
\label{hamces2sq}
H^{\text{NLO}}_{C_{\text{ES}_1^2}^2 \text{S}_1^3} &=&  - \frac{9 G^2 m_{2}}{2 m_{1}{}^2 r{}^5} \vec{S}_{1}\times\vec{n}\cdot\vec{p}_{2} ( \vec{S}_{1}\cdot\vec{n})^{2} + 	\frac{3 G^2 m_{2}{}^2}{m_{1}{}^3 r{}^5} \vec{S}_{1}\times\vec{n}\cdot\vec{p}_{1} ( \vec{S}_{1}\cdot\vec{n})^{2} ,
\eea
\bea
H^{\text{NLO}}_{\text{BS}_1^3 } &=& 	\frac{G m_{2}}{m_{1}{}^5 r{}^4} \Big[ 5 \vec{S}_{1}\times\vec{n}\cdot\vec{p}_{1} \vec{S}_{1}\cdot\vec{n} \vec{p}_{1}\cdot\vec{n} \vec{p}_{1}\cdot\vec{S}_{1} - \vec{S}_{1}\times\vec{n}\cdot\vec{p}_{1} ( \vec{p}_{1}\cdot\vec{S}_{1})^{2} \Big] \nn\\ 
&& - 	\frac{G}{2 m_{1}{}^4 r{}^4} \Big[ \vec{S}_{1}\times\vec{n}\cdot\vec{p}_{1} S_{1}^2 \big( \vec{p}_{1}\cdot\vec{p}_{2} + 5 \vec{p}_{1}\cdot\vec{n} \vec{p}_{2}\cdot\vec{n} \big) + \vec{S}_{1}\times\vec{n}\cdot\vec{p}_{2} S_{1}^2 p_{1}^2 \nn\\ 
&& - \vec{S}_{1}\times\vec{p}_{1}\cdot\vec{p}_{2} S_{1}^2 \vec{p}_{1}\cdot\vec{n} + 5 \vec{S}_{1}\times\vec{n}\cdot\vec{p}_{1} \vec{S}_{1}\cdot\vec{n} \vec{p}_{2}\cdot\vec{n} \vec{p}_{1}\cdot\vec{S}_{1} \nn\\ 
&& + 15 \vec{S}_{1}\times\vec{n}\cdot\vec{p}_{2} \vec{S}_{1}\cdot\vec{n} \vec{p}_{1}\cdot\vec{n} \vec{p}_{1}\cdot\vec{S}_{1} - 6 \vec{S}_{1}\times\vec{p}_{1}\cdot\vec{p}_{2} \vec{S}_{1}\cdot\vec{n} \vec{p}_{1}\cdot\vec{S}_{1} \nn\\ 
&& + 10 \vec{S}_{1}\times\vec{n}\cdot\vec{p}_{1} \vec{S}_{1}\cdot\vec{n} \vec{p}_{1}\cdot\vec{n} \vec{p}_{2}\cdot\vec{S}_{1} - \vec{S}_{1}\times\vec{n}\cdot\vec{p}_{1} \vec{p}_{1}\cdot\vec{S}_{1} \vec{p}_{2}\cdot\vec{S}_{1} - 5 \vec{S}_{1}\times\vec{n}\cdot\vec{p}_{1} \big( \vec{p}_{1}\cdot\vec{p}_{2} \nn\\ 
&& + 7 \vec{p}_{1}\cdot\vec{n} \vec{p}_{2}\cdot\vec{n} \big) ( \vec{S}_{1}\cdot\vec{n})^{2} - 5 \vec{S}_{1}\times\vec{n}\cdot\vec{p}_{2} p_{1}^2 ( \vec{S}_{1}\cdot\vec{n})^{2} + 5 \vec{S}_{1}\times\vec{p}_{1}\cdot\vec{p}_{2} \vec{p}_{1}\cdot\vec{n} ( \vec{S}_{1}\cdot\vec{n})^{2} \nn\\ 
&& - 3 \vec{S}_{1}\times\vec{n}\cdot\vec{p}_{2} ( \vec{p}_{1}\cdot\vec{S}_{1})^{2} \Big] + 	\frac{G}{2 m_{1}{}^3 m_{2} r{}^4} \Big[ \vec{S}_{1}\times\vec{n}\cdot\vec{p}_{1} S_{1}^2 p_{2}^2 + \vec{S}_{1}\times\vec{n}\cdot\vec{p}_{2} S_{1}^2 \big( \vec{p}_{1}\cdot\vec{p}_{2} \nn\\ 
&& + 5 \vec{p}_{1}\cdot\vec{n} \vec{p}_{2}\cdot\vec{n} \big) - \vec{S}_{1}\times\vec{p}_{1}\cdot\vec{p}_{2} S_{1}^2 \vec{p}_{2}\cdot\vec{n} + 10 \vec{S}_{1}\times\vec{n}\cdot\vec{p}_{2} \vec{S}_{1}\cdot\vec{n} \vec{p}_{2}\cdot\vec{n} \vec{p}_{1}\cdot\vec{S}_{1} \nn\\ 
&& + 10 \vec{S}_{1}\times\vec{n}\cdot\vec{p}_{2} \vec{S}_{1}\cdot\vec{n} \vec{p}_{1}\cdot\vec{n} \vec{p}_{2}\cdot\vec{S}_{1} - 2 \vec{S}_{1}\times\vec{p}_{1}\cdot\vec{p}_{2} \vec{S}_{1}\cdot\vec{n} \vec{p}_{2}\cdot\vec{S}_{1} \nn\\ 
&& - 2 \vec{S}_{1}\times\vec{n}\cdot\vec{p}_{2} \vec{p}_{1}\cdot\vec{S}_{1} \vec{p}_{2}\cdot\vec{S}_{1} - 5 \vec{S}_{1}\times\vec{n}\cdot\vec{p}_{1} p_{2}^2 ( \vec{S}_{1}\cdot\vec{n})^{2} - 5 \vec{S}_{1}\times\vec{n}\cdot\vec{p}_{2} \big( \vec{p}_{1}\cdot\vec{p}_{2} \nn\\ 
&& + 7 \vec{p}_{1}\cdot\vec{n} \vec{p}_{2}\cdot\vec{n} \big) ( \vec{S}_{1}\cdot\vec{n})^{2} + 5 \vec{S}_{1}\times\vec{p}_{1}\cdot\vec{p}_{2} \vec{p}_{2}\cdot\vec{n} ( \vec{S}_{1}\cdot\vec{n})^{2} \Big]\nn\\ && - 	\frac{G^2 m_{2}{}^2}{m_{1}{}^3 r{}^5} \Big[ 7 \vec{S}_{1}\times\vec{n}\cdot\vec{p}_{1} S_{1}^2 - 36 \vec{S}_{1}\times\vec{n}\cdot\vec{p}_{1} ( \vec{S}_{1}\cdot\vec{n})^{2} \Big] + 	\frac{5 G^2}{2 m_{1} r{}^5} \Big[ \vec{S}_{1}\times\vec{n}\cdot\vec{p}_{2} S_{1}^2 \nn\\ 
&& - 5 \vec{S}_{1}\times\vec{n}\cdot\vec{p}_{2} ( \vec{S}_{1}\cdot\vec{n})^{2} \Big] - 	\frac{G^2 m_{2}}{6 m_{1}{}^2 r{}^5} \Big[ 17 \vec{S}_{1}\times\vec{n}\cdot\vec{p}_{1} S_{1}^2 - 46 \vec{S}_{1}\times\vec{n}\cdot\vec{p}_{2} S_{1}^2 \nn\\ 
&& - 87 \vec{S}_{1}\times\vec{n}\cdot\vec{p}_{1} ( \vec{S}_{1}\cdot\vec{n})^{2} + 234 \vec{S}_{1}\times\vec{n}\cdot\vec{p}_{2} ( \vec{S}_{1}\cdot\vec{n})^{2} \Big],
\eea
\bea
H^{\text{NLO}}_{\text{S}_1^2 \text{S}_2}  &=&- 	\frac{3 G}{8 m_{1}{}^4 r{}^4} \Big[ 2 \vec{S}_{2}\times\vec{n}\cdot\vec{p}_{1} S_{1}^2 \big( 9 p_{1}^2 -10 ( \vec{p}_{1}\cdot\vec{n})^{2} \big) \nn\\ 
&& - \vec{S}_{1}\times\vec{n}\cdot\vec{p}_{1} \vec{S}_{1}\cdot\vec{S}_{2} \big( 17 p_{1}^2 -30 ( \vec{p}_{1}\cdot\vec{n})^{2} \big) + 2 \vec{p}_{1}\cdot\vec{S}_{1} \vec{S}_{1}\times\vec{n}\cdot\vec{S}_{2} p_{1}^2 \nn\\ 
&& - \vec{S}_{1}\cdot\vec{n} \vec{S}_{1}\times\vec{p}_{1}\cdot\vec{S}_{2} p_{1}^2 + 40 \vec{S}_{2}\times\vec{n}\cdot\vec{p}_{1} \vec{S}_{1}\cdot\vec{n} \vec{p}_{1}\cdot\vec{n} \vec{p}_{1}\cdot\vec{S}_{1} \nn\\ 
&& + 5 \vec{S}_{1}\times\vec{n}\cdot\vec{p}_{1} \vec{S}_{1}\cdot\vec{n} p_{1}^2 \vec{S}_{2}\cdot\vec{n} - 30 \vec{S}_{1}\times\vec{n}\cdot\vec{p}_{1} \vec{S}_{1}\cdot\vec{n} \vec{p}_{1}\cdot\vec{n} \vec{p}_{1}\cdot\vec{S}_{2} \nn\\ 
&& + 14 \vec{S}_{1}\times\vec{n}\cdot\vec{p}_{1} \vec{p}_{1}\cdot\vec{S}_{1} \vec{p}_{1}\cdot\vec{S}_{2} - 20 \vec{S}_{2}\times\vec{n}\cdot\vec{p}_{1} p_{1}^2 ( \vec{S}_{1}\cdot\vec{n})^{2} - 16 \vec{S}_{2}\times\vec{n}\cdot\vec{p}_{1} ( \vec{p}_{1}\cdot\vec{S}_{1})^{2} \Big] \nn\\ 
&& + 	\frac{3 G}{16 m_{1}{}^3 m_{2} r{}^4} \Big[ 8 \vec{S}_{2}\times\vec{n}\cdot\vec{p}_{1} S_{1}^2 \big( 2 \vec{p}_{1}\cdot\vec{p}_{2} -5 \vec{p}_{1}\cdot\vec{n} \vec{p}_{2}\cdot\vec{n} \big) \nn\\ 
&& + \vec{S}_{2}\times\vec{n}\cdot\vec{p}_{2} S_{1}^2 \big( 16 p_{1}^2 -5 ( \vec{p}_{1}\cdot\vec{n})^{2} \big) - 16 \vec{S}_{2}\times\vec{p}_{1}\cdot\vec{p}_{2} S_{1}^2 \vec{p}_{1}\cdot\vec{n} \nn\\ 
&& - 10 \vec{S}_{1}\times\vec{n}\cdot\vec{p}_{1} \vec{S}_{1}\cdot\vec{S}_{2} \big( \vec{p}_{1}\cdot\vec{p}_{2} -8 \vec{p}_{1}\cdot\vec{n} \vec{p}_{2}\cdot\vec{n} \big) \nn\\ 
&& - 8 \vec{S}_{1}\times\vec{n}\cdot\vec{p}_{2} \vec{S}_{1}\cdot\vec{S}_{2} \big( 2 p_{1}^2 -5 ( \vec{p}_{1}\cdot\vec{n})^{2} \big) + 16 \vec{S}_{1}\times\vec{p}_{1}\cdot\vec{p}_{2} \vec{S}_{1}\cdot\vec{S}_{2} \vec{p}_{1}\cdot\vec{n} \nn\\ 
&& - 16 \vec{p}_{1}\cdot\vec{S}_{1} \vec{S}_{1}\times\vec{n}\cdot\vec{S}_{2} \vec{p}_{1}\cdot\vec{p}_{2} + 8 \vec{S}_{1}\cdot\vec{n} \vec{S}_{1}\times\vec{p}_{1}\cdot\vec{S}_{2} \vec{p}_{1}\cdot\vec{p}_{2} \nn\\ 
&& - 8 \vec{p}_{2}\cdot\vec{S}_{1} \vec{S}_{1}\times\vec{p}_{1}\cdot\vec{S}_{2} \vec{p}_{1}\cdot\vec{n} + 12 \vec{p}_{1}\cdot\vec{S}_{1} \vec{S}_{1}\times\vec{p}_{2}\cdot\vec{S}_{2} \vec{p}_{1}\cdot\vec{n} \nn\\ 
&& + 40 \vec{S}_{2}\times\vec{n}\cdot\vec{p}_{1} \vec{S}_{1}\cdot\vec{n} \vec{p}_{2}\cdot\vec{n} \vec{p}_{1}\cdot\vec{S}_{1} + 30 \vec{S}_{2}\times\vec{n}\cdot\vec{p}_{2} \vec{S}_{1}\cdot\vec{n} \vec{p}_{1}\cdot\vec{n} \vec{p}_{1}\cdot\vec{S}_{1} \nn\\ 
&& + 4 \vec{S}_{2}\times\vec{p}_{1}\cdot\vec{p}_{2} \vec{S}_{1}\cdot\vec{n} \vec{p}_{1}\cdot\vec{S}_{1} + 40 \vec{S}_{2}\times\vec{n}\cdot\vec{p}_{1} \vec{S}_{1}\cdot\vec{n} \vec{p}_{1}\cdot\vec{n} \vec{p}_{2}\cdot\vec{S}_{1} \nn\\ 
&& - 32 \vec{S}_{2}\times\vec{n}\cdot\vec{p}_{1} \vec{p}_{1}\cdot\vec{S}_{1} \vec{p}_{2}\cdot\vec{S}_{1} - 40 \vec{S}_{1}\times\vec{n}\cdot\vec{p}_{1} \vec{S}_{1}\cdot\vec{n} \vec{p}_{1}\cdot\vec{p}_{2} \vec{S}_{2}\cdot\vec{n} \nn\\ 
&& - 40 \vec{S}_{1}\times\vec{p}_{1}\cdot\vec{p}_{2} \vec{S}_{1}\cdot\vec{n} \vec{p}_{1}\cdot\vec{n} \vec{S}_{2}\cdot\vec{n} + 8 \vec{S}_{1}\times\vec{p}_{1}\cdot\vec{p}_{2} \vec{p}_{1}\cdot\vec{S}_{1} \vec{S}_{2}\cdot\vec{n} \nn\\ 
&& - 80 \vec{S}_{1}\times\vec{n}\cdot\vec{p}_{1} \vec{S}_{1}\cdot\vec{n} \vec{p}_{2}\cdot\vec{n} \vec{p}_{1}\cdot\vec{S}_{2} - 40 \vec{S}_{1}\times\vec{n}\cdot\vec{p}_{2} \vec{S}_{1}\cdot\vec{n} \vec{p}_{1}\cdot\vec{n} \vec{p}_{1}\cdot\vec{S}_{2} \nn\\ 
&& + 16 \vec{S}_{1}\times\vec{n}\cdot\vec{p}_{2} \vec{p}_{1}\cdot\vec{S}_{1} \vec{p}_{1}\cdot\vec{S}_{2} + 26 \vec{S}_{1}\times\vec{n}\cdot\vec{p}_{1} \vec{p}_{2}\cdot\vec{S}_{1} \vec{p}_{1}\cdot\vec{S}_{2} \nn\\ 
&& + 8 \vec{S}_{1}\times\vec{n}\cdot\vec{p}_{1} \vec{p}_{1}\cdot\vec{S}_{1} \vec{p}_{2}\cdot\vec{S}_{2} - 40 \vec{S}_{2}\times\vec{n}\cdot\vec{p}_{1} \vec{p}_{1}\cdot\vec{p}_{2} ( \vec{S}_{1}\cdot\vec{n})^{2} \nn\\ 
&& - 25 \vec{S}_{2}\times\vec{n}\cdot\vec{p}_{2} p_{1}^2 ( \vec{S}_{1}\cdot\vec{n})^{2} - 16 \vec{S}_{2}\times\vec{n}\cdot\vec{p}_{2} ( \vec{p}_{1}\cdot\vec{S}_{1})^{2} \Big] \nn\\ 
&& + 	\frac{3 G}{4 m_{1}{}^2 m_{2}{}^2 r{}^4} \Big[ 2 \vec{S}_{2}\times\vec{n}\cdot\vec{p}_{1} S_{1}^2 \big( p_{2}^2 -5 ( \vec{p}_{2}\cdot\vec{n})^{2} \big) - 3 \vec{S}_{2}\times\vec{n}\cdot\vec{p}_{2} S_{1}^2 \vec{p}_{1}\cdot\vec{p}_{2} \nn\\ 
&& - \vec{S}_{1}\times\vec{n}\cdot\vec{p}_{1} \vec{S}_{1}\cdot\vec{S}_{2} \big( p_{2}^2 -10 ( \vec{p}_{2}\cdot\vec{n})^{2} \big) \nn\\ 
&& + \vec{S}_{1}\times\vec{n}\cdot\vec{p}_{2} \vec{S}_{1}\cdot\vec{S}_{2} \big( 3 \vec{p}_{1}\cdot\vec{p}_{2} -10 \vec{p}_{1}\cdot\vec{n} \vec{p}_{2}\cdot\vec{n} \big) + \vec{S}_{1}\times\vec{p}_{1}\cdot\vec{p}_{2} \vec{S}_{1}\cdot\vec{S}_{2} \vec{p}_{2}\cdot\vec{n} \nn\\ 
&& + 2 \vec{p}_{1}\cdot\vec{S}_{1} \vec{S}_{1}\times\vec{n}\cdot\vec{S}_{2} \big( p_{2}^2 -5 ( \vec{p}_{2}\cdot\vec{n})^{2} \big) - \vec{S}_{1}\cdot\vec{n} \vec{S}_{1}\times\vec{p}_{1}\cdot\vec{S}_{2} \big( p_{2}^2 -5 ( \vec{p}_{2}\cdot\vec{n})^{2} \big) \nn\\ 
&& - 5 \vec{S}_{2}\times\vec{n}\cdot\vec{p}_{2} \vec{S}_{1}\cdot\vec{n} \vec{p}_{1}\cdot\vec{n} \vec{p}_{2}\cdot\vec{S}_{1} + 3 \vec{S}_{2}\times\vec{n}\cdot\vec{p}_{2} \vec{p}_{1}\cdot\vec{S}_{1} \vec{p}_{2}\cdot\vec{S}_{1} \nn\\ 
&& + 5 \vec{S}_{1}\times\vec{n}\cdot\vec{p}_{1} \vec{S}_{1}\cdot\vec{n} \big( p_{2}^2 -7 ( \vec{p}_{2}\cdot\vec{n})^{2} \big) \vec{S}_{2}\cdot\vec{n} + 5 \vec{S}_{1}\times\vec{n}\cdot\vec{p}_{1} \vec{p}_{2}\cdot\vec{S}_{1} \vec{p}_{2}\cdot\vec{n} \vec{S}_{2}\cdot\vec{n} \nn\\ 
&& - \vec{S}_{1}\times\vec{p}_{1}\cdot\vec{p}_{2} \vec{p}_{2}\cdot\vec{S}_{1} \vec{S}_{2}\cdot\vec{n} + 10 \vec{S}_{1}\times\vec{n}\cdot\vec{p}_{2} \vec{S}_{1}\cdot\vec{n} \vec{p}_{2}\cdot\vec{n} \vec{p}_{1}\cdot\vec{S}_{2} \nn\\ 
&& - 3 \vec{S}_{1}\times\vec{n}\cdot\vec{p}_{2} \vec{p}_{2}\cdot\vec{S}_{1} \vec{p}_{1}\cdot\vec{S}_{2} + 10 \vec{S}_{1}\times\vec{n}\cdot\vec{p}_{1} \vec{S}_{1}\cdot\vec{n} \vec{p}_{2}\cdot\vec{n} \vec{p}_{2}\cdot\vec{S}_{2} \nn\\ 
&& - 2 \vec{S}_{1}\times\vec{n}\cdot\vec{p}_{1} \vec{p}_{2}\cdot\vec{S}_{1} \vec{p}_{2}\cdot\vec{S}_{2} + 5 \vec{S}_{2}\times\vec{n}\cdot\vec{p}_{2} \vec{p}_{1}\cdot\vec{p}_{2} ( \vec{S}_{1}\cdot\vec{n})^{2} \Big]\nn\\ && - 	\frac{G^2 m_{2}}{2 m_{1}{}^2 r{}^5} \Big[ 27 \vec{S}_{1}\cdot\vec{n} \vec{S}_{1}\times\vec{n}\cdot\vec{S}_{2} \vec{p}_{1}\cdot\vec{n} + 42 \vec{S}_{2}\times\vec{n}\cdot\vec{p}_{1} S_{1}^2 + 81 \vec{S}_{1}\times\vec{n}\cdot\vec{p}_{1} \vec{S}_{1}\cdot\vec{n} \vec{S}_{2}\cdot\vec{n} \nn\\ 
&& - 45 \vec{S}_{1}\times\vec{n}\cdot\vec{p}_{1} \vec{S}_{1}\cdot\vec{S}_{2} + 35 \vec{p}_{1}\cdot\vec{S}_{1} \vec{S}_{1}\times\vec{n}\cdot\vec{S}_{2} - 25 \vec{S}_{1}\cdot\vec{n} \vec{S}_{1}\times\vec{p}_{1}\cdot\vec{S}_{2} \nn\\ 
&& - 3 \vec{S}_{2}\times\vec{n}\cdot\vec{p}_{1} ( \vec{S}_{1}\cdot\vec{n})^{2} \Big] - 	\frac{3 G^2}{4 m_{2} r{}^5} \Big[ 6 \vec{S}_{1}\cdot\vec{n} \vec{S}_{1}\times\vec{n}\cdot\vec{S}_{2} \vec{p}_{2}\cdot\vec{n} - \vec{S}_{2}\times\vec{n}\cdot\vec{p}_{2} S_{1}^2 \nn\\ 
&& + 28 \vec{S}_{1}\times\vec{n}\cdot\vec{p}_{2} \vec{S}_{1}\cdot\vec{n} \vec{S}_{2}\cdot\vec{n} - 20 \vec{S}_{1}\times\vec{n}\cdot\vec{p}_{2} \vec{S}_{1}\cdot\vec{S}_{2} + 6 \vec{p}_{2}\cdot\vec{S}_{1} \vec{S}_{1}\times\vec{n}\cdot\vec{S}_{2} \nn\\ 
&& - 8 \vec{S}_{1}\cdot\vec{n} \vec{S}_{1}\times\vec{p}_{2}\cdot\vec{S}_{2} + 13 \vec{S}_{2}\times\vec{n}\cdot\vec{p}_{2} ( \vec{S}_{1}\cdot\vec{n})^{2} \Big] \nn\\ 
&& + 	\frac{G^2}{4 m_{1} r{}^5} \Big[ 6 \vec{S}_{1}\cdot\vec{n} \vec{S}_{1}\times\vec{n}\cdot\vec{S}_{2} \big( 3 \vec{p}_{1}\cdot\vec{n} + 8 \vec{p}_{2}\cdot\vec{n} \big) - 145 \vec{S}_{2}\times\vec{n}\cdot\vec{p}_{1} S_{1}^2 \nn\\ 
&& - 28 \vec{S}_{2}\times\vec{n}\cdot\vec{p}_{2} S_{1}^2 - 240 \vec{S}_{1}\times\vec{n}\cdot\vec{p}_{1} \vec{S}_{1}\cdot\vec{n} \vec{S}_{2}\cdot\vec{n} - 126 \vec{S}_{1}\times\vec{n}\cdot\vec{p}_{2} \vec{S}_{1}\cdot\vec{n} \vec{S}_{2}\cdot\vec{n} \nn\\ 
&& + 144 \vec{S}_{1}\times\vec{n}\cdot\vec{p}_{1} \vec{S}_{1}\cdot\vec{S}_{2} + 90 \vec{S}_{1}\times\vec{n}\cdot\vec{p}_{2} \vec{S}_{1}\cdot\vec{S}_{2} - 126 \vec{p}_{1}\cdot\vec{S}_{1} \vec{S}_{1}\times\vec{n}\cdot\vec{S}_{2} \nn\\ 
&& - 49 \vec{p}_{2}\cdot\vec{S}_{1} \vec{S}_{1}\times\vec{n}\cdot\vec{S}_{2} + 33 \vec{S}_{1}\cdot\vec{n} \vec{S}_{1}\times\vec{p}_{1}\cdot\vec{S}_{2} + 3 \vec{S}_{1}\cdot\vec{n} \vec{S}_{1}\times\vec{p}_{2}\cdot\vec{S}_{2} \nn\\ 
&& + 57 \vec{S}_{2}\times\vec{n}\cdot\vec{p}_{1} ( \vec{S}_{1}\cdot\vec{n})^{2} - 6 \vec{S}_{2}\times\vec{n}\cdot\vec{p}_{2} ( \vec{S}_{1}\cdot\vec{n})^{2} \Big],
\eea
\bea
H^{\text{NLO}}_{(\text{ES}_1^2 ) \text{S}_2}&=& 	\frac{15 G}{16 m_{1} m_{2}{}^3 r{}^4} \Big[ \vec{S}_{2}\times\vec{n}\cdot\vec{p}_{2} S_{1}^2 p_{2}^2 + 2 \vec{S}_{1}\cdot\vec{n} \vec{S}_{1}\times\vec{p}_{2}\cdot\vec{S}_{2} p_{2}^2 - 5 \vec{S}_{2}\times\vec{n}\cdot\vec{p}_{2} p_{2}^2 ( \vec{S}_{1}\cdot\vec{n})^{2} \Big] \nn\\ 
&& + 	\frac{3 G}{m_{1}{}^4 r{}^4} \Big[ \vec{S}_{2}\times\vec{n}\cdot\vec{p}_{1} S_{1}^2 \big( p_{1}^2 -5 ( \vec{p}_{1}\cdot\vec{n})^{2} \big) - \vec{p}_{1}\cdot\vec{S}_{1} \vec{S}_{1}\times\vec{p}_{1}\cdot\vec{S}_{2} \vec{p}_{1}\cdot\vec{n} \nn\\ 
&& + 5 \vec{S}_{2}\times\vec{n}\cdot\vec{p}_{1} \vec{S}_{1}\cdot\vec{n} \vec{p}_{1}\cdot\vec{n} \vec{p}_{1}\cdot\vec{S}_{1} - \vec{S}_{2}\times\vec{n}\cdot\vec{p}_{1} ( \vec{p}_{1}\cdot\vec{S}_{1})^{2} \Big] \nn\\ 
&& - 	\frac{3 G}{8 m_{1}{}^3 m_{2} r{}^4} \Big[ 4 \vec{S}_{2}\times\vec{n}\cdot\vec{p}_{1} S_{1}^2 \big( 3 \vec{p}_{1}\cdot\vec{p}_{2} -5 \vec{p}_{1}\cdot\vec{n} \vec{p}_{2}\cdot\vec{n} \big) \nn\\ 
&& + \vec{S}_{2}\times\vec{n}\cdot\vec{p}_{2} S_{1}^2 \big( 8 p_{1}^2 -35 ( \vec{p}_{1}\cdot\vec{n})^{2} \big) + 18 \vec{S}_{2}\times\vec{p}_{1}\cdot\vec{p}_{2} S_{1}^2 \vec{p}_{1}\cdot\vec{n} \nn\\ 
&& + 8 \vec{S}_{1}\cdot\vec{n} \vec{S}_{1}\times\vec{p}_{1}\cdot\vec{S}_{2} \big( \vec{p}_{1}\cdot\vec{p}_{2} + 5 \vec{p}_{1}\cdot\vec{n} \vec{p}_{2}\cdot\vec{n} \big) - 8 \vec{p}_{1}\cdot\vec{S}_{1} \vec{S}_{1}\times\vec{p}_{1}\cdot\vec{S}_{2} \vec{p}_{2}\cdot\vec{n} \nn\\ 
&& - 8 \vec{p}_{2}\cdot\vec{S}_{1} \vec{S}_{1}\times\vec{p}_{1}\cdot\vec{S}_{2} \vec{p}_{1}\cdot\vec{n} + 10 \vec{S}_{1}\cdot\vec{n} \vec{S}_{1}\times\vec{p}_{2}\cdot\vec{S}_{2} ( \vec{p}_{1}\cdot\vec{n})^{2} \nn\\ 
&& - 10 \vec{p}_{1}\cdot\vec{S}_{1} \vec{S}_{1}\times\vec{p}_{2}\cdot\vec{S}_{2} \vec{p}_{1}\cdot\vec{n} + 40 \vec{S}_{2}\times\vec{n}\cdot\vec{p}_{1} \vec{S}_{1}\cdot\vec{n} \vec{p}_{2}\cdot\vec{n} \vec{p}_{1}\cdot\vec{S}_{1} \nn\\ 
&& + 50 \vec{S}_{2}\times\vec{n}\cdot\vec{p}_{2} \vec{S}_{1}\cdot\vec{n} \vec{p}_{1}\cdot\vec{n} \vec{p}_{1}\cdot\vec{S}_{1} + 14 \vec{S}_{2}\times\vec{p}_{1}\cdot\vec{p}_{2} \vec{S}_{1}\cdot\vec{n} \vec{p}_{1}\cdot\vec{S}_{1} \nn\\ 
&& + 40 \vec{S}_{2}\times\vec{n}\cdot\vec{p}_{1} \vec{S}_{1}\cdot\vec{n} \vec{p}_{1}\cdot\vec{n} \vec{p}_{2}\cdot\vec{S}_{1} - 8 \vec{S}_{2}\times\vec{n}\cdot\vec{p}_{1} \vec{p}_{1}\cdot\vec{S}_{1} \vec{p}_{2}\cdot\vec{S}_{1} \nn\\ 
&& - 20 \vec{S}_{2}\times\vec{n}\cdot\vec{p}_{1} \big( \vec{p}_{1}\cdot\vec{p}_{2} + 7 \vec{p}_{1}\cdot\vec{n} \vec{p}_{2}\cdot\vec{n} \big) ( \vec{S}_{1}\cdot\vec{n})^{2} - 35 \vec{S}_{2}\times\vec{n}\cdot\vec{p}_{2} ( \vec{p}_{1}\cdot\vec{n})^{2} ( \vec{S}_{1}\cdot\vec{n})^{2} \nn\\ 
&& - 50 \vec{S}_{2}\times\vec{p}_{1}\cdot\vec{p}_{2} \vec{p}_{1}\cdot\vec{n} ( \vec{S}_{1}\cdot\vec{n})^{2} - 8 \vec{S}_{2}\times\vec{n}\cdot\vec{p}_{2} ( \vec{p}_{1}\cdot\vec{S}_{1})^{2} \Big] \nn\\ 
&& + 	\frac{3 G}{2 m_{1}{}^2 m_{2}{}^2 r{}^4} \Big[ \vec{S}_{2}\times\vec{n}\cdot\vec{p}_{2} S_{1}^2 \big( 2 \vec{p}_{1}\cdot\vec{p}_{2} -5 \vec{p}_{1}\cdot\vec{n} \vec{p}_{2}\cdot\vec{n} \big) + 4 \vec{S}_{2}\times\vec{p}_{1}\cdot\vec{p}_{2} S_{1}^2 \vec{p}_{2}\cdot\vec{n} \nn\\ 
&& + 10 \vec{S}_{1}\cdot\vec{n} \vec{S}_{1}\times\vec{p}_{2}\cdot\vec{S}_{2} \vec{p}_{1}\cdot\vec{n} \vec{p}_{2}\cdot\vec{n} - 2 \vec{p}_{1}\cdot\vec{S}_{1} \vec{S}_{1}\times\vec{p}_{2}\cdot\vec{S}_{2} \vec{p}_{2}\cdot\vec{n} \nn\\ 
&& - 2 \vec{p}_{2}\cdot\vec{S}_{1} \vec{S}_{1}\times\vec{p}_{2}\cdot\vec{S}_{2} \vec{p}_{1}\cdot\vec{n} + 10 \vec{S}_{2}\times\vec{n}\cdot\vec{p}_{2} \vec{S}_{1}\cdot\vec{n} \vec{p}_{2}\cdot\vec{n} \vec{p}_{1}\cdot\vec{S}_{1} \nn\\ 
&& + 10 \vec{S}_{2}\times\vec{n}\cdot\vec{p}_{2} \vec{S}_{1}\cdot\vec{n} \vec{p}_{1}\cdot\vec{n} \vec{p}_{2}\cdot\vec{S}_{1} + 4 \vec{S}_{2}\times\vec{p}_{1}\cdot\vec{p}_{2} \vec{S}_{1}\cdot\vec{n} \vec{p}_{2}\cdot\vec{S}_{1} \nn\\ 
&& - 2 \vec{S}_{2}\times\vec{n}\cdot\vec{p}_{2} \vec{p}_{1}\cdot\vec{S}_{1} \vec{p}_{2}\cdot\vec{S}_{1} - 35 \vec{S}_{2}\times\vec{n}\cdot\vec{p}_{2} \vec{p}_{1}\cdot\vec{n} \vec{p}_{2}\cdot\vec{n} ( \vec{S}_{1}\cdot\vec{n})^{2} \nn\\ 
&& - 10 \vec{S}_{2}\times\vec{p}_{1}\cdot\vec{p}_{2} \vec{p}_{2}\cdot\vec{n} ( \vec{S}_{1}\cdot\vec{n})^{2} \Big]\nn\\ && +  	\frac{G^2 m_{2}}{m_{1}{}^2 r{}^5} \Big[ 66 \vec{S}_{1}\cdot\vec{n} \vec{S}_{1}\times\vec{n}\cdot\vec{S}_{2} \vec{p}_{1}\cdot\vec{n} - 24 \vec{S}_{2}\times\vec{n}\cdot\vec{p}_{1} S_{1}^2 - 8 \vec{p}_{1}\cdot\vec{S}_{1} \vec{S}_{1}\times\vec{n}\cdot\vec{S}_{2} \nn\\ 
&& - 47 \vec{S}_{1}\cdot\vec{n} \vec{S}_{1}\times\vec{p}_{1}\cdot\vec{S}_{2} + 99 \vec{S}_{2}\times\vec{n}\cdot\vec{p}_{1} ( \vec{S}_{1}\cdot\vec{n})^{2} \Big] \nn\\ 
&& + 	\frac{G^2}{4 m_{2} r{}^5} \Big[ 6 \vec{S}_{1}\cdot\vec{n} \vec{S}_{1}\times\vec{n}\cdot\vec{S}_{2} \vec{p}_{2}\cdot\vec{n} + 39 \vec{S}_{2}\times\vec{n}\cdot\vec{p}_{2} S_{1}^2 - 2 \vec{p}_{2}\cdot\vec{S}_{1} \vec{S}_{1}\times\vec{n}\cdot\vec{S}_{2} \nn\\ 
&& + 58 \vec{S}_{1}\cdot\vec{n} \vec{S}_{1}\times\vec{p}_{2}\cdot\vec{S}_{2} - 177 \vec{S}_{2}\times\vec{n}\cdot\vec{p}_{2} ( \vec{S}_{1}\cdot\vec{n})^{2} \Big] \nn\\ 
&& - 	\frac{G^2}{4 m_{1} r{}^5} \Big[ 6 \vec{S}_{1}\cdot\vec{n} \vec{S}_{1}\times\vec{n}\cdot\vec{S}_{2} \big( \vec{p}_{1}\cdot\vec{n} + 42 \vec{p}_{2}\cdot\vec{n} \big) + 76 \vec{S}_{2}\times\vec{n}\cdot\vec{p}_{1} S_{1}^2 \nn\\ 
&& - 57 \vec{S}_{2}\times\vec{n}\cdot\vec{p}_{2} S_{1}^2 + 5 \vec{p}_{1}\cdot\vec{S}_{1} \vec{S}_{1}\times\vec{n}\cdot\vec{S}_{2} - 22 \vec{p}_{2}\cdot\vec{S}_{1} \vec{S}_{1}\times\vec{n}\cdot\vec{S}_{2} \nn\\ 
&& + 103 \vec{S}_{1}\cdot\vec{n} \vec{S}_{1}\times\vec{p}_{1}\cdot\vec{S}_{2} - 131 \vec{S}_{1}\cdot\vec{n} \vec{S}_{1}\times\vec{p}_{2}\cdot\vec{S}_{2} - 312 \vec{S}_{2}\times\vec{n}\cdot\vec{p}_{1} ( \vec{S}_{1}\cdot\vec{n})^{2} \nn\\ 
&& + 252 \vec{S}_{2}\times\vec{n}\cdot\vec{p}_{2} ( \vec{S}_{1}\cdot\vec{n})^{2} \Big].
\eea

\section{COM Generator of the N$^3$LO Spin-Orbit Sector} 
\label{thirdsubleadinggenwspin}

We recall that the solution for $\vec{G}^{\text{N$^3$LO}}_{\text{SO}}$ is written as:
\bea
\vec{G}^{\text{N}^3\text{LO}}_{\text{SO}} &=& 
H^{\text{N}^2\text{LO}}_{\text{SO}} \frac{\left( \vec{x}_1 + \vec{x}_2 \right)}{2} 
+  \left( \vec{Y}^{\text{N}^3\text{LO}}_{\text{S}_1}  + \left(1 \leftrightarrow 2 \right) \right),
\eea
and we find:
\bea
\vec{Y}^{\text{N}^3\text{LO}}_{\text{S}_1 } &=&  	\frac{G m_{2}}{32 m_{1}{}^5 r} \Big[ 7 \vec{S}_{1}\times\vec{n}\cdot\vec{p}_{1} p_{1}^{4} \vec{n} + 6 \vec{S}_{1}\times\vec{p}_{1} p_{1}^{4} \Big] - 	\frac{G}{32 m_{1}{}^3 m_{2} r} \Big[ \vec{S}_{1}\times\vec{n}\cdot\vec{p}_{1} \big( 11 p_{1}^2 p_{2}^2 \vec{n} \nn\\ 
&& - 16 ( \vec{p}_{1}\cdot\vec{p}_{2})^{2} \vec{n} + 42 \vec{p}_{2}\cdot\vec{n} \vec{p}_{1}\cdot\vec{p}_{2} \vec{p}_{1} - 39 \vec{p}_{1}\cdot\vec{n} p_{2}^2 \vec{p}_{1} - 2 p_{1}^2 \vec{p}_{2}\cdot\vec{n} \vec{p}_{2} \nn\\ 
&& + 24 \vec{p}_{1}\cdot\vec{n} \vec{p}_{1}\cdot\vec{p}_{2} \vec{p}_{2} -12 p_{1}^2 ( \vec{p}_{2}\cdot\vec{n})^{2} \vec{n} + 21 \vec{p}_{1}\cdot\vec{n} ( \vec{p}_{2}\cdot\vec{n})^{2} \vec{p}_{1} - 24 \vec{p}_{2}\cdot\vec{n} ( \vec{p}_{1}\cdot\vec{n})^{2} \vec{p}_{2} \big) \nn\\ 
&& + 8 \vec{S}_{1}\times\vec{n}\cdot\vec{p}_{2} \big( 3 p_{1}^2 \vec{p}_{1}\cdot\vec{p}_{2} \vec{n} - p_{1}^2 \vec{p}_{2}\cdot\vec{n} \vec{p}_{1} - 2 \vec{p}_{1}\cdot\vec{n} p_{1}^2 \vec{p}_{2} \big) + 2 \vec{S}_{1}\times\vec{p}_{1}\cdot\vec{p}_{2} \big( 13 \vec{p}_{1}\cdot\vec{p}_{2} \vec{p}_{1} \nn\\ 
&& - 5 p_{1}^2 \vec{p}_{2} -9 \vec{p}_{1}\cdot\vec{n} \vec{p}_{2}\cdot\vec{n} \vec{p}_{1} + 16 ( \vec{p}_{1}\cdot\vec{n})^{2} \vec{p}_{2} \big) - 2 \vec{S}_{1}\times\vec{n} \big( 6 p_{1}^2 \vec{p}_{2}\cdot\vec{n} \vec{p}_{1}\cdot\vec{p}_{2} \nn\\ 
&& - 13 \vec{p}_{1}\cdot\vec{n} p_{1}^2 p_{2}^2 + 3 \vec{p}_{1}\cdot\vec{n} p_{1}^2 ( \vec{p}_{2}\cdot\vec{n})^{2} \big) + \vec{S}_{1}\times\vec{p}_{1} \big( 2 p_{1}^2 p_{2}^2 - 46 ( \vec{p}_{1}\cdot\vec{p}_{2})^{2} \nn\\ 
&& + 68 \vec{p}_{1}\cdot\vec{n} \vec{p}_{2}\cdot\vec{n} \vec{p}_{1}\cdot\vec{p}_{2} - 29 p_{2}^2 ( \vec{p}_{1}\cdot\vec{n})^{2} - 12 p_{1}^2 ( \vec{p}_{2}\cdot\vec{n})^{2} -9 ( \vec{p}_{1}\cdot\vec{n})^{2} ( \vec{p}_{2}\cdot\vec{n})^{2} \big) \nn\\ 
&& + 4 \vec{S}_{1}\times\vec{p}_{2} \big( 11 p_{1}^2 \vec{p}_{1}\cdot\vec{p}_{2} -5 \vec{p}_{1}\cdot\vec{n} p_{1}^2 \vec{p}_{2}\cdot\vec{n} \big) \Big] + 	\frac{G}{32 m_{1}{}^4 r} \Big[ \vec{S}_{1}\times\vec{n}\cdot\vec{p}_{1} \big( 6 p_{1}^2 \vec{p}_{1}\cdot\vec{p}_{2} \vec{n} \nn\\ 
&& - 24 p_{1}^2 \vec{p}_{2}\cdot\vec{n} \vec{p}_{1} + 7 \vec{p}_{1}\cdot\vec{n} p_{1}^2 \vec{p}_{2} + 3 \vec{p}_{1}\cdot\vec{n} p_{1}^2 \vec{p}_{2}\cdot\vec{n} \vec{n} \big) + 15 \vec{S}_{1}\times\vec{n}\cdot\vec{p}_{2} \big( - p_{1}^{4} \vec{n} \nn\\ 
&& + \vec{p}_{1}\cdot\vec{n} p_{1}^2 \vec{p}_{1} \big) - 4 \vec{S}_{1}\times\vec{p}_{1}\cdot\vec{p}_{2} p_{1}^2 \vec{p}_{1} - \vec{S}_{1}\times\vec{n} \big( 15 \vec{p}_{1}\cdot\vec{n} p_{1}^2 \vec{p}_{1}\cdot\vec{p}_{2} - 22 \vec{p}_{2}\cdot\vec{n} p_{1}^{4} \big) \nn\\ 
&& + 4 \vec{S}_{1}\times\vec{p}_{1} \big( 8 p_{1}^2 \vec{p}_{1}\cdot\vec{p}_{2} -3 \vec{p}_{1}\cdot\vec{n} p_{1}^2 \vec{p}_{2}\cdot\vec{n} \big) - \vec{S}_{1}\times\vec{p}_{2} \big( 14 p_{1}^{4} -15 p_{1}^2 ( \vec{p}_{1}\cdot\vec{n})^{2} \big) \Big] \nn\\ 
&& + 	\frac{G}{32 m_{1}{}^2 m_{2}{}^2 r} \Big[ \vec{S}_{1}\times\vec{n}\cdot\vec{p}_{1} \big( 22 \vec{p}_{1}\cdot\vec{p}_{2} p_{2}^2 \vec{n} - \vec{p}_{2}\cdot\vec{n} p_{2}^2 \vec{p}_{1} - 2 \vec{p}_{2}\cdot\vec{n} \vec{p}_{1}\cdot\vec{p}_{2} \vec{p}_{2} \nn\\ 
&& + 3 \vec{p}_{1}\cdot\vec{n} \vec{p}_{2}\cdot\vec{n} p_{2}^2 \vec{n} - 63 \vec{p}_{1}\cdot\vec{p}_{2} ( \vec{p}_{2}\cdot\vec{n})^{2} \vec{n} + 12 ( \vec{p}_{2}\cdot\vec{n})^{3} \vec{p}_{1} - 3 \vec{p}_{1}\cdot\vec{n} ( \vec{p}_{2}\cdot\vec{n})^{2} \vec{p}_{2} \big) \nn\\ 
&& - \vec{S}_{1}\times\vec{n}\cdot\vec{p}_{2} \big( 38 p_{1}^2 p_{2}^2 \vec{n} - 32 ( \vec{p}_{1}\cdot\vec{p}_{2})^{2} \vec{n} + 34 \vec{p}_{2}\cdot\vec{n} \vec{p}_{1}\cdot\vec{p}_{2} \vec{p}_{1} - 30 \vec{p}_{1}\cdot\vec{n} p_{2}^2 \vec{p}_{1} \nn\\ 
&& - 42 p_{1}^2 \vec{p}_{2}\cdot\vec{n} \vec{p}_{2} + 32 \vec{p}_{1}\cdot\vec{n} \vec{p}_{1}\cdot\vec{p}_{2} \vec{p}_{2} + 3 \vec{p}_{1}\cdot\vec{n} \vec{p}_{2}\cdot\vec{n} \vec{p}_{1}\cdot\vec{p}_{2} \vec{n} - 15 p_{1}^2 ( \vec{p}_{2}\cdot\vec{n})^{2} \vec{n} \nn\\ 
&& - 9 \vec{p}_{1}\cdot\vec{n} ( \vec{p}_{2}\cdot\vec{n})^{2} \vec{p}_{1} + 21 \vec{p}_{2}\cdot\vec{n} ( \vec{p}_{1}\cdot\vec{n})^{2} \vec{p}_{2} \big) - \vec{S}_{1}\times\vec{p}_{1}\cdot\vec{p}_{2} \big( 43 p_{2}^2 \vec{p}_{1} \nn\\ 
&& - 32 \vec{p}_{1}\cdot\vec{p}_{2} \vec{p}_{2} -54 \vec{p}_{1}\cdot\vec{n} p_{2}^2 \vec{n} - 54 ( \vec{p}_{2}\cdot\vec{n})^{2} \vec{p}_{1} + 21 \vec{p}_{1}\cdot\vec{n} \vec{p}_{2}\cdot\vec{n} \vec{p}_{2} \nn\\ 
&& + 60 \vec{p}_{1}\cdot\vec{n} ( \vec{p}_{2}\cdot\vec{n})^{2} \vec{n} \big) + \vec{S}_{1}\times\vec{n} \big( 3 p_{1}^2 \vec{p}_{2}\cdot\vec{n} p_{2}^2 -8 p_{1}^2 ( \vec{p}_{2}\cdot\vec{n})^{3} -5 ( \vec{p}_{1}\cdot\vec{n})^{2} ( \vec{p}_{2}\cdot\vec{n})^{3} \big) \nn\\ 
&& - \vec{S}_{1}\times\vec{p}_{1} \big( 23 \vec{p}_{1}\cdot\vec{p}_{2} p_{2}^2 -4 \vec{p}_{1}\cdot\vec{n} \vec{p}_{2}\cdot\vec{n} p_{2}^2 - 8 \vec{p}_{1}\cdot\vec{p}_{2} ( \vec{p}_{2}\cdot\vec{n})^{2} -11 \vec{p}_{1}\cdot\vec{n} ( \vec{p}_{2}\cdot\vec{n})^{3} \big) \nn\\ 
&& - \vec{S}_{1}\times\vec{p}_{2} \big( 41 p_{1}^2 p_{2}^2 - 42 ( \vec{p}_{1}\cdot\vec{p}_{2})^{2} + 57 \vec{p}_{1}\cdot\vec{n} \vec{p}_{2}\cdot\vec{n} \vec{p}_{1}\cdot\vec{p}_{2} - 41 p_{2}^2 ( \vec{p}_{1}\cdot\vec{n})^{2} \nn\\ 
&& - 46 p_{1}^2 ( \vec{p}_{2}\cdot\vec{n})^{2} + 21 ( \vec{p}_{1}\cdot\vec{n})^{2} ( \vec{p}_{2}\cdot\vec{n})^{2} \big) \Big] + 	\frac{G}{32 m_{1} m_{2}{}^3 r} \Big[ \vec{S}_{1}\times\vec{n}\cdot\vec{p}_{1} \big( -2 p_{2}^{4} \vec{n} \nn\\ 
&& + 2 \vec{p}_{2}\cdot\vec{n} p_{2}^2 \vec{p}_{2} + 33 p_{2}^2 ( \vec{p}_{2}\cdot\vec{n})^{2} \vec{n} - 9 ( \vec{p}_{2}\cdot\vec{n})^{3} \vec{p}_{2} -15 ( \vec{p}_{2}\cdot\vec{n})^{4} \vec{n} \big) \nn\\ 
&& + \vec{S}_{1}\times\vec{n}\cdot\vec{p}_{2} \big( 2 \vec{p}_{1}\cdot\vec{p}_{2} p_{2}^2 \vec{n} + 8 \vec{p}_{2}\cdot\vec{n} p_{2}^2 \vec{p}_{1} - 10 \vec{p}_{1}\cdot\vec{n} p_{2}^2 \vec{p}_{2} -9 \vec{p}_{1}\cdot\vec{p}_{2} ( \vec{p}_{2}\cdot\vec{n})^{2} \vec{n} \nn\\ 
&& + 9 \vec{p}_{1}\cdot\vec{n} ( \vec{p}_{2}\cdot\vec{n})^{2} \vec{p}_{2} \big) + \vec{S}_{1}\times\vec{p}_{1}\cdot\vec{p}_{2} \big( 10 p_{2}^2 \vec{p}_{2} -18 \vec{p}_{2}\cdot\vec{n} p_{2}^2 \vec{n} - 9 ( \vec{p}_{2}\cdot\vec{n})^{2} \vec{p}_{2} \nn\\ 
&& + 45 ( \vec{p}_{2}\cdot\vec{n})^{3} \vec{n} \big) - \vec{S}_{1}\times\vec{n} \big( 11 \vec{p}_{1}\cdot\vec{p}_{2} ( \vec{p}_{2}\cdot\vec{n})^{3} + 5 \vec{p}_{1}\cdot\vec{n} ( \vec{p}_{2}\cdot\vec{n})^{4} \big) - 2 \vec{S}_{1}\times\vec{p}_{1} \big( 2 p_{2}^{4} \nn\\ 
&& + 4 p_{2}^2 ( \vec{p}_{2}\cdot\vec{n})^{2} -5 ( \vec{p}_{2}\cdot\vec{n})^{4} \big) - \vec{S}_{1}\times\vec{p}_{2} \big( 8 \vec{p}_{1}\cdot\vec{p}_{2} p_{2}^2 + 16 \vec{p}_{1}\cdot\vec{n} \vec{p}_{2}\cdot\vec{n} p_{2}^2 \nn\\ 
&& + 21 \vec{p}_{1}\cdot\vec{p}_{2} ( \vec{p}_{2}\cdot\vec{n})^{2} -13 \vec{p}_{1}\cdot\vec{n} ( \vec{p}_{2}\cdot\vec{n})^{3} \big) \Big] + 	\frac{G}{32 m_{2}{}^4 r} \Big[ \vec{S}_{1}\times\vec{n} \big( 3 \vec{p}_{2}\cdot\vec{n} p_{2}^{4} \nn\\ 
&& + 10 p_{2}^2 ( \vec{p}_{2}\cdot\vec{n})^{3} -5 ( \vec{p}_{2}\cdot\vec{n})^{5} \big) + \vec{S}_{1}\times\vec{p}_{2} \big( 5 p_{2}^{4} + 6 p_{2}^2 ( \vec{p}_{2}\cdot\vec{n})^{2} -19 ( \vec{p}_{2}\cdot\vec{n})^{4} \big) \Big] \nn\\ 
&& - 	\frac{G^2 m_{2}{}^2}{48 m_{1}{}^3 r{}^2} \Big[ 2 \vec{S}_{1}\times\vec{n}\cdot\vec{p}_{1} \big( p_{1}^2 \vec{n} - 28 \vec{p}_{1}\cdot\vec{n} \vec{p}_{1} -4 ( \vec{p}_{1}\cdot\vec{n})^{2} \vec{n} \big) + 30 \vec{S}_{1}\times\vec{n} \vec{p}_{1}\cdot\vec{n} p_{1}^2 \nn\\ 
&& + \vec{S}_{1}\times\vec{p}_{1} \big( 355 p_{1}^2 + 16 ( \vec{p}_{1}\cdot\vec{n})^{2} \big) \Big] + 	\frac{G^2 m_{2}}{96 m_{1}{}^2 r{}^2} \Big[ \vec{S}_{1}\times\vec{n}\cdot\vec{p}_{1} \big( 1626 p_{1}^2 \vec{n} + 632 \vec{p}_{1}\cdot\vec{p}_{2} \vec{n} \nn\\ 
&& + 282 \vec{p}_{1}\cdot\vec{n} \vec{p}_{1} + 426 \vec{p}_{2}\cdot\vec{n} \vec{p}_{1} + 196 \vec{p}_{1}\cdot\vec{n} \vec{p}_{2} -670 \vec{p}_{1}\cdot\vec{n} \vec{p}_{2}\cdot\vec{n} \vec{n} + 1287 ( \vec{p}_{1}\cdot\vec{n})^{2} \vec{n} \big) \nn\\ 
&& - \vec{S}_{1}\times\vec{n}\cdot\vec{p}_{2} \big( 155 p_{1}^2 \vec{n} - 176 \vec{p}_{1}\cdot\vec{n} \vec{p}_{1} + 32 ( \vec{p}_{1}\cdot\vec{n})^{2} \vec{n} \big) \nn\\ 
&& + 2 \vec{S}_{1}\times\vec{p}_{1}\cdot\vec{p}_{2} \big( 256 \vec{p}_{1} -241 \vec{p}_{1}\cdot\vec{n} \vec{n} \big) + \vec{S}_{1}\times\vec{n} \big( 1929 \vec{p}_{1}\cdot\vec{n} p_{1}^2 \nn\\ 
&& + 384 p_{1}^2 \vec{p}_{2}\cdot\vec{n} -47 ( \vec{p}_{1}\cdot\vec{n})^{3} - 320 \vec{p}_{2}\cdot\vec{n} ( \vec{p}_{1}\cdot\vec{n})^{2} \big) - \vec{S}_{1}\times\vec{p}_{1} \big( 219 p_{1}^2 - 3136 \vec{p}_{1}\cdot\vec{p}_{2} \nn\\ 
&& + 358 \vec{p}_{1}\cdot\vec{n} \vec{p}_{2}\cdot\vec{n} + 744 ( \vec{p}_{1}\cdot\vec{n})^{2} \big) + 8 \vec{S}_{1}\times\vec{p}_{2} \big( 49 p_{1}^2 + 38 ( \vec{p}_{1}\cdot\vec{n})^{2} \big) \Big] \nn\\ 
&& - 	\frac{G^2}{96 m_{1} r{}^2} \Big[ \vec{S}_{1}\times\vec{n}\cdot\vec{p}_{1} \big( 6300 \vec{p}_{1}\cdot\vec{p}_{2} \vec{n} + 2894 p_{2}^2 \vec{n} - 4262 \vec{p}_{2}\cdot\vec{n} \vec{p}_{1} - 1372 \vec{p}_{1}\cdot\vec{n} \vec{p}_{2} \nn\\ 
&& + 6343 \vec{p}_{2}\cdot\vec{n} \vec{p}_{2} + 2512 \vec{p}_{1}\cdot\vec{n} \vec{p}_{2}\cdot\vec{n} \vec{n} - 111 ( \vec{p}_{2}\cdot\vec{n})^{2} \vec{n} \big) - \vec{S}_{1}\times\vec{n}\cdot\vec{p}_{2} \big( 3384 p_{1}^2 \vec{n} \nn\\ 
&& + 1413 \vec{p}_{1}\cdot\vec{p}_{2} \vec{n} - 3888 \vec{p}_{1}\cdot\vec{n} \vec{p}_{1} + 2332 \vec{p}_{2}\cdot\vec{n} \vec{p}_{1} + 1073 \vec{p}_{1}\cdot\vec{n} \vec{p}_{2} -1960 \vec{p}_{1}\cdot\vec{n} \vec{p}_{2}\cdot\vec{n} \vec{n} \nn\\ 
&& - 224 ( \vec{p}_{1}\cdot\vec{n})^{2} \vec{n} \big) - \vec{S}_{1}\times\vec{p}_{1}\cdot\vec{p}_{2} \big( 4972 \vec{p}_{1} - 1323 \vec{p}_{2} -5458 \vec{p}_{1}\cdot\vec{n} \vec{n} + 1283 \vec{p}_{2}\cdot\vec{n} \vec{n} \big) \nn\\ 
&& + \vec{S}_{1}\times\vec{n} \big( 3282 p_{1}^2 \vec{p}_{2}\cdot\vec{n} - 3105 \vec{p}_{2}\cdot\vec{n} \vec{p}_{1}\cdot\vec{p}_{2} + 28 \vec{p}_{2}\cdot\vec{n} ( \vec{p}_{1}\cdot\vec{n})^{2} \nn\\ 
&& + 1951 \vec{p}_{1}\cdot\vec{n} ( \vec{p}_{2}\cdot\vec{n})^{2} \big) + 2 \vec{S}_{1}\times\vec{p}_{1} \big( 789 \vec{p}_{1}\cdot\vec{p}_{2} + 1034 p_{2}^2 -1993 \vec{p}_{1}\cdot\vec{n} \vec{p}_{2}\cdot\vec{n} \nn\\ 
&& - 108 ( \vec{p}_{2}\cdot\vec{n})^{2} \big) - \vec{S}_{1}\times\vec{p}_{2} \big( 3276 p_{1}^2 - 3429 \vec{p}_{1}\cdot\vec{p}_{2} + 941 \vec{p}_{1}\cdot\vec{n} \vec{p}_{2}\cdot\vec{n} \nn\\ 
&& - 2420 ( \vec{p}_{1}\cdot\vec{n})^{2} \big) \Big] - 	\frac{G^2 m_{1}}{48 m_{2}{}^2 r{}^2} \Big[ 2 \vec{S}_{1}\times\vec{n} \big( 24 \vec{p}_{2}\cdot\vec{n} p_{2}^2 -83 ( \vec{p}_{2}\cdot\vec{n})^{3} \big) \nn\\ 
&& - 15 \vec{S}_{1}\times\vec{p}_{2} \big( 13 p_{2}^2 -15 ( \vec{p}_{2}\cdot\vec{n})^{2} \big) \Big] + 	\frac{G^2}{96 m_{2} r{}^2} \Big[ \vec{S}_{1}\times\vec{n}\cdot\vec{p}_{1} \big( 3561 p_{2}^2 \vec{n} \nn\\ 
&& - 1827 \vec{p}_{2}\cdot\vec{n} \vec{p}_{2} -284 ( \vec{p}_{2}\cdot\vec{n})^{2} \vec{n} \big) - \vec{S}_{1}\times\vec{n}\cdot\vec{p}_{2} \big( 2343 \vec{p}_{1}\cdot\vec{p}_{2} \vec{n} - 984 p_{2}^2 \vec{n} \nn\\ 
&& + 320 \vec{p}_{2}\cdot\vec{n} \vec{p}_{1} - 935 \vec{p}_{1}\cdot\vec{n} \vec{p}_{2} - 3180 \vec{p}_{2}\cdot\vec{n} \vec{p}_{2} -560 \vec{p}_{1}\cdot\vec{n} \vec{p}_{2}\cdot\vec{n} \vec{n} - 900 ( \vec{p}_{2}\cdot\vec{n})^{2} \vec{n} \big) \nn\\ 
&& - \vec{S}_{1}\times\vec{p}_{1}\cdot\vec{p}_{2} \big( 1459 \vec{p}_{2} -2773 \vec{p}_{2}\cdot\vec{n} \vec{n} \big) + \vec{S}_{1}\times\vec{n} \big( 1713 \vec{p}_{2}\cdot\vec{n} \vec{p}_{1}\cdot\vec{p}_{2} \nn\\ 
&& - 3057 \vec{p}_{2}\cdot\vec{n} p_{2}^2 -1453 \vec{p}_{1}\cdot\vec{n} ( \vec{p}_{2}\cdot\vec{n})^{2} + 583 ( \vec{p}_{2}\cdot\vec{n})^{3} \big) \nn\\ 
&& + 2 \vec{S}_{1}\times\vec{p}_{1} \big( 516 p_{2}^2 -631 ( \vec{p}_{2}\cdot\vec{n})^{2} \big) - \vec{S}_{1}\times\vec{p}_{2} \big( 2805 \vec{p}_{1}\cdot\vec{p}_{2} \nn\\ 
&& - 2589 p_{2}^2 -1313 \vec{p}_{1}\cdot\vec{n} \vec{p}_{2}\cdot\vec{n} + 777 ( \vec{p}_{2}\cdot\vec{n})^{2} \big) \Big] - 	\frac{G^3 m_{2}{}^3}{1800 m_{1} r{}^3} \Big[ 17349 \vec{S}_{1}\times\vec{n}\cdot\vec{p}_{1} \vec{n} \nn\\ 
&& + 42474 \vec{S}_{1}\times\vec{n} \vec{p}_{1}\cdot\vec{n} - 15950 \vec{S}_{1}\times\vec{p}_{1} \Big] + 	\frac{G^3 m_{1}{}^2}{720 r{}^3} \Big[ 9744 \vec{S}_{1}\times\vec{n}\cdot\vec{p}_{2} \vec{n} \nn\\ 
&& + 13449 \vec{S}_{1}\times\vec{n} \vec{p}_{2}\cdot\vec{n} - 9665 \vec{S}_{1}\times\vec{p}_{2} \Big] - 	\frac{G^3 m_{2}{}^2}{14400 r{}^3} \Big[ 150 \vec{S}_{1}\times\vec{n}\cdot\vec{p}_{1} {(479 - 657 \pi^2)} \vec{n} \nn\\ 
&& + 484176 \vec{S}_{1}\times\vec{n}\cdot\vec{p}_{2} \vec{n} + 3 \vec{S}_{1}\times\vec{n} \big( {(182400 - 20025 \pi^2)} \vec{p}_{1}\cdot\vec{n} + 101392 \vec{p}_{2}\cdot\vec{n} \big) \nn\\ 
&& - 25 {(21320 + 513 \pi^2)} \vec{S}_{1}\times\vec{p}_{1} + 151800 \vec{S}_{1}\times\vec{p}_{2} \Big] - 	\frac{G^3 m_{1} m_{2}}{14400 r{}^3} \Big[ 667716 \vec{S}_{1}\times\vec{n}\cdot\vec{p}_{1} \vec{n} \nn\\ 
&& + 150 \vec{S}_{1}\times\vec{n}\cdot\vec{p}_{2} {(1001 - 639 \pi^2)} \vec{n} + 3 \vec{S}_{1}\times\vec{n} \big( 239372 \vec{p}_{1}\cdot\vec{n} \nn\\ 
&& - {(107500 + 44775 \pi^2)} \vec{p}_{2}\cdot\vec{n} \big) - 226500 \vec{S}_{1}\times\vec{p}_{1} + 25 {(21896 + 513 \pi^2)} \vec{S}_{1}\times\vec{p}_{2} \Big]. \nn\\
\eea

\bibliographystyle{jhep}
\bibliography{gwbibtex}

\providecommand{\href}[2]{#2}\begingroup\raggedright\begin{thebibliography}{10}

\bibitem{LIGOScientific:2018mvr}
{\scshape LIGO Scientific, Virgo} collaboration, B.~P. Abbott et~al.,
  \emph{{GWTC-1: A Gravitational-Wave Transient Catalog of Compact Binary
  Mergers Observed by LIGO and Virgo during the First and Second Observing
  Runs}}, \href{https://doi.org/10.1103/PhysRevX.9.031040}{\emph{Phys. Rev.}
  {\bfseries X9} (2019) 031040}
  [\href{https://arxiv.org/abs/1811.12907}{{\ttfamily 1811.12907}}].

\bibitem{LIGOScientific:2020ibl}
{\scshape LIGO Scientific, Virgo} collaboration, R.~Abbott et~al.,
  \emph{{GWTC-2: Compact Binary Coalescences Observed by LIGO and Virgo During
  the First Half of the Third Observing Run}},
  \href{https://doi.org/10.1103/PhysRevX.11.021053}{\emph{Phys. Rev. X}
  {\bfseries 11} (2021) 021053}
  [\href{https://arxiv.org/abs/2010.14527}{{\ttfamily 2010.14527}}].

\bibitem{LIGOScientific:2021djp}
{\scshape LIGO Scientific, VIRGO, KAGRA} collaboration, R.~Abbott et~al.,
  \emph{{GWTC-3: Compact Binary Coalescences Observed by LIGO and Virgo During
  the Second Part of the Third Observing Run}},
  \href{https://arxiv.org/abs/2111.03606}{{\ttfamily 2111.03606}}.

\bibitem{LIGOScientific:2014pky}
{\scshape LIGO Scientific} collaboration, J.~Aasi et~al., \emph{{Advanced
  LIGO}}, \href{https://doi.org/10.1088/0264-9381/32/7/074001}{\emph{Class.
  Quant. Grav.} {\bfseries 32} (2015) 074001}
  [\href{https://arxiv.org/abs/1411.4547}{{\ttfamily 1411.4547}}].

\bibitem{VIRGO:2014yos}
{\scshape VIRGO} collaboration, F.~Acernese et~al., \emph{{Advanced Virgo: a
  second-generation interferometric gravitational wave detector}},
  \href{https://doi.org/10.1088/0264-9381/32/2/024001}{\emph{Class. Quant.
  Grav.} {\bfseries 32} (2015) 024001}
  [\href{https://arxiv.org/abs/1408.3978}{{\ttfamily 1408.3978}}].

\bibitem{KAGRA:2020tym}
{\scshape KAGRA} collaboration, T.~Akutsu et~al., \emph{{Overview of KAGRA:
  Detector design and construction history}},
  \href{https://arxiv.org/abs/2005.05574}{{\ttfamily 2005.05574}}.

\bibitem{Abbott:2016blz}
{\scshape LIGO, VIRGO} collaboration, B.~P. Abbott et~al., \emph{{Observation
  of Gravitational Waves from a Binary Black Hole Merger}},
  \href{https://doi.org/10.1103/PhysRevLett.116.061102}{\emph{Phys. Rev. Lett.}
  {\bfseries 116} (2016) 061102}
  [\href{https://arxiv.org/abs/1602.03837}{{\ttfamily 1602.03837}}].

\bibitem{TheLIGOScientific:2017qsa}
{\scshape Virgo, LIGO Scientific} collaboration, B.~Abbott et~al.,
  \emph{{GW170817: Observation of Gravitational Waves from a Binary Neutron
  Star Inspiral}},
  \href{https://doi.org/10.1103/PhysRevLett.119.161101}{\emph{Phys. Rev. Lett.}
  {\bfseries 119} (2017) 161101}
  [\href{https://arxiv.org/abs/1710.05832}{{\ttfamily 1710.05832}}].

\bibitem{LIGOScientific:2021qlt}
{\scshape LIGO Scientific, KAGRA, VIRGO} collaboration, R.~Abbott et~al.,
  \emph{{Observation of Gravitational Waves from Two Neutron
  Star\textendash{}Black Hole Coalescences}},
  \href{https://doi.org/10.3847/2041-8213/ac082e}{\emph{Astrophys. J. Lett.}
  {\bfseries 915} (2021) L5}
  [\href{https://arxiv.org/abs/2106.15163}{{\ttfamily 2106.15163}}].

\bibitem{Blanchet:2013haa}
L.~Blanchet, \emph{{Gravitational Radiation from Post-Newtonian Sources and
  Inspiralling Compact Binaries}},
  \href{https://doi.org/10.12942/lrr-2014-2}{\emph{Living Rev. Rel.} {\bfseries
  17} (2014) 2} [\href{https://arxiv.org/abs/1310.1528}{{\ttfamily
  1310.1528}}].

\bibitem{Buonanno:1998gg}
A.~Buonanno and T.~Damour, \emph{{Effective one-body approach to general
  relativistic two-body dynamics}},
  \href{https://doi.org/10.1103/PhysRevD.59.084006}{\emph{Phys.Rev.} {\bfseries
  D59} (1999) 084006} [\href{https://arxiv.org/abs/gr-qc/9811091}{{\ttfamily
  gr-qc/9811091}}].

\bibitem{Bini:2019nra}
D.~Bini, T.~Damour and A.~Geralico, \emph{{Novel approach to binary dynamics:
  application to the fifth post-Newtonian level}},
  \href{https://doi.org/10.1103/PhysRevLett.123.231104}{\emph{Phys. Rev. Lett.}
  {\bfseries 123} (2019) 231104}
  [\href{https://arxiv.org/abs/1909.02375}{{\ttfamily 1909.02375}}].

\bibitem{Bini:2020wpo}
D.~Bini, T.~Damour and A.~Geralico, \emph{{Binary dynamics at the fifth and
  fifth-and-a-half post-Newtonian orders}},
  \href{https://arxiv.org/abs/2003.11891}{{\ttfamily 2003.11891}}.

\bibitem{Bini:2020uiq}
D.~Bini, T.~Damour, A.~Geralico, S.~Laporta and P.~Mastrolia,
  \emph{{Gravitational dynamics at $O(G^6)$: perturbative gravitational
  scattering meets experimental mathematics}},
  \href{https://arxiv.org/abs/2008.09389}{{\ttfamily 2008.09389}}.

\bibitem{Goldberger:2004jt}
W.~D. Goldberger and I.~Z. Rothstein, \emph{{An Effective field theory of
  gravity for extended objects}},
  \href{https://doi.org/10.1103/PhysRevD.73.104029}{\emph{Phys.Rev.} {\bfseries
  D73} (2006) 104029} [\href{https://arxiv.org/abs/hep-th/0409156}{{\ttfamily
  hep-th/0409156}}].

\bibitem{Blumlein:2020pyo}
J.~Bl\"umlein, A.~Maier, P.~Marquard and G.~Sch\"afer, \emph{{The fifth-order
  post-Newtonian Hamiltonian dynamics of two-body systems from an effective
  field theory approach: potential contributions}},
  \href{https://doi.org/10.1016/j.nuclphysb.2021.115352}{\emph{Nucl. Phys. B}
  {\bfseries 965} (2021) 115352}
  [\href{https://arxiv.org/abs/2010.13672}{{\ttfamily 2010.13672}}].

\bibitem{Goldberger:2022ebt}
W.~D. Goldberger, \emph{{Effective field theories of gravity and compact binary
  dynamics: A Snowmass 2021 whitepaper}},  in \emph{{2022 Snowmass Summer
  Study}}, 6, 2022, \href{https://arxiv.org/abs/2206.14249}{{\ttfamily
  2206.14249}}.

\bibitem{Antonelli:2020aeb}
A.~Antonelli, C.~Kavanagh, M.~Khalil, J.~Steinhoff and J.~Vines,
  \emph{{Gravitational spin-orbit coupling through third-subleading
  post-Newtonian order: from first-order self-force to arbitrary mass ratios}},
  \href{https://doi.org/10.1103/PhysRevLett.125.011103}{\emph{Phys. Rev. Lett.}
  {\bfseries 125} (2020) 011103}
  [\href{https://arxiv.org/abs/2003.11391}{{\ttfamily 2003.11391}}].

\bibitem{Antonelli:2020ybz}
A.~Antonelli, C.~Kavanagh, M.~Khalil, J.~Steinhoff and J.~Vines,
  \emph{{Gravitational spin-orbit and aligned spin$_1$-spin$_2$ couplings
  through third-subleading post-Newtonian orders}},
  \href{https://doi.org/10.1103/PhysRevD.102.124024}{\emph{Phys. Rev. D}
  {\bfseries 102} (2020) 124024}
  [\href{https://arxiv.org/abs/2010.02018}{{\ttfamily 2010.02018}}].

\bibitem{Levi:2015msa}
M.~Levi and J.~Steinhoff, \emph{{Spinning gravitating objects in the effective
  field theory in the post-Newtonian scheme}},
  \href{https://doi.org/10.1007/JHEP09(2015)219}{\emph{JHEP} {\bfseries 09}
  (2015) 219} [\href{https://arxiv.org/abs/1501.04956}{{\ttfamily
  1501.04956}}].

\bibitem{Levi:2017kzq}
M.~Levi and J.~Steinhoff, \emph{{EFTofPNG: A package for high precision
  computation with the Effective Field Theory of Post-Newtonian Gravity}},
  \href{https://doi.org/10.1088/1361-6382/aa941e}{\emph{Class. Quant. Grav.}
  {\bfseries 34} (2017) 244001}
  [\href{https://arxiv.org/abs/1705.06309}{{\ttfamily 1705.06309}}].

\bibitem{Levi:2020kvb}
M.~Levi, A.~J. Mcleod and M.~Von~Hippel, \emph{{N$^3$LO gravitational
  spin-orbit coupling at order $G^4$}},
  \href{https://doi.org/10.1007/JHEP07(2021)115}{\emph{JHEP} {\bfseries 07}
  (2021) 115} [\href{https://arxiv.org/abs/2003.02827}{{\ttfamily
  2003.02827}}].

\bibitem{Kim:2022pou}
J.-W. Kim, M.~Levi and Z.~Yin, \emph{{N$^3$LO Spin-Orbit Interaction via the
  EFT of Spinning Gravitating Objects}},
  \href{https://arxiv.org/abs/2208.14949}{{\ttfamily 2208.14949}}.

\bibitem{Mandal:2022nty}
M.~K. Mandal, P.~Mastrolia, R.~Patil and J.~Steinhoff, \emph{{Gravitational
  Spin-Orbit Hamiltonian at NNNLO in the post-Newtonian framework}},
  \href{https://arxiv.org/abs/2209.00611}{{\ttfamily 2209.00611}}.

\bibitem{Barker:1975ae}
B.~Barker and R.~O'Connell, \emph{{Gravitational Two-Body Problem with
  Arbitrary Masses, Spins, and Quadrupole Moments}},
  \href{https://doi.org/10.1103/PhysRevD.12.329}{\emph{Phys.Rev.} {\bfseries
  D12} (1975) 329}.

\bibitem{Levi:2014gsa}
M.~Levi and J.~Steinhoff, \emph{{Leading order finite size effects with spins
  for inspiralling compact binaries}},
  \href{https://doi.org/10.1007/JHEP06(2015)059}{\emph{JHEP} {\bfseries 06}
  (2015) 059} [\href{https://arxiv.org/abs/1410.2601}{{\ttfamily 1410.2601}}].

\bibitem{Bekaert:2022poo}
X.~Bekaert, N.~Boulanger, A.~Campoleoni, M.~Chiodaroli, D.~Francia,
  M.~Grigoriev et~al., \emph{{Snowmass White Paper: Higher Spin Gravity and
  Higher Spin symmetry}},  \href{https://arxiv.org/abs/2205.01567}{{\ttfamily
  2205.01567}}.

\bibitem{Levi:2019kgk}
M.~Levi, S.~Mougiakakos and M.~Vieira, \emph{{Gravitational cubic-in-spin
  interaction at the next-to-leading post-Newtonian order}},
  \href{https://doi.org/10.1007/JHEP01(2021)036}{\emph{JHEP} {\bfseries 01}
  (2021) 036} [\href{https://arxiv.org/abs/1912.06276}{{\ttfamily
  1912.06276}}].

\bibitem{Levi:2018nxp}
M.~Levi, \emph{{Effective Field Theories of Post-Newtonian Gravity: A
  comprehensive review}},
  \href{https://doi.org/10.1088/1361-6633/ab12bc}{\emph{Rept. Prog. Phys.}
  {\bfseries 83} (2020) 075901}
  [\href{https://arxiv.org/abs/1807.01699}{{\ttfamily 1807.01699}}].

\bibitem{Levi:2011eq}
M.~Levi, \emph{{Binary dynamics from spin1-spin2 coupling at fourth
  post-Newtonian order}},
  \href{https://doi.org/10.1103/PhysRevD.85.064043}{\emph{Phys.Rev.} {\bfseries
  D85} (2012) 064043} [\href{https://arxiv.org/abs/1107.4322}{{\ttfamily
  1107.4322}}].

\bibitem{Levi:2015uxa}
M.~Levi and J.~Steinhoff, \emph{{Next-to-next-to-leading order gravitational
  spin-orbit coupling via the effective field theory for spinning objects in
  the post-Newtonian scheme}},
  \href{https://doi.org/10.1088/1475-7516/2016/01/011}{\emph{JCAP} {\bfseries
  1601} (2016) 011} [\href{https://arxiv.org/abs/1506.05056}{{\ttfamily
  1506.05056}}].

\bibitem{Levi:2015ixa}
M.~Levi and J.~Steinhoff, \emph{{Next-to-next-to-leading order gravitational
  spin-squared potential via the effective field theory for spinning objects in
  the post-Newtonian scheme}},
  \href{https://doi.org/10.1088/1475-7516/2016/01/008}{\emph{JCAP} {\bfseries
  1601} (2016) 008} [\href{https://arxiv.org/abs/1506.05794}{{\ttfamily
  1506.05794}}].

\bibitem{Levi:2016ofk}
M.~Levi and J.~Steinhoff, \emph{{Complete conservative dynamics for
  inspiralling compact binaries with spins at the fourth post-Newtonian
  order}}, \href{https://doi.org/10.1088/1475-7516/2021/09/029}{\emph{JCAP}
  {\bfseries 09} (2021) 029}
  [\href{https://arxiv.org/abs/1607.04252}{{\ttfamily 1607.04252}}].

\bibitem{Levi:2020uwu}
M.~Levi, A.~J. Mcleod and M.~Von~Hippel, \emph{{N$^{3}$LO gravitational
  quadratic-in-spin interactions at G$^{4}$}},
  \href{https://doi.org/10.1007/JHEP07(2021)116}{\emph{JHEP} {\bfseries 07}
  (2021) 116} [\href{https://arxiv.org/abs/2003.07890}{{\ttfamily
  2003.07890}}].

\bibitem{Kim:2021rfj}
J.-W. Kim, M.~Levi and Z.~Yin, \emph{{Quadratic-in-spin interactions at fifth
  post-Newtonian order probe new physics}},
  \href{https://doi.org/10.1016/j.physletb.2022.137410}{\emph{Phys. Lett. B}
  {\bfseries 834} (2022) 137410}
  [\href{https://arxiv.org/abs/2112.01509}{{\ttfamily 2112.01509}}].

\bibitem{Kim:2022bwv}
J.-W. Kim, M.~Levi and Z.~Yin, \emph{{N$^3$LO Quadratic-in-Spin Interactions
  for Generic Compact Binaries}},
  \href{https://arxiv.org/abs/2209.09235}{{\ttfamily 2209.09235}}.

\bibitem{Levi:2022rrq}
M.~Levi and Z.~Yin, \emph{{Completing the Fifth PN Precision Frontier via the
  EFT of Spinning Gravitating Objects}},
  \href{https://arxiv.org/abs/2211.14018}{{\ttfamily 2211.14018}}.

\bibitem{Mandal:2022ufb}
M.~K. Mandal, P.~Mastrolia, R.~Patil and J.~Steinhoff, \emph{{Gravitational
  Quadratic-in-Spin Hamiltonian at NNNLO in the post-Newtonian framework}},
  \href{https://arxiv.org/abs/2210.09176}{{\ttfamily 2210.09176}}.

\bibitem{Guevara:2018wpp}
A.~Guevara, A.~Ochirov and J.~Vines, \emph{{Scattering of Spinning Black Holes
  from Exponentiated Soft Factors}},
  \href{https://doi.org/10.1007/JHEP09(2019)056}{\emph{JHEP} {\bfseries 09}
  (2019) 056} [\href{https://arxiv.org/abs/1812.06895}{{\ttfamily
  1812.06895}}].

\bibitem{Chen:2021qkk}
W.-M. Chen, M.-Z. Chung, Y.-t. Huang and J.-W. Kim, \emph{{The 2PM Hamiltonian
  for binary Kerr to quartic in spin}},
  \href{https://arxiv.org/abs/2111.13639}{{\ttfamily 2111.13639}}.

\bibitem{Bern:2022kto}
Z.~Bern, D.~Kosmopoulos, A.~Luna, R.~Roiban and F.~Teng, \emph{{Binary Dynamics
  Through the Fifth Power of Spin at $\mathcal{O}(G^2)$}},
  \href{https://arxiv.org/abs/2203.06202}{{\ttfamily 2203.06202}}.

\bibitem{Levi:2020lfn}
M.~Levi and F.~Teng, \emph{{NLO gravitational quartic-in-spin interaction}},
  \href{https://doi.org/10.1007/JHEP01(2021)066}{\emph{JHEP} {\bfseries 01}
  (2021) 066} [\href{https://arxiv.org/abs/2008.12280}{{\ttfamily
  2008.12280}}].

\bibitem{Hanson:1974qy}
A.~J. Hanson and T.~Regge, \emph{{The Relativistic Spherical Top}},
  \href{https://doi.org/10.1016/0003-4916(74)90046-3}{\emph{Annals Phys.}
  {\bfseries 87} (1974) 498}.

\bibitem{Bailey:1975fe}
I.~Bailey and W.~Israel, \emph{{Lagrangian Dynamics of Spinning Particles and
  Polarized Media in General Relativity}},
  \href{https://doi.org/10.1007/BF01609434}{\emph{Commun.Math.Phys.} {\bfseries
  42} (1975) 65}.

\bibitem{Porto:2005ac}
R.~A. Porto, \emph{{Post-Newtonian corrections to the motion of spinning bodies
  in NRGR}}, \href{https://doi.org/10.1103/PhysRevD.73.104031}{\emph{Phys.Rev.}
  {\bfseries D73} (2006) 104031}
  [\href{https://arxiv.org/abs/gr-qc/0511061}{{\ttfamily gr-qc/0511061}}].

\bibitem{Tulczyjew:1959b}
W.~Tulczyjew, \emph{{Motion of multipole particles in general relativity
  theory}}, {\emph{Acta Phys.Polon.} {\bfseries 18} (1959) 393}.

\bibitem{Yee:1993ya}
K.~Yee and M.~Bander, \emph{{Equations of motion for spinning particles in
  external electromagnetic and gravitational fields}},
  \href{https://doi.org/10.1103/PhysRevD.48.2797}{\emph{Phys.Rev.} {\bfseries
  D48} (1993) 2797} [\href{https://arxiv.org/abs/hep-th/9302117}{{\ttfamily
  hep-th/9302117}}].

\bibitem{Porto:2008jj}
R.~A. Porto and I.~Z. Rothstein, \emph{{Next to Leading Order Spin(1)Spin(1)
  Effects in the Motion of Inspiralling Compact Binaries}},
  \href{https://doi.org/10.1103/PhysRevD.78.044013}{\emph{Phys.Rev.} {\bfseries
  D78} (2008) 044013} [\href{https://arxiv.org/abs/0804.0260}{{\ttfamily
  0804.0260}}].

\bibitem{Levi:2010zu}
M.~Levi, \emph{{Next to Leading Order gravitational Spin-Orbit coupling in an
  Effective Field Theory approach}},
  \href{https://doi.org/10.1103/PhysRevD.82.104004}{\emph{Phys.Rev.} {\bfseries
  D82} (2010) 104004} [\href{https://arxiv.org/abs/1006.4139}{{\ttfamily
  1006.4139}}].

\bibitem{Levi:2008nh}
M.~Levi, \emph{{Next to Leading Order gravitational Spin1-Spin2 coupling with
  Kaluza-Klein reduction}},
  \href{https://doi.org/10.1103/PhysRevD.82.064029}{\emph{Phys.Rev.} {\bfseries
  D82} (2010) 064029} [\href{https://arxiv.org/abs/0802.1508}{{\ttfamily
  0802.1508}}].

\bibitem{Pryce:1948pf}
M.~Pryce, \emph{{The Mass center in the restricted theory of relativity and its
  connection with the quantum theory of elementary particles}},
  \href{https://doi.org/10.1098/rspa.1948.0103}{\emph{Proc.Roy.Soc.Lond.}
  {\bfseries A195} (1948) 62}.

\bibitem{Newton:1949cq}
T.~Newton and E.~P. Wigner, \emph{{Localized States for Elementary Systems}},
  \href{https://doi.org/10.1103/RevModPhys.21.400}{\emph{Rev.Mod.Phys.}
  {\bfseries 21} (1949) 400}.

\bibitem{Levi:2014sba}
M.~Levi and J.~Steinhoff, \emph{{Equivalence of ADM Hamiltonian and Effective
  Field Theory approaches at next-to-next-to-leading order spin1-spin2 coupling
  of binary inspirals}},
  \href{https://doi.org/10.1088/1475-7516/2014/12/003}{\emph{JCAP} {\bfseries
  1412} (2014) 003} [\href{https://arxiv.org/abs/1408.5762}{{\ttfamily
  1408.5762}}].

\bibitem{Morales:2021}
R.~Morales, \emph{{High-Precision Gravity Observables: From EFTs to Particle
  Amplitudes}},  Master's thesis, U.~of Copenhagen, May 2021.

\bibitem{Bel:1980}
L.~Bel and J.~Martin, \emph{{Predictive relativistic mechanics of systems of N
  particles with spin}}, {\emph{Ann.~Inst.~H.~Poincar\'e Phys.~Th\'eor.}
  {\bfseries 33} (1980) 409}.

\bibitem{Edison:2022cdu}
A.~Edison and M.~Levi, \emph{{A tale of tails through generalized unitarity}},
  \href{https://doi.org/10.1016/j.physletb.2022.137634}{\emph{Phys. Lett. B}
  {\bfseries 837} (2023) 137634}
  [\href{https://arxiv.org/abs/2202.04674}{{\ttfamily 2202.04674}}].

\bibitem{Damour:1988mr}
T.~Damour and G.~Sch{\"a}fer, \emph{{Higher Order Relativistic Periastron
  Advances and Binary Pulsars}},
  \href{https://doi.org/10.1007/BF02828697}{\emph{Nuovo Cim.} {\bfseries B101}
  (1988) 127}.

\end{thebibliography}\endgroup

\end{document}